\documentclass[onecolumn,showpacs,superscriptaddress,preprintnumbers,nofootinbib,notitlepage]{revtex4}

\usepackage{bm}
\usepackage{feynmp}
\usepackage{simplewick}
\usepackage{graphics,graphicx}
\usepackage{amsmath,amssymb}
\usepackage{color}
\usepackage{array,mathtools}
\usepackage{booktabs,multirow}
\usepackage{bbold}
\usepackage[normalem]{ulem}

\newcommand{\bea}{\begin{eqnarray}}
\newcommand{\eea}{\end{eqnarray}}
\newcommand{\GeV}{\mathrm{GeV}}
\newcommand{\beq}{\begin{equation}}
\newcommand{\stat}{\mathrm{stat}}
\newcommand{\syst}{\mathrm{syst}}
\newcommand{\fm}{\rm fm}
\newcommand{\eeq}{\end{equation}}
\newcommand{\GEVP}{\mathrm{GEVP}}

\newcommand{\MeV}{{\rm MeV}}
\makeatother

\begin{document}
\preprint{\tt LPT Orsay 16-18}
\preprint{\tt MITP/16-033}

\vspace*{22mm}

\title{Density distributions in the $B$ meson}

\author{Beno\^it~Blossier}
\affiliation{Laboratoire de Physique Th\'eorique\footnote[1]{Unit\'e Mixte de Recherche 8627 du Centre National de la Recherche Scientifique}, CNRS, Univ. Paris-Sud et Universit\'e Paris-Saclay,  B\^atiment 210,  91405 Orsay Cedex, France}
\author{Antoine~G\'erardin}
\affiliation{Laboratoire de Physique Corpusculaire de Clermont-Ferrand\footnote[3]{Unit\'e Mixte de Recherche 6533 CNRS/IN2P3 -- Universit\'e Blaise Pascal}, Campus des C\'ezeaux, 24 avenue des Landais, BP 80026, 63171 Aubi\`ere Cedex, France}
\affiliation{PRISMA Cluster of Excellence and Institut fur Kernphysik, University of Mainz, Becher-Weg 45, 55099 Mainz, Germany}

\begin{abstract}
We report on a two-flavor lattice QCD study of the axial, charge and matter distributions of the $B$ meson and its first radial excitation. As our framework is the static limit of Heavy Quark Effective Theory (HQET), taking their Fourier transform gives access to several form factors at the kinematical point $q^2=0$ while our previous computations in that framework were performed at 
$q^2_{\rm max}$. Moreover they provide some useful information on the nature of an excited state, i.e. a radial excitation of a quark-antiquark bound state or a multihadron state.

\end{abstract}

\maketitle

\section{\label{Introduction}Introduction}

\begin{figure}[t]
	\begin{minipage}[c]{0.49\linewidth}
	\centering 
	\includegraphics*[width=4.5cm]{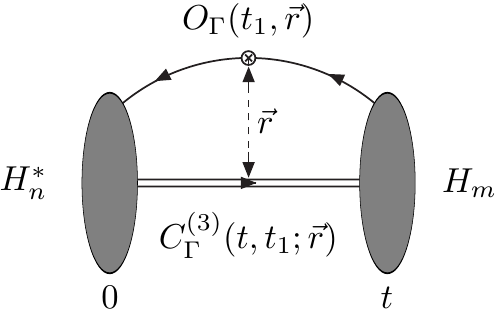}
	\end{minipage}
	\caption{\label{figdensity}Three-point correlation function computed to extract the density distribution $f_{\Gamma}(\vec{r})$.}	
\label{figA0}
\end{figure}

In experiments a significant amount of events are produced with excited hadronic states that decay strongly into ground states. They complicate the data analysis, cuts are introduced on invariant masses, tail of distributions are sometimes modelized using empirical recipes: one may wonder whether the theory can bring any help. Form factors $f^{B \to \pi}_+$ and $f^{D \to \pi}_+$ of the semileptonic decays $B \to \pi l \nu$ and $D \to \pi l \nu$ have received a lot of attention as those processes are used to extract the CKM matrix elements $V_{ub}$ and $V_{cd}$. The most popular parametrizations in the literature \cite{BecirevicKT, BoydCF, BoydSQ, BoydKZ, CapriniMU} are based on the scaling laws of Heavy Quark Effective Theory (HQET) and a unitarity argument. More recently, a three-pole parametrization of $f^{D \to \pi}_+$ revealed to match well with data \cite{BecirevicKAA}: $f^{D\to \pi}_+(q^2)=\frac{\gamma_0}{m^2_0-q^2} + \frac{\gamma_1}{m^2_1-q^2} + \frac{\gamma_{\rm eff}}{m^2_{\rm eff}-q^2}$ where $\gamma_0$ and $\gamma_1$ are proportional to the couplings $g_{D^*D\pi}$ and $g_{D^{*\prime} D\pi}$, respectively, in addition to a ``superconvergence" constraint $\sum_n \gamma_n=0$. Assuming the smoothness of results in $1/m_b$, $1/m_c$, it is tempting to test the hypothesis of a negative $g_{D^{*\prime} D \pi}$ versus its counterpart in the $B$ sector, $g_{B^{*\prime} B \pi}$. In a previous paper \cite{BlossierQMA} a first step was followed in that direction, in the static limit of HQET. But our computation was done at the kinematical point $q^2_{\rm max}$: \textit{the final target is at $q^2=0$}. To do this, an elegant technique is to measure the density distributions $f^{B B^{*\prime}}_{\Gamma}(r)$  \cite{BecirevicYA} where $r$ is the distance between the light-light current and the static quark, as sketched on Fig.~\ref{figdensity}. Their Fourier transform gives the corresponding form factors at every $q^2$, which is the subject of our present work. The plan of the paper is the following: in Section \ref{sec2} we describe our analysis method, our results are presented in Section \ref{sec3} and we explain in Section \ref{sec4} how an unexpected coupling of a $\bar{q} b$ interpolating field to a multihadron state is presumably observed on distributions.

\section{\label{sec2}Extraction of the pion couplings at $q^2=0$}

The $g_{H_n^{*} H^{\phantom{}}_m \pi}$ coupling is defined by the following matrix element 
\begin{equation}
\langle H^{\phantom{}}_m(p) \pi(q)|H_n^{*}(p^{\prime},\lambda) \rangle=g_{H_n^{*} H^{\phantom{}}_m \pi} \  q \cdot \epsilon(p^{\prime},\lambda) \,.
\end{equation}
where $\epsilon(p^{\prime}, \lambda)$ is the polarization vector of the vector meson, $q=p^{\prime}-p$ is the transferred momentum and $H_m$ is the $m^{\rm th}$ radially excited state of a pseudoscalar heavy-light meson while $H_n^{*}$ is the $n^{\rm th}$ radially excited state of the vector heavy-light meson. The transition amplitude of interest is parametrized by 
\begin{multline}
\langle H^{\phantom{}}_m(p) |{\cal A}^{\mu}| H_n^{*}(p^{\prime},\lambda) \rangle = 2m_{H_n^{*}}A^{mn}_0(q^2) \frac{\epsilon(p^{\prime},\lambda)\cdot q}{q^2} q^\mu +(m_{H^{\phantom{}}_m} + m_{H_n^*}) A^{mn}_1(q^2)\left(\epsilon^{\mu}(p^{\prime},\lambda) -  \frac{\epsilon(p^{\prime},\lambda) \cdot q}{q^2} \, q^\mu \right) \\
+A^{mn}_2(q^2) \frac{\epsilon(p^{\prime},\lambda) \cdot q}{m_{H^{\phantom{}}_m} + m_{H_n^*}} \left[(p+p^{\prime})^\mu + \frac{m^2_{H^{\phantom{}}_m} - m^2_{H_n^*}}{q^2}q^\mu \right] \,,
\end{multline}
with ${\cal A}^{\mu}(x) = \overline{d}(x) \gamma^\mu \gamma^5 u(x)$. Taking the divergence of the current $q_\mu {\cal A}^\mu$ we are left with
\begin{equation}
\langle H^{\phantom{}}_m(p) |q_{\mu} {\cal A}^{\mu} | H_n^{*}(p^{\prime},\lambda) \rangle = 2 \, m_{H_n^{*}} \, A^{mn}_0(q^2) \ q \cdot \epsilon(p^{\prime}, \lambda) \,.
\end{equation}
Then, using the Partially Conserved Axial Current (PCAC) relation and the LSZ reduction formula, the $g_{H_n^{*} H^{\phantom{}}_m \pi}$ coupling is obtained from the form factor $A^{mn}_0$ at $q^2=0$
\begin{equation}
g_{H_n^{*} H^{\phantom{}}_m \pi}    =  \frac{ 2 \, m_{H_n^{*}}A^{mn}_0(0)}{f_{\pi}}  \,,
\end{equation}
where $f_{\pi} = 130.4~\MeV$ is the pion decay constant. Finally, in the vector meson rest frame, the form factor is given by the matrix element\footnote{ We use the relation $\sum_{\lambda} \epsilon_{\mu}(k,\lambda) \, \epsilon^{*}_{\nu}(k,\lambda) =  - g_{\mu\nu} + \frac{k_{\mu}k_{\nu}}{m^2}$ }
\begin{equation}
A^{mn}_0(q^2) =  - \sum_{\lambda} \frac{  \langle H^{\phantom{}}_m(p) |q_{\mu} {\cal A}^{\mu} | H_n^{*}(p^{\prime},\lambda) \rangle  }{ 2 m_{H_n^*} \, q_i}  \, \epsilon^{*}_i(p^{\prime}, \lambda)  \quad {\rm (no\ sum\ over\ }i) \,.
\label{eq:formfac}
\end{equation}
However, in lattice simulations with static heavy quarks, only the kinematical point $q^2=q^2_{\rm max}=\Delta^2$, where $\Delta=m_{H_n^*}-m_{H^{\phantom{}}_m}$, is directly accessible. To extract the form factor at $q^2=0$ we follow the ideas presented in \cite{BecirevicYA} and compute the \emph{axial density distributions} $f_{\gamma_{i} \gamma_5}^{(mn)}(\vec{r})$ where the axial current, acting like a probe, is inserted at a distance $r$ from the static heavy quark. Within our conventions, distributions are written using covariant indices and their exact definition is given in next lines. Finally, in the static limit of HQET, the form factor of interest at $q^2=0$ is obtained by taking the Fourier transform of the axial density distributions with a spatial momentum $|q_i| = \Delta$ in the direction $i$ 
\begin{equation}
A^{mn}_0(0) = - \frac{q_0}{q_i} \int d^3 r\, f_{\gamma_{0} \gamma_5}^{(mn)}(\vec{r}) \, e^{i \vec{q}\cdot {\vec r}} + \int d^3 r \, f_{\gamma_{i} \gamma_5}^{(mn)}(\vec{r}) \, e^{i \vec{q}\cdot {\vec r}} \quad {\rm (no\ sum\ over\ }i)\,.
\end{equation}

\noindent On the lattice of spatial volume $V$, we are interested in the $N \times N$ matrices of two-point and three-point correlation functions 
\begin{align}
C_{\mathcal{P}, ij}^{(2)}(t) &= \frac{1}{V} \langle \, \sum_{\vec{x},\vec{y}} \mathcal{P}^{(i)}(\vec{x},t) \, \mathcal{P}^{(j)\dag}(\vec{y},0) \, \rangle \,, \label{eq:two_point_P}\\
C_{\mathcal{V}, ij}^{(2)}(t) &= \frac{1}{3V} \sum_{k=1}^{3} \langle \, \sum_{\vec{x},\vec{y}} \mathcal{V}_k^{(i)}(\vec{x}, t) \, \mathcal{V}_k^{(j)\dag}(\vec{y}, 0) \, \rangle  \,,  
\label{eq:two_point_V} 
\end{align}
\begin{align}
C_{\gamma_{\mu} \gamma_5,  ij}^{(3)}(t, t_1;\vec{r}) &= \sum_{\vec{y}} \langle \, \mathcal{P}^{(i)}(\vec{y},t) \, \mathcal{A}_{\mu}(\vec{x}+\vec{r},t_1) \, \mathcal{V}_{k}^{(j)\dag}(\vec{x},0) \, \rangle_{\, \vec{x} {\rm \ fixed}} \,, 
\label{eq:three_point}
\end{align}
where the interpolating operators for the pseudoscalar and vector static-light mesons are defined by 
\begin{equation*}
\mathcal{P}^{(i)}(x) = \overline{h}(x) \, \gamma_5 \, d^{(i)}(x) \ \,, \ \mathcal{V}^{(i)}_{k}(x) = \overline{h}(x) \, \Gamma_k \, u^{(i)}(x) \ \,, \  \Gamma_k = \gamma_k\,, \nabla_{k} \,.
\end{equation*}
Here, $\nabla_i$ is the symmetrized covariant derivative and $u^{(i)}$ and $d^{(i)}$ are smeared light quark fields. Since it is important to keep trace of the distance $r$ between the current insertion and the heavy quark line, smearing is only applied in contractions with the heavy quark propagator. Finally, as a consequence of the heavy quark symmetry, the two-point correlation functions (\ref{eq:two_point_P}) and (\ref{eq:two_point_V}) are equal for $\Gamma_k = \gamma_k$. Using the spectral decomposition and the normalization of states $\langle H_{n} | H_{m} \rangle = \delta_{nm}$, the asymptotic behavior of the two-point correlation functions is
\begin{align*}
C_{\mathcal{P}, \, ij}^{(2)}(t) &\xrightarrow[t \gg a ]{} Z^{(i)}_{H_1} \, Z^{(j)}_{H_1}\ e^{-E_{H_1} t}  \,, \\
C_{\mathcal{V}, \, ij}^{(2)}(t) &\xrightarrow[t \gg a ]{} Z^{(i)}_{H_{1}^{*}} \, Z^{(j)}_{H_{1}^{*}} \ e^{-E_{H_{1}^{*}} t} \,.
\end{align*}
The overlap factors $Z_{H_{1}^{(*)}}^{(i)}$ are defined by $Z^{(i)}_P = \langle 0 | \mathcal{P}^{(i)} | P(p) \rangle$ and $Z^{(i)}_V \, \epsilon_{k}(p,\lambda) = \langle 0 | \mathcal{V}^{(i)}_{k} | V(p,\lambda) \rangle$ 
where $P$ is a pseudoscalar meson and $V$ is a vector meson with polarization $\lambda$. Similarly, for the three-point correlation  function (\ref{eq:three_point}), one has 
\begin{equation*}
\sum_{\vec{r}} C_{\gamma_{\mu} \gamma_5, ij}^{(3)}(t, t_1;\vec{r}) \xrightarrow[t \gg t_1 \gg a ]{}  \sum_{\lambda}  Z_H^{(i)} \, Z_{H^*}^{(j)}   \langle \, H | \mathcal{A}_{\mu} | H^{*}(\lambda) \, \rangle \, \epsilon_k(\lambda) \, e^{-E_{H_1} (t-t_1)} \, e^{-E_{H_{1}^*} t_1} \,,
\end{equation*}
and
\begin{equation*}
C_{\gamma_{\mu} \gamma_5, ij}^{(3)}(t, t_1;\vec{r}) = \sum_{n,m} \, Z_{H^{\phantom{}}_m}^{(i)}  \, Z_{H_n^{*}}^{(j)} \,  f^{(mn)}_{\gamma_{\mu} \gamma_5}(\vec{r}) \, e^{-E_{H_m^{\phantom{}}} (t-t_1)} \, e^{-E_{H_n^*} t_1} \,,
\end{equation*}
where $f_{\gamma_{i} \gamma_5}^{(mn)}(\vec{r})$ are the axial density distributions. Therefore, the radial distributions of the axial densities for the ground state pseudoscalar and vector mesons can be extracted from the asymptotic behavior of the ratio
\begin{equation}
\mathcal{R}_{\gamma_{\mu} \gamma_5}(t,t_1,\vec{r}) = \frac{C_{\gamma_{\mu} \gamma_5, ij}^{(3)}(t,t_1;\vec{r})}{ \left( C_{\mathcal{P}, ii}^{(2)}(t) \ C_{\mathcal{V}, jj}^{(2)}(t)  \right)^{1/2} } \  \xrightarrow[t \gg t_1 \gg a ]{} \ a^3 f^{(11)}_{\gamma_{\mu} \gamma_5}(\vec{r})  \,.
\label{eq:ratio}
\end{equation} 
Solving the Generalized Eigenvalue Problems~\cite{MichaelNE, LuscherCK, BlossierKD}
\begin{align}
C_{\mathcal{P}}^{(2)}(t) \, v_n(t,t_0) &= \lambda_n(t,t_0) \, C_{\mathcal{P}}^{(2)}(t_0) \, v_n(t,t_0) \,, \\
C_{\mathcal{V}}^{(2)}(t) \, w_n(t,t_0) &= \widetilde{\lambda}_n(t,t_0) \, C_{\mathcal{V}}^{(2)}(t_0) \, w_n(t,t_0) \,,
\label{eq:gevp}
\end{align}
where $v_n(t,t_0)$, $w_n(t,t_0)$ and $\lambda_n(t,t_0)$, $\widetilde{\lambda}_n(t,t_0)$ are respectively the generalized eigenvectors\footnote{The global phase of the eigenvectors is fixed by imposing the positivity of the decay constants $f_{H_n^{(*)}} = \langle 0 | \mathcal{O}_L | H_n^{(*)} \rangle$ where $\mathcal{O}=\mathcal{P},\mathcal{V}_k$ and $L$ refers to the local interpolating field.} and eigenvalues of the pseudoscalar and vector correlators, and following the method used in \cite{BlossierQMA}, we define the GEVP ratio 
\begin{equation}
\mathcal{R}_{mn}^{\GEVP}(t,t_1;\vec{r}) =  \frac{  \left( v_m(t_2), C_{\gamma_{\mu} \gamma_5}^{(3)}(t_1+t_2,t_1;\vec{r}) \, w_n(t_1) \right) \lambda_m(t_2+a)^{-t_2/(2a)} \, \widetilde{\lambda}_n(t_1+a)^{-t_1/(2a)}}{ \left(v_m(t_2), C_{\mathcal{P}}^{(2)}(t_2) v_m(t_2) \right)^{1/2} \left(w_n(t_1), C_{\mathcal{V}}^{(2)}(t_1) w_n(t_1) \right)^{1/2}  } \,,
\label{eq:Agevp}
\end{equation}
where $t=t_1+t_2$ and the shorthand notations $v_n(t) = v_n(t+a,t)$ and $\lambda_n(t) = \lambda_n(t+a,t)$ are used. For $n=m=1$, this ratio converges to $a^3 f^{(11)}_{\gamma_{\mu} \gamma_5}$ at large time but with a reduced contribution from excited states compared to the previous ratio method. More generally, this method also allows us to extract the radial distributions involving excited states:
\begin{equation*}
\mathcal{R}_{mn}^{\GEVP}(t,t_1;\vec{r}) = a^3  f_{\gamma_{\mu} \gamma_5}^{(mn)}(\vec{r})  + \mathcal{O}\left(  e^{-\Delta_{N+1,m}t_2}, e^{-\Delta_{N+1,n}t_1} \right) \,,
\end{equation*}
where $\Delta_{nm} = E_n - E_m$ is the energy difference between the $n^{\rm th}$ and $m^{\rm th}$ excited states. All these estimators can be further improved by using the sGEVP method \cite{BulavaYZ} where the three-point correlation function is summed over the insertion time $t_1$ 
\begin{equation}
\mathcal{R}_{mn}^{\rm sGEVP}(t,t_0;\vec{r}) = - \partial_t \left( \frac{ \left| \left(v_m(t,t_0), \left[  K(t,t_0;\vec{r})/\widetilde{\lambda}_n(t,t_0) - K(t_0,t_0;\vec{r}) \right] w_n(t,t_0) \right) \right| }{  \left(v_m(t,t_0),C^{(2)}_{\mathcal{P}}(t_0)v_m(t,t_0)\right)^{1/2}  \left(w_n(t,t_0),C^{(2)}_{\mathcal{V}}(t_0)w_n(t,t_0)\right)^{1/2} } e^{\Sigma_{mn}(t_0,t_0) t_0/2} \right) \,.
\label{eq:Asgevp}
\end{equation}
with $\partial_t f(t) = (f(t+a)-f(t))/a$. Here, $\Sigma_{mn}(t,t_0) = E_n(t,t_0)-E_m(t,t_0)$ is the effective energy difference computed at each time $t$ between the $m^{\text{th}}$ and $n^{\text{th}}$ radially excited states and
\begin{equation*}
K_{ij}(t,t_0;\vec{r}) = a \sum_{t_1} e^{-(t-t_1)\Sigma_{mn}(t,t_0)} C_{ij}^{(3)}(t,t_1;\vec{r}) \,,
\end{equation*}
is the summed three-point correlation function. We recall that the advantage of this estimator is the faster suppression of higher excited state contribution when $t_0>t/2$~\cite{BlossierQMA}
\begin{align*}
\mathcal{R}_{mn}^{\rm sGEVP}(t,t_0;\vec{r}) &= a^3 f_{\gamma_{\mu} \gamma_5}^{(mn)}(\vec{r}) + \mathcal{O}\left( t \, e^{-\Delta_{N+1,n}t} \right) \quad \ \, n>m \,, \\
&= a^3  f_{\gamma_{\mu} \gamma_5}^{(mn)}(\vec{r}) + \mathcal{O}\left( e^{-\Delta_{N+1,m}t} \right) \quad \quad n<m \,,
\end{align*}
where $t=t_1+t_2$. Moreover, the estimator $\mathcal{R}_{mn}^{\rm sGEVP}(t,t_0;\vec{r})$ only requires the knowledge of both three and two-point correlation functions up to time $t$ whereas the estimator $\mathcal{R}_{mn}^{\GEVP}(t,t_0;\vec{r})$ involves the three-point correlation function at twice the time of the two-point correlation functions and is therefore statistically noisier at large $t$.

\section{Lattice computation\label{sec3}}
 
\renewcommand{\arraystretch}{1.1}
\begin{table}[t]
\begin{center}
\begin{tabular}{lcc@{\hskip 02em}c@{\hskip 02em}c@{\hskip 01em}c@{\hskip 01em}c@{\hskip 01em}c@{\hskip 01em}c@{\hskip 01em}c}
\hline
	\toprule
	id	&	$\quad\beta\quad$	&	$(L/a)^3\times (T/a)$ 		&	$\kappa$		&	$a~(\rm fm)$	&	$m_{\pi}~(\MeV)$	& $Lm_{\pi}$ 	& $\#$ cfgs	&	$\kappa_G$	&	$\{R_1,R_2,R_3\}$\\
\hline
	\midrule 
	A5	&	$5.2$	&  	$32^3\times64$	&	$0.13594$	& 	$0.075$  	& 	$330$	&4& $200$	& 0.1	& $\{15,60,155\}$ \\ 
	B6	&			&	$48^3\times96$	&	$0.13597$	&			&	$280$	&5.2& $200$	&&\\ 
	\midrule
	D5	&	$5.3$	&	$24^3\times48$	& 	$0.13625$	& 	$0.065$  	& 	$450$	&3.6	& $150$	& 0.1	& $\{22,90,225\}$\\  
	E5	&			&	$32^3\times64$	& 	$0.13625$	& 		  	& 	$440$	&4.7	& $200$	&&\\  
	F6	&			& 	$48^3\times96$	&	$0.13635$	& 			& 	$310$	&5	& $200$	&&\\    
	\midrule
	N6	&	$5.5$	&	$48^3\times96$	&	$0.13667$	& 	$0.048$  	& 	$340$	&4	& $200$	& 0.1	& $\{33,135,338\}$	\\  
	\midrule
\hline
	\midrule
	Q1	&	$6.2885$	&	$24^3\times48$	&	$0.13498$	& 	$0.06$  	& 	$\times$	&$\times$	& $100$&0.1&\{22,90,225\}	\\
	Q2	&			&	$32^3\times64$	&	$0.13498$	& 		  	& 	$\times$	&$\times$	& $100$&&	\\  
	\bottomrule
\hline
\end{tabular} 
\end{center}
\caption{Parameters of the simulations: bare coupling $\beta = 6/g_0^2$, lattice resolution, hopping parameter $\kappa$, lattice spacing $a$ in physical units, pion mass and number of gauge configurations. The smeared quark field are defined as $\psi^{(i)}_l (x)= (1+ \kappa_G a^2 \Delta)^{R_i} \psi_l (x)$ where $\kappa_G=0.1$ and $\Delta$ is the covariant Laplacian made with APE-blocked links. Sets D5, Q1 and Q2 are not used to extrapolate our results at the physical point: they are used to study finite volume and quenching effects. The quark mass for Q1 and Q2 is tuned to the strange quark mass.}
\label{tabsim}
\end{table}

This work is based on a subset of the CLS ensembles, made of $N_f=2$ nonperturbatively $\mathcal{O}(a)$-improved Wilson-Clover fermions \cite{SheikholeslamiIJ, LuscherUG}, the plaquette gauge action \cite{WilsonSK} for gluon fields and generated using either the DD-HMC algorithm \cite{LuscherQA, LuscherRX, LuscherES, Luscherweb} or the MP-HMC algorithm \cite{MarinkovicEG}. The static quark is discretized through HYP2 \cite{HasenfratzHP, DellaMorteYC}. We have also simulated two quenched ensembles ($Q$) to study the influence of the sea quarks. We collect in Table~\ref{tabsim} our simulation parameters. Three lattice spacings ($0.05\,\mathrm{fm} \lesssim a \lesssim 0.08\,\mathrm{fm}$) are considered with pion masses in the range $[280\,, 440]~\MeV$. Finally, the statistical error is estimated from the jackknife procedure. We denote by $G_h(x,y)$ the static quark propagator and by $G^{ij}_l(x,y)$ the light quark propagator with $j$ iterations of Gaussian smearing applied at the source and $i$ iterations of Gaussian smearing applied at the sink, smearing parameters are collected in Table \ref{tabsim}. The static quark propagator is explicitly given (in lattice units) by
\begin{equation*}
G_h(x,y) = \theta(x_0-y_0) \delta_{\vec{x},\vec{y}} \mathcal{P}(y,x)^{\dag} P_+ \,,
\end{equation*}
where $\mathcal{P}(x,x)=1$, $\mathcal{P}(x,y+a\hat{0})=\mathcal{P}(x,y) U^{\rm HYP}_{0}(y)$ is a HYP-smeared Wilson line and $P_+ = (1/2) (1+\gamma_0)$.
To take advantage of translational invariance, the light quark propagators in two-point correlation functions are computed using $U(1)$ stochastic sources with full-time dilution 
\begin{align*}
C_{\mathcal{P}, ij}^{(2)}(t) &=  \frac{1}{V}  \sum_{\vec{x},\vec{y}} \langle \, \mathrm{Tr}\left[ G_l^{ij}(x,y)  \gamma_5 G_h(y,x) \gamma_5  \right] \, \rangle \, \Big|_{x_0=t\,,y_0=0} \,, \\
C_{\mathcal{V}, ij}^{(2)}(t) &= \frac{1}{3V}  \sum_{k=1}^3 \sum_{\vec{x},\vec{y}} \langle \, \mathrm{Tr}\left[ G_l^{ij}(x,y)  \Gamma_k G_h(y,x) \Gamma_k  \right] \, \rangle \, \Big|_{x_0=t\,,y_0=0} \,,
\end{align*}
where $\langle \cdots \rangle$ stands for the average over gauge field configurations and stochastic sources. We also average the results over $k=1,2,3$ for the vector correlation functions. In the case of three-point correlation function, we use $N_s$ point-sources $\eta_s$, located at $(t_1+t_x, \vec{x}+\vec{r})$, and compute the light quark propagator by inverting the Dirac operator on both sources $\eta_s$ and $\gamma_{\mu} \eta_s$. Denoting respectively the solution vectors by $\psi_s$ and $\widetilde{\psi}_s$, we obtain
\begin{align*}
C_{\gamma_{\mu} \gamma_5, ij}^{(3)}(t, t_1;\vec{r}) &= \frac{a}{T} \sum_{t_x}  \langle \, \mathcal{P}^{(j)}(t+t_x;\vec{x}) \mathcal{A}_{\mu}(t_1+t_x;\vec{x}+\vec{r})  \mathcal{V}^{(i)\dag}_{k}(t_x;\vec{x}) \, \rangle \\
&= \frac{a}{T} \sum_{t_x} \langle \, \mathrm{Tr}\left[ G_l^{j0}(z,y) \gamma_{\mu} \gamma_5 G_l^{0i}(y,x) \gamma_k G_h(x,z) \gamma_5 \right] \rangle \Big|_{\vec{y}=\vec{x}+\vec{r}, \vec{z}=\vec{x}} \\
&= - \frac{a}{T N_s} \sum_{t_x,s} \langle \, \mathrm{Tr}\left[ \psi_s^{(i)\dag}(x) P_+ \mathcal{P}^{\dag}(z,x) \gamma_k \widetilde{\psi}_s^{(j)}(z)  \right] \rangle \Big|_{\vec{y}=\vec{x}+\vec{r}\,,\ \vec{z}=\vec{x}} \,.
\end{align*}
A usual point source of the form $\delta^{ab} \delta_{\alpha\beta} \delta_{xy}$ ($a, b$ color indices, $\alpha, \beta$ spinor indices) would require twelve inversions of the Dirac operator per light quark propagator. Moreover, such sources do not take advantage of the full gauge information unless different positions of the source are used. Instead, we consider $N_s$ stochastic point-sources which have nonzero values at a single spacetime site $x$ and at every color-spin component ($a$, $\alpha$) ($\eta_{\alpha}^{a}(y)_s = 0$ if $y\neq x$) and satisfying the condition 
\begin{equation}
\lim_{N_s\rightarrow \infty} \frac{1}{N_s} \sum_{s=1}^{N_s} \eta_{\alpha}^a(x)_s \left[ \eta_{\beta}^b(y)_s \right]^{*} = \delta^{ab}\delta_{\alpha\beta} \delta_{xy} \,,
\label{stoch_sources}
\end{equation}
where each component is normalized to one, $\eta_{\alpha}^a(x)_{[s]}^{*}\ \eta_{\alpha}^a(x)_{[s]} = 1$ (no summation). This can be implemented by using $U(1)$ noise for each color and spinor index on site $x$. Therefore, only one inversion per light propagator is required, which allows us to perform the computation with different point source positions at a reasonably small computational cost. Having different spatial positions of the probe is expected to decrease the gauge noise while the stochastic noise is kept under control by using a sufficient number of stochastic point-sources. In practice, this number can be small since the stochastic average commutes with the gauge average and we have taken $N_s=T/a$. 

\subsection{Energy levels from the two-point correlation functions} 

\begin{figure}[t] 

	\begin{minipage}[c]{0.45\linewidth}
	\centering 
	\includegraphics*[width=0.9\linewidth]{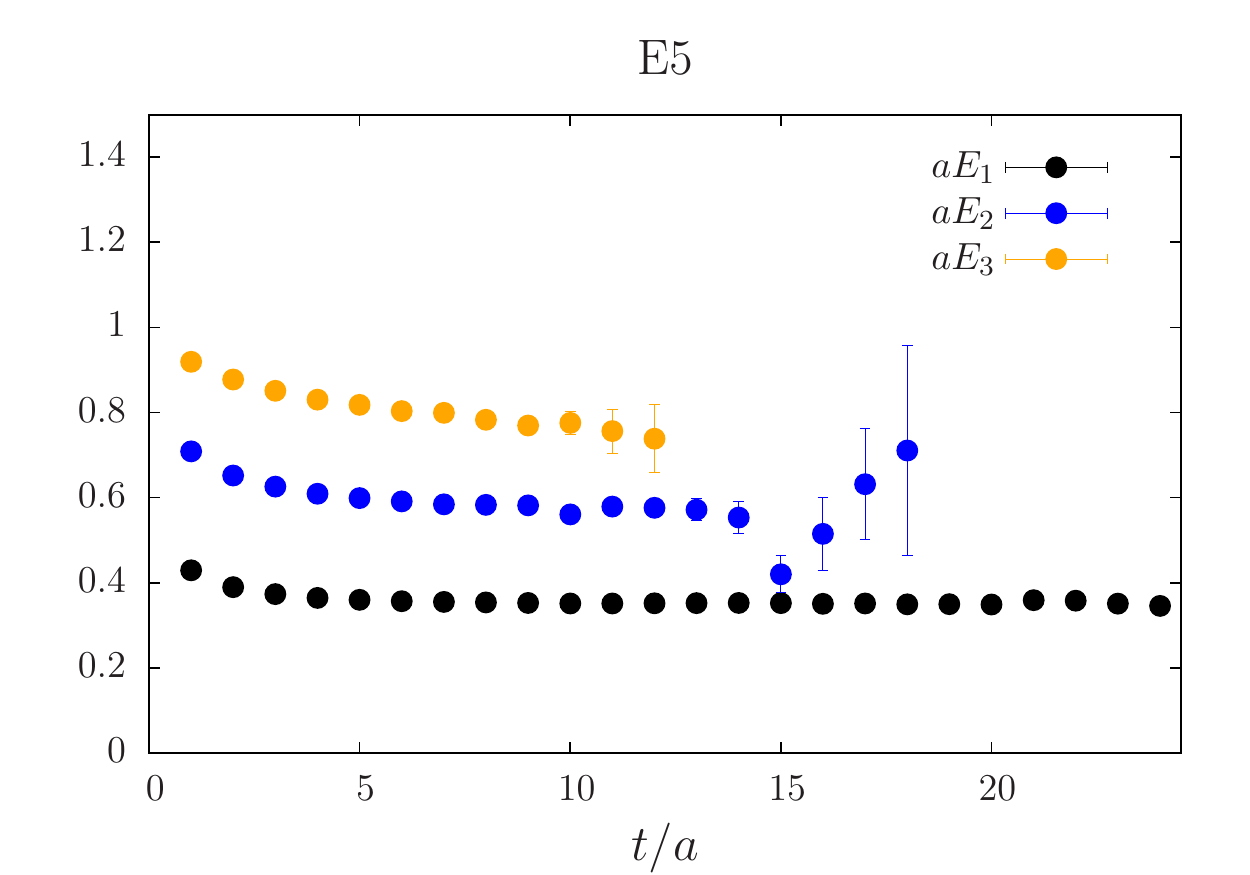}
	\end{minipage}
	\begin{minipage}[c]{0.45\linewidth}
	\centering 
	\includegraphics*[width=0.9\linewidth]{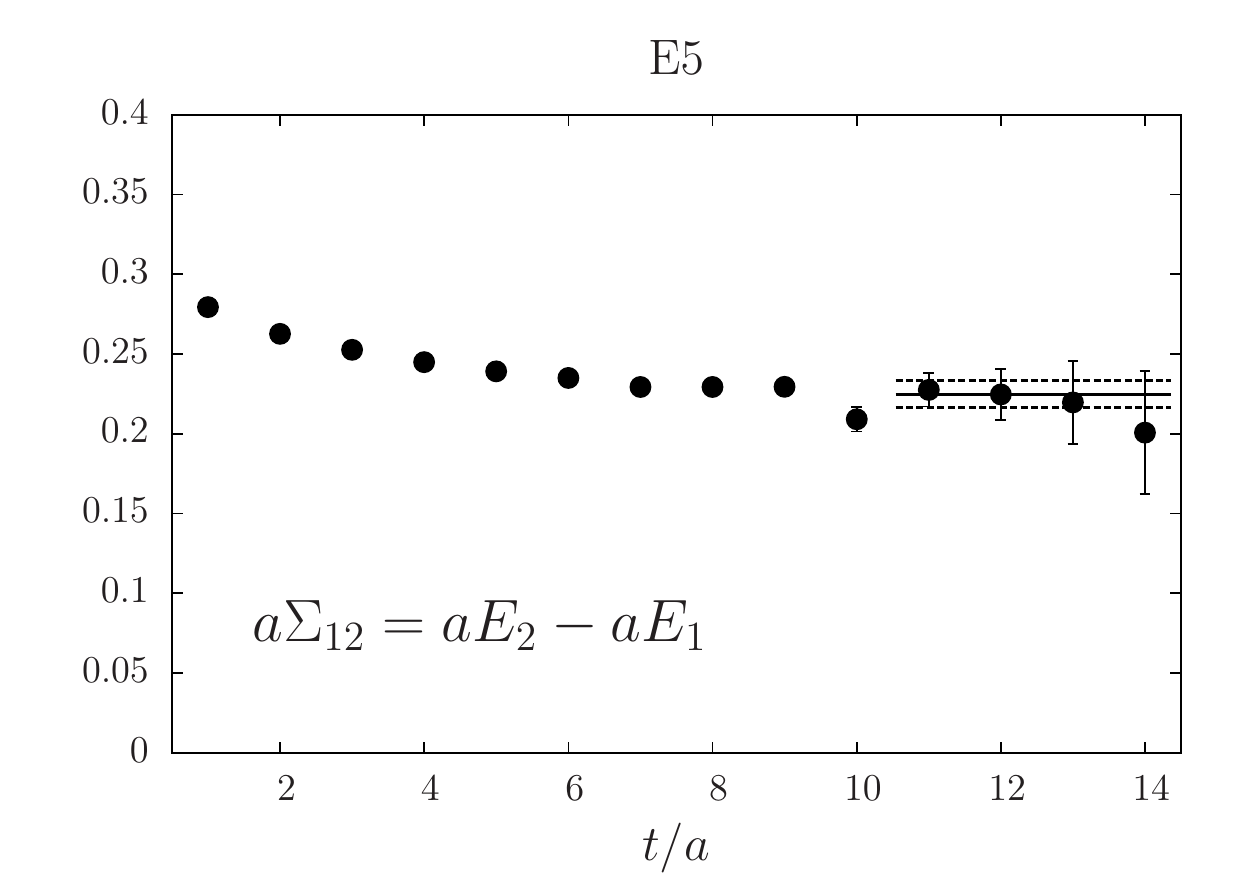}
	\end{minipage}

	\caption{(\textit{left}) Effective mass plot extracted from a $3\times 3$ GEVP for the lattice ensemble E5. (\textit{right}) Energy difference $\Sigma_{12} = E_2-E_1$ between the first radial excitation and the ground state. The value of $t_0$ is chosen such that $t>t_0/2$ to reduce the contamination of higher excited states. We also plot the plateau in the chosen fit interval.}
\label{fig:gevp}
\end{figure}
To compute the axial density distribution from sGEVP method using Eq.~(\ref{eq:Asgevp}), we need to evaluate $\Sigma_{12} = E_2 - E_1$, the mass splitting between the first radial excitation of the vector meson ($H^{*\prime}$) and the ground state pseudoscalar meson ($H$). We solve a $3\times3$ GEVP given by Eq.~(\ref{eq:gevp}) and the effective masses of the ground state ($n=1$) and first excited state ($n=2$) are estimated from the generalized eigenvalues $\lambda_n(t,t_0)$ according to
\begin{equation*}
a E_n^{\rm eff}(t,t_0) = \log \frac{ \lambda_n(t,t_0) }{ \lambda_n(t+a,t_0) } \quad , \quad t_0 > t/2 \,, 
\end{equation*}
and fitted to a plateau at large $t$ where the contribution of higher excited states has been shown analytically and numerically to be negligible \cite{BernardoniNQA}. Results for the lattice ensemble E5 are depicted in Fig.~\ref{fig:gevp} and values of $\Sigma_{12}$ for each lattice ensemble are collected in Table~\ref{tab:sum_rule}.

\subsection{Spatial component of the axial density distribution} 

\begin{figure}[t] 
	\begin{minipage}[c]{0.28\linewidth}
	\centering 
	\includegraphics*[width=\linewidth]{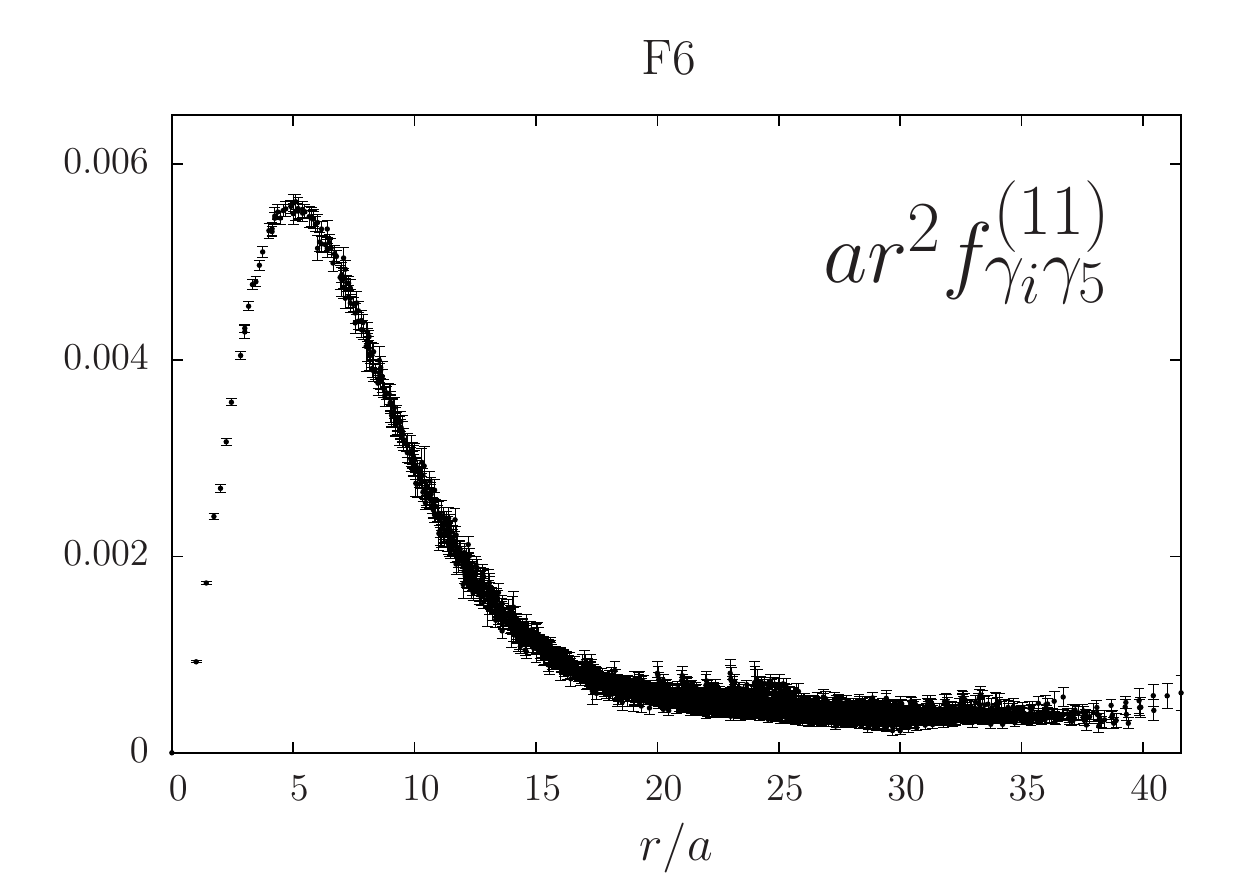}
	\end{minipage}
	\begin{minipage}[c]{0.28\linewidth}
	\centering 
	\includegraphics*[width=\linewidth]{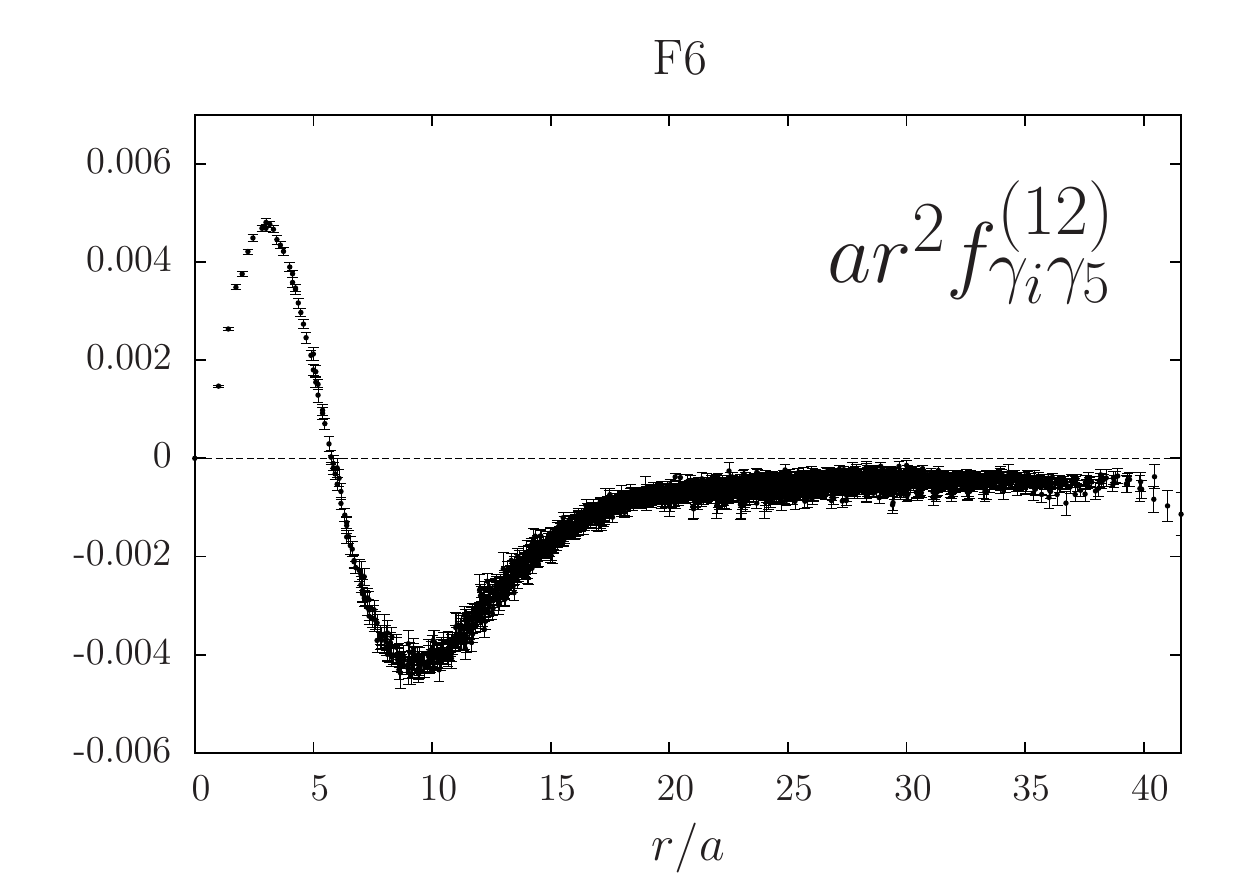}
	\end{minipage}
	\begin{minipage}[c]{0.28\linewidth}
	\centering 
	\includegraphics*[width=\linewidth]{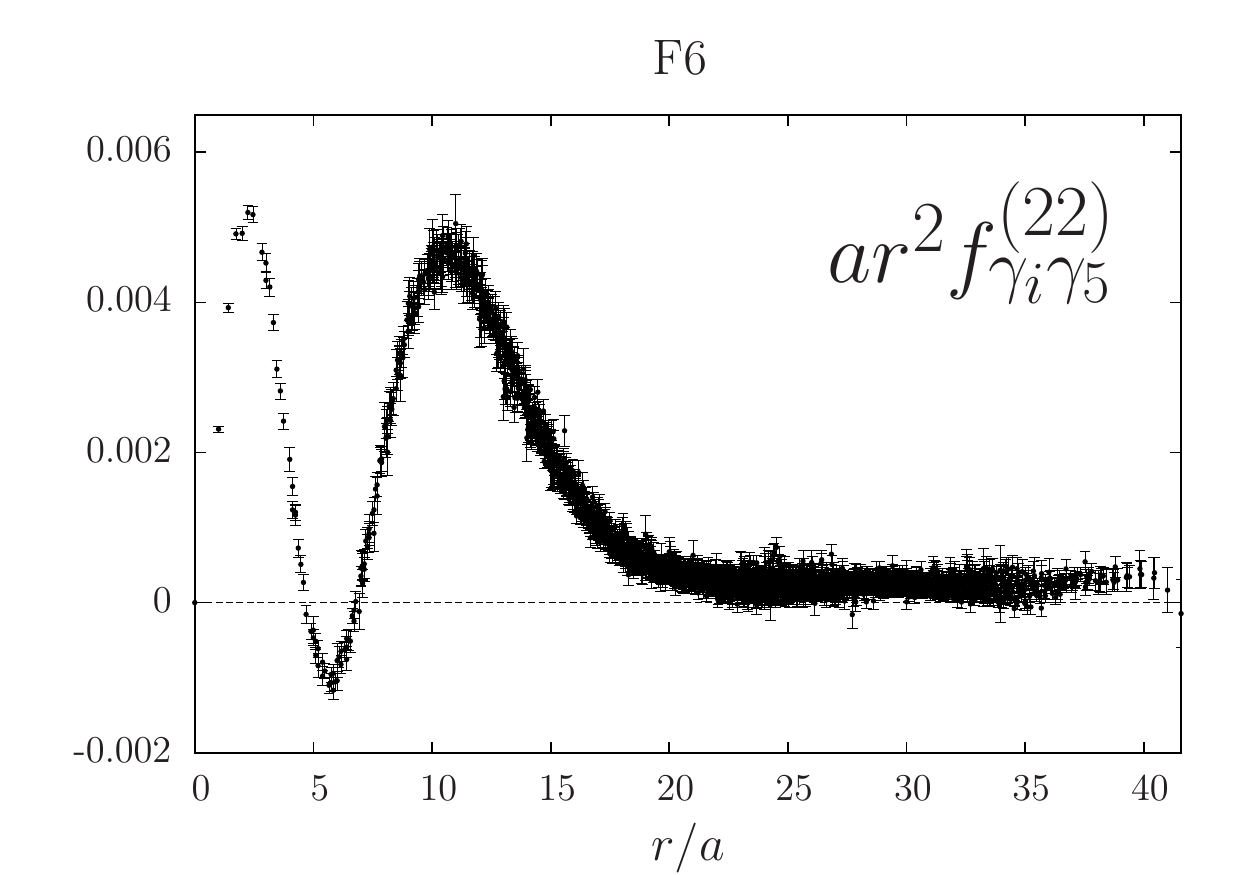}
	\end{minipage}
	\caption{Spatial component of the radial distributions of the axial density $ar^2\, f^{(11)}_{\gamma_i\gamma_5}(r/a)$, $ar^2\, f^{(12)}_{\gamma_i\gamma_5}(r/a)$ and $ar^2\, f^{(22)}_{\gamma_i\gamma_5}(r/a)$ for the lattice ensemble F6.}	
\label{fig:distrib_space}
\end{figure}

The axial density distributions are estimated using the sGEVP method (Eq.~(\ref{eq:Asgevp})) with $t_0>t/2$. Since the spatial component of the distributions depends only on $r$, we have averaged the raw data over the cubic isometry group H(3) but not over the symmetry group SO(3) of the continuum theory. Therefore, we have $(N+1)(N+2)(N+3)/6$ independent points where $N=L/(2a)$. Results for the ground state and first excited state are depicted in Fig.~\ref{fig:distrib_space}. We have checked that compatible results, within statistical error bars, are obtained using the GEVP method (Eq.~(\ref{eq:Agevp})) or the ratio method (Eq.~(\ref{eq:ratio})), that is applicable only for the ground state radial distribution. From these plots, two main observations can be made. First, the spatial component of the distributions $r^2\, f^{(nm)}_{\gamma_i\gamma_5}(r)$ do not converge to zero at large values of $r$. Secondly, a fishbone structure is observed -- it appears also at the finest lattice spacing. Several possible explanations are as follows:
\begin{itemize}
\item The GEVP wrongly isolates the ground state and the first excited state. This may affect the shape of the radial distributions. However,  similar results are obtained for the ground state distributions using the ratio method and for all choices of interpolating operators as discussed in Section~\ref{sec:excited_state}. Moreover, as shown in Appendix~\ref{app:charge_matter_distrib}, results for the charge (vector) distributions are in perfect agreement with the value of $Z_V$ determined using a completely different method~\cite{DellaMorteRD, Fritzsch:2012wq}.
\item The shift from zero at large $r$ could also be a sign that our interpolating operators couple to a two-body system. This is very unlikely since the shift is also visible for the ground state. Within our lattice setup, we are near the $B_1^* \pi$ threshold (see Table~\ref{tab:threshold}) which has the same quantum numbers as the vector meson. This issue is discussed in Section~\ref{sec4}.
\item The lattices used to extrapolate our results at the physical point all satisfy the condition $Lm_{\pi}>4$ and volume effects are expected to be small. However, the fishbone structure at large $r$ could also be explained by volume effects: assuming that the static source is at $\vec{x}=0$, lattice points $(x_1, x_2, x_3)$ with small $r^{[4]}\equiv \sum_{i=1}^3 x^4_i$ are more affected by volume effects, compared to other points belonging to the same orbit $r^2\equiv \sum_{i=1}^3 x^2_i$, since they are closer to their periodic images. Indeed, from previous plots, the radial distribution, multiplied by $r^2$, does not vanish at $r=L/2$ and the overlap of the tails of the distributions cannot be neglected. We also observe that maxima of the fishbone structure appear first at $r=L/2$ and then at $r=L/2\times \sqrt{2}$, where the overlap is expected to be large. In ref.~\cite{BecirevicZZA}, the radial distribution of the axial density has been computed for the ground state, yet at a coarse lattice and far from the chiral limit ($m_{\pi} \sim 750~\MeV$): however, their results show that the tail of the distribution is still sizable at $r\approx 1~{\rm fm}$. This issue has been discussed in \cite{BurkardtPW} in the case of hadron correlation functions. This issue is discussed in Section~\ref{sec:volume_effects}.
\item In lattice QCD, lattice artefacts may appear at finite lattice spacing due to the breaking of the continuum O(3) symmetry group down to the subgroup H(3). Cubic artefacts are seen in momentum space for instance in \cite{BlossierKTA} and show a similar fishbone structure at large $p$. However, in the latter case, artefacts are of the form $a^2 p^{[4]} / p^2$, $a^2 p^{[6]} / p^4$, $a^2 p^{[8]} / p^6$ and are enhanced at large momenta. Using the same argument as in \cite{deSotoHT} and based on dimensional analysis, we can show that in our case we expect lattice artefacts for $a^3 f^{mn}_{\gamma_i \gamma_5}$ of the form $a^2 r^{[4]} / r^6$ and  $a^2 r^{[6]} / r^8$ which are not enhanced at large radii: more details are given in Appendix~\ref{sec:artefacts}. 
\end{itemize}

\subsection{Excited states contribution} 
\label{sec:excited_state}

\begin{figure}[t]
	\centering 
	\includegraphics*[width=0.4\linewidth]{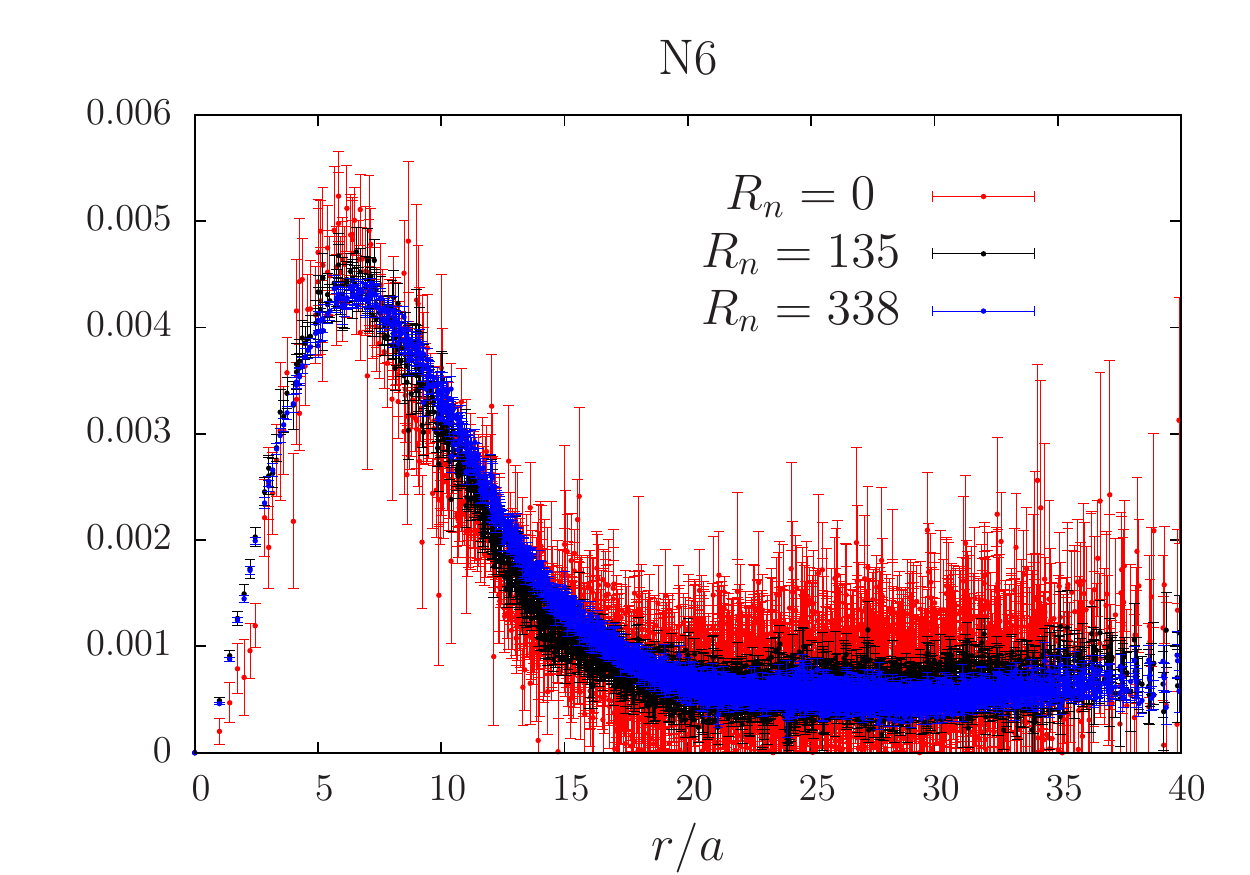}
	\caption{Spatial component of the radial distribution $a r^2 f_{\gamma_i \gamma_5}^{(11)}(r/a)$ obtained using the ratio method (Eq.~(\ref{eq:ratio})) for different numbers of Gaussian smearing iterations $R_n$, for the lattice ensemble N6. For each curve, $t$ is chosen such that the radial distribution has reached a plateau ($t/a = 17,14,10$ for $R_n=0,135,338$ respectively) .}	
\label{fig:cmp_smr}
\end{figure}

Gaussian smearing is used to reduce the contamination from excited states to the correlators we analyse. It is applied on the heavy-quark propagator entering the contractions but not on the probe which must stay local. We have checked that our results are indeed independent of the number of iterations in the procedure to obtain a Gaussian smearing. In Fig.~\ref{fig:cmp_smr}, the ground state radial distributions computed by the ratio method (Eq.~(\ref{eq:ratio})) are plotted for the local interpolating fields and for two non-local interpolating fields with different levels of Gaussian smearing, that correspond to $R_n = 135$ and $338$ iterations, respectively, on the ensemble N6. The time $t$ is chosen such that the radial distribution has reached a plateau ($t/a = 17,14,10$ for $R_n=0,135,338$ respectively). In particular, for the local interpolating field where the contribution from excited states is important, $t$ has to be chosen large, increasing the statistical error. For the smeared interpolating fields, we obtain compatible results but the contribution of excited states is significantly reduced and the plateau is reached earlier where statistical errors are still small. 

\subsection{Treatment of volume effects} 
\label{sec:volume_effects}

\begin{figure}[t]

	\begin{minipage}[c]{0.28\linewidth}
	\centering 
	\includegraphics*[width=0.9\linewidth]{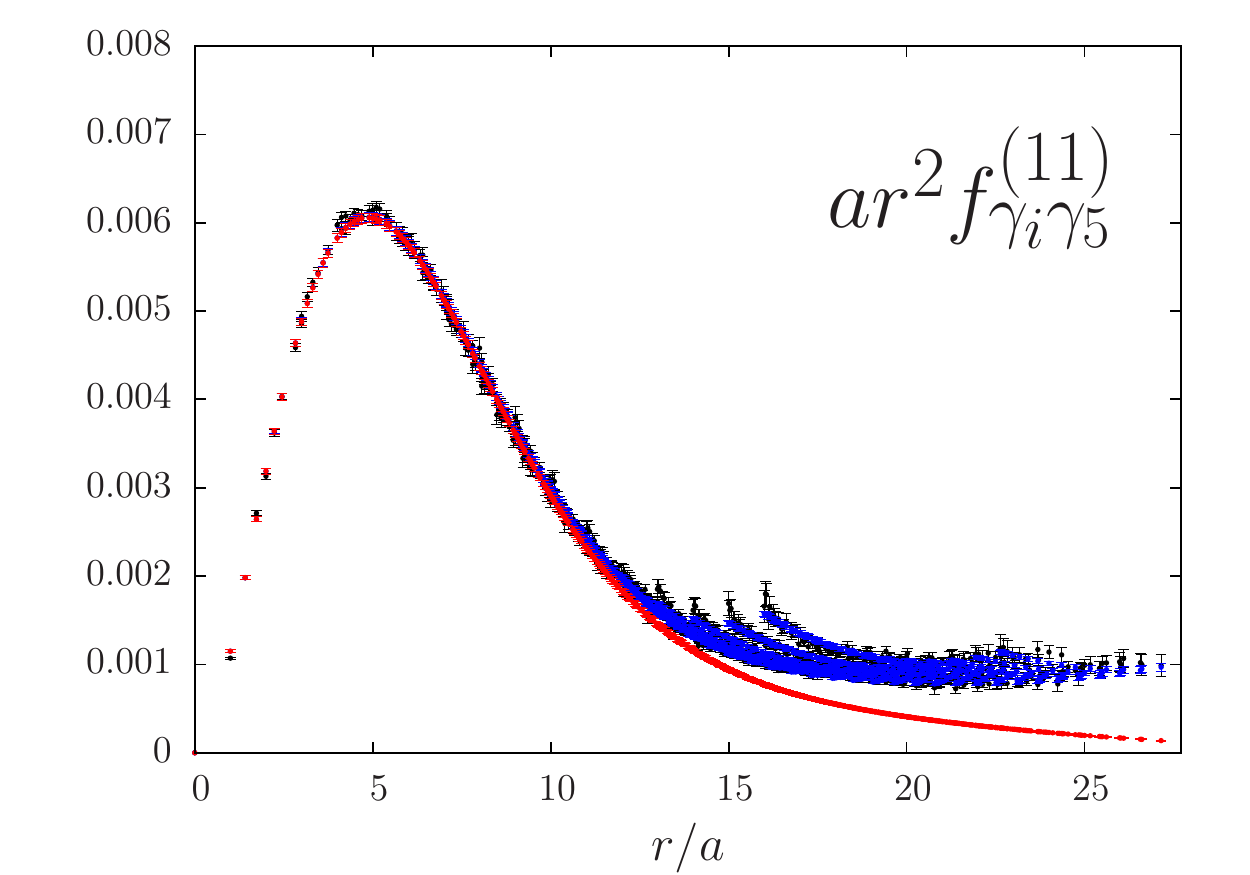}
	\end{minipage}
	\begin{minipage}[c]{0.28\linewidth}
	\centering 
	\includegraphics*[width=0.9\linewidth]{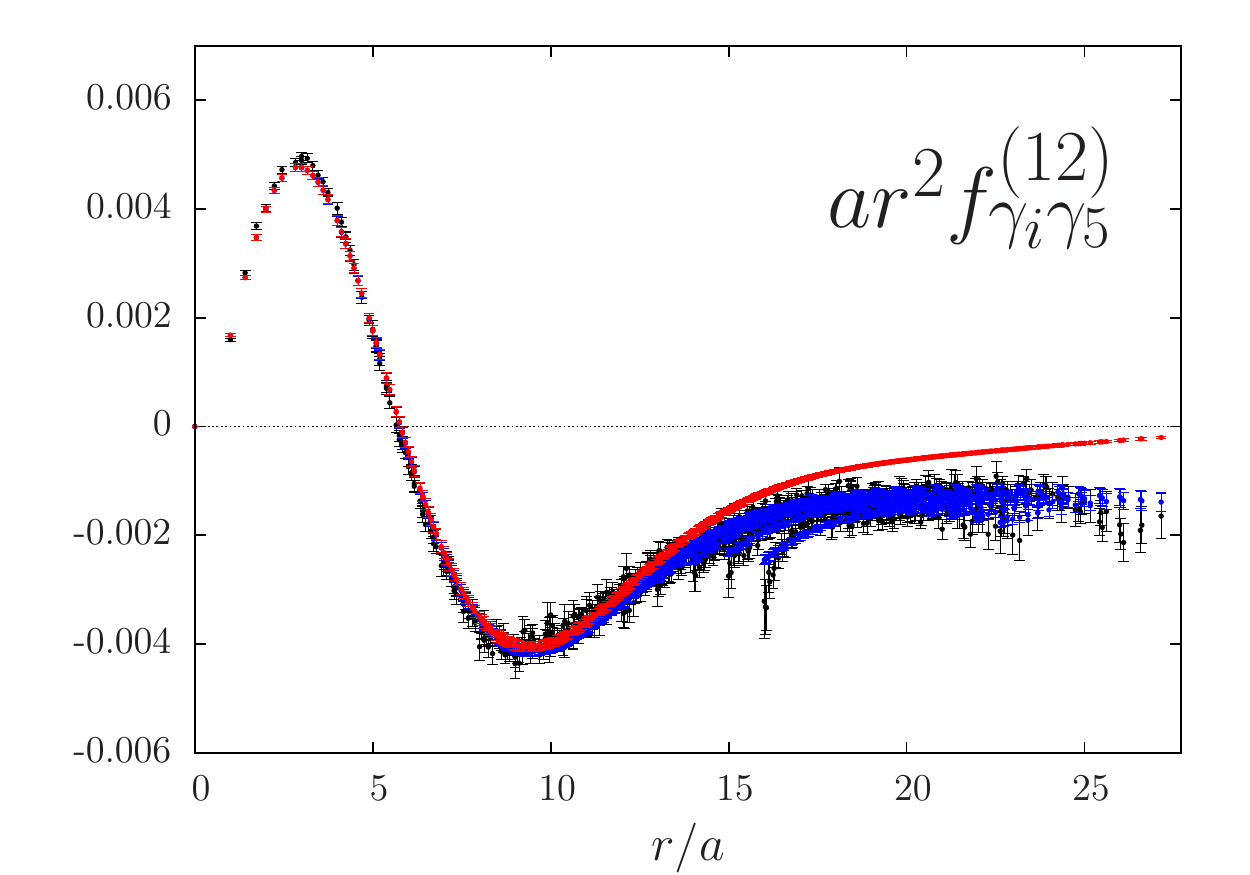}
	\end{minipage}
	\centering
	\begin{minipage}[c]{0.28\linewidth}
	\centering 
	\includegraphics*[width=0.9\linewidth]{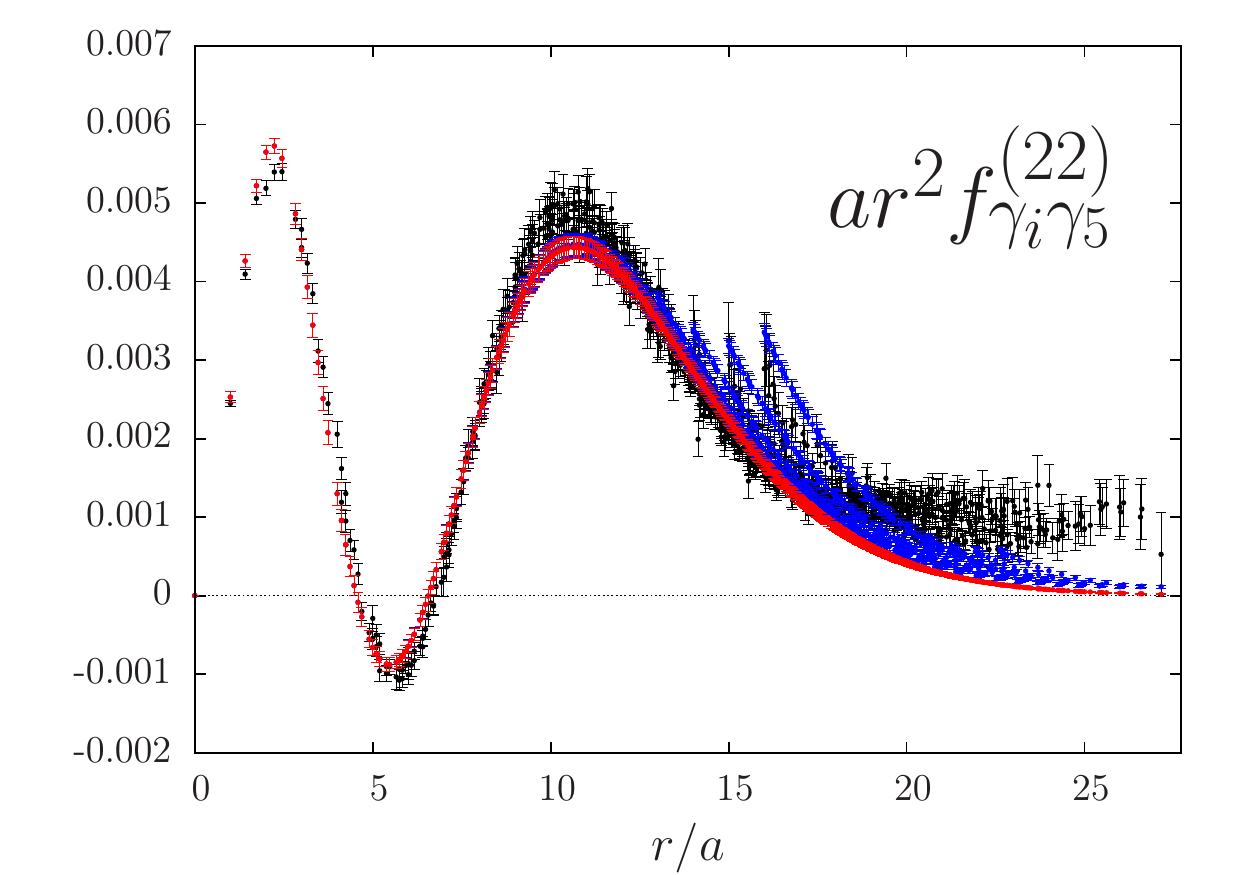}
	\end{minipage}
	\caption{Curing of volumes effects on the axial spatial density distributions for the lattice ensemble E5. 
	Raw data are in black, results of the fit in blue and results after image corrections in red.
	\label{fig:vol_effects}}
\end{figure}

On the lattice with periodic boundary conditions in space directions, one expects to compute~\cite{BurkardtPW} 
\begin{equation}
a^3 f^{\rm lat}_{\gamma_{i}\gamma_5}(\vec{r}) = \sum_{\vec{n}} a^3 \widetilde{f}_{\gamma_{i}\gamma_5}(\vec{r} + \vec{n} L)  \quad , \quad n_i \in \mathbb{Z} \,,
\label{eq:f_lat}
\end{equation}
where $\widetilde{f}_{\gamma_{i}\gamma_5}(\vec{r})$ can still differ from the infinite volume distribution due to interactions with periodic images ($n_i \neq 0)$. If the lattice is sufficiently large such that the overlap of the tail of the distribution with its periodic images is small, interactions between periodic images can be neglected and $\widetilde{f}_{\gamma_{i}\gamma_5}(\vec{r}) \approx f_{\gamma_{i}\gamma_5}(\vec{r})$, even in the overlap region. In what follows this is assumed to be a good approximation and only the nearest image contribution is considered: the tails just add with each other without deformation and this assumption is discussed later. From plots shown in Fig.~\ref{fig:distrib_space}, we see that radial distributions differ significantly from zero at $r=L/2$: the overlap of the tails cannot be neglected. Then, we can remove the contribution of periodic images by fitting our raw data with some given function using Eq.~(\ref{eq:f_lat}) to reproduce the fishbone structure. The following fit ansatz has been considered
\begin{align}
f^{(mn)}_{\gamma_{i}\gamma_5}(\vec{r}) &= P_{mn}(r) \, \exp \left( - \left( r/r_0 \right)^{\alpha}  \right)  \,,
\label{eq:fit_vol}
\end{align}
where $P_{mn}$ is a polynomial function and where only the nearest image contributions ($n_i \in \{0,1\} $) are considered. In practice, $P_{11}, P_{12}$ and $P_{22}$ are of degree $2, 3$ and $4$ respectively. This form is motivated by quark models \cite{BecirevicZZA} and the small number of parameters. It also reproduces results for the scalar and vector densities discussed in Appendix~\ref{app:charge_matter_distrib}, where volume effects are negligible, and the fishbone structure of our quenched results where the data are more precise (see Section~\ref{Quenched_results}). Results for the lattice ensemble E5 are depicted in Fig.~\ref{fig:vol_effects} and the corresponding $\chi^2$ are respectively $\chi^2_{11}/{\rm d.o.f} = 1.07$, $\chi^2_{12}/{\rm d.o.f} = 1.27$ and $\chi^2_{22}/{\rm d.o.f} = 4.2$. The quality of the fit is good, especially for $f^{11}_{\gamma_i \gamma_5}$ and $f^{12}_{\gamma_i \gamma_5}$ and the radial distributions indeed converge to zero at large radii after image corrections. \\

\begin{figure}[t]

	\begin{minipage}[c]{0.49\linewidth}
	\centering 
	\includegraphics*[width=0.9\linewidth]{plots/extrap_axial_E5_00.pdf}
	\end{minipage}
	\begin{minipage}[c]{0.49\linewidth}
	\centering 
	\includegraphics*[width=0.9\linewidth]{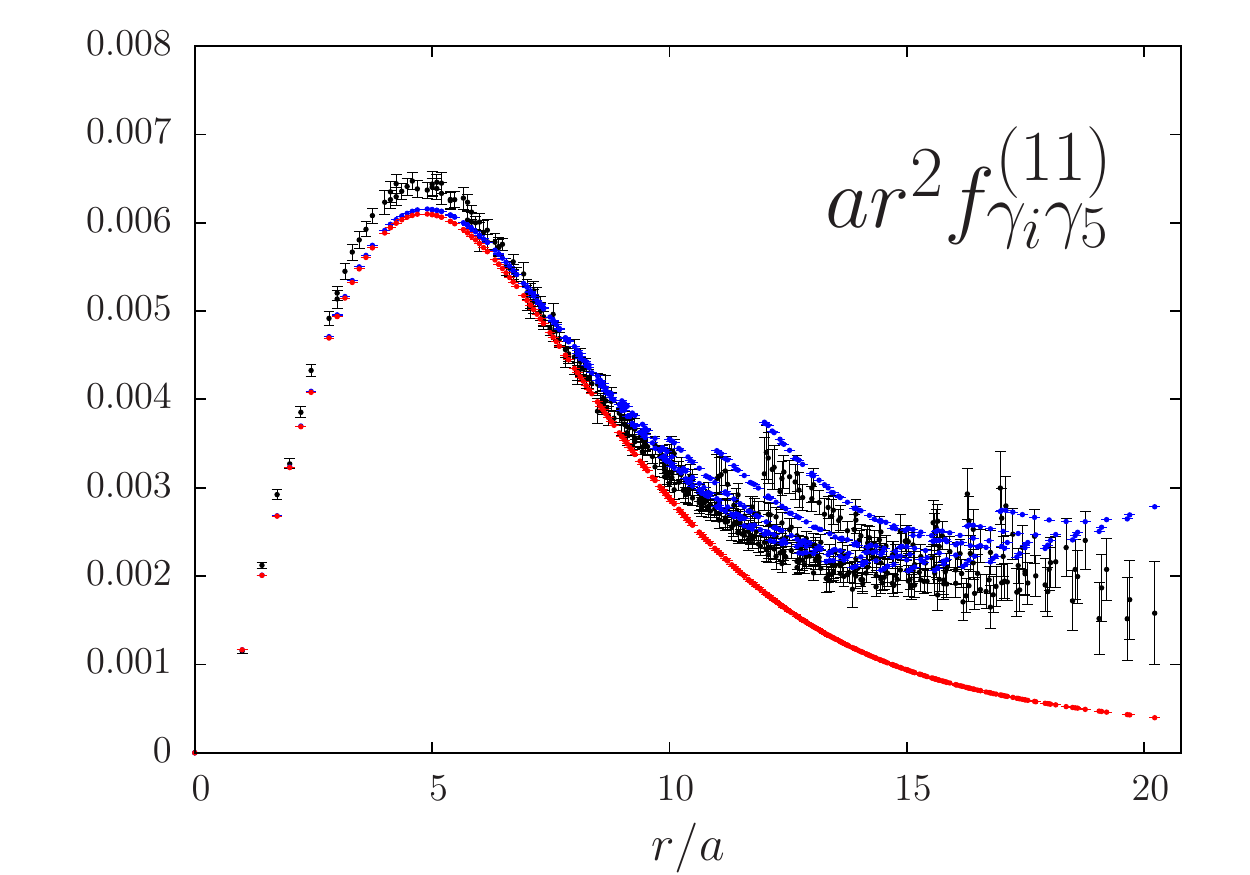}
	\end{minipage}
	
	\caption{(\textit{left}) Results of the fit for the lattice ensemble E5 ($L/a=32$) using the method described in Sec.~\ref{sec:volume_effects}. Raw data are in black, results of the fit in blue and results after image corrections in red. (\textit{right}) Results obtained for the lattice ensemble D5 ($L/a=24$) using the fit parameters of the ensemble E5 that has the same value of $\beta$ and a similar pion mass. It means that the blue and red points of the right panel are not obtained by a fit of the black points.
	\label{fig:vol_effects_2}}
\end{figure}

As explained before, two kinds of volume effects are expected. First, the overlap of the tails of the distribution with periodic images and, secondly, a deformation of the distribution when the overlap is too large. This second effect was neglected since the overlap of the tails of the distributions is small (Fig.~\ref{fig:vol_effects}) and because all our ensembles satisfy the condition $Lm_{\pi}>4$. In particular, the fact that the images correction procedure works well is an indication that this second source of volume effects is indeed small. However, to check the validity of this assertion, we have performed an analysis on the CLS ensemble D5, which is close to E5 ($m_{\pi} = 450~\MeV$ and $a=0.065~{\rm fm}$) but with a smaller lattice ($L/a=24$). Since the volume is smaller, the overlap of the tails is more important and more subject to deformation. We have checked that using the fit parameters of E5 we can reproduce the radial distribution for D5 to a good precision (see Fig.~\ref{fig:vol_effects_2}): it makes us confident that our assumption is indeed correct within our statistical accuracy. In particular, as can be seen in Table~\ref{tab:ro}, the best fit values of $r_0$ for the ensembles D5 and E5 are perfectly compatible.\\

Finally, it is important to notice that the first kind of volume effect is irrelevant in the computation of $g_{nm}$ or any form factor, at discrete lattice momenta, as long as the distribution vanishes before $r=L$. Indeed, the contribution coming from periodic images -- which leads to the fishbone structure -- compensates exactly the missing part of the tail of the distribution for $r>L/2$. For example, in one space dimension (this is easily generalized to three dimensions) and using Eq.~(\ref{eq:f_lat}), one has
\begin{multline*}
\int_{0}^{L/2} f^{\rm lat}_{\gamma_{i}\gamma_5}(r) \, e^{iq \, r} \, \mathrm{d}r = \int_{0}^{L/2} \left( f_{\gamma_{i}\gamma_5}(r) + f_{\gamma_{i}\gamma_5}(r-L) \right) \, e^{i q \, r} \, \mathrm{d}r \\
=\int_{0}^{L/2} f_{\gamma_{i}\gamma_5}(r) \, e^{i q \, r} \, \mathrm{d}r + \int_{0}^{L/2} f_{\gamma_{i}\gamma_5}(r-L) \, e^{i q \, (r-L)} \, \mathrm{d}r\\
= \int_{0}^{L/2} f_{\gamma_{i}\gamma_5}(r) \, e^{i q \, r} \, \mathrm{d}r + \int_{0}^{L/2} f_{\gamma_{i}\gamma_5}(L-r) \, e^{i q \, (L-r)} \, \mathrm{d}r
=\int_{0}^{L} f_{\gamma_{i}\gamma_5}(r) \, e^{i q \, r} \, \mathrm{d}r
 =\int_{0}^{\infty} f_{\gamma_{i}\gamma_5}(r) \, e^{i q \, r} \, \mathrm{d}r \,,
\end{multline*}
where in the second line we use the fact that lattice momenta are discrete on the lattice, $q = \frac{2\pi}{L} n$, with $n \in \mathbb{Z}$,
and in the third line we use the parity properties of the integrand and we integrate up to infinity since the distribution is assumed to vanish for $r>L$. However, it should be noted that it affects the computation of the moments of the distribution or form factors at non-lattice discrete momenta, as discussed later. In this case, one needs to use the fitted function extracted from Eq.~(\ref{eq:f_lat}).

\renewcommand{\arraystretch}{1.3} 
\begin{table}[t]
	\begin{center}
	\begin{tabular}{l@{\quad}l@{\quad}c@{\quad}c@{\quad}c@{\quad}c@{\quad}c@{\quad}c@{\quad}c}
	\hline
	$(mn)$	&				& 	A5		&	B6		&	D5	&	E5		&	F6		&	N6		\\ 
	\hline 
	$(11)$ 	&	$r_0~[\fm]$	&	$0.26(1)$	&	$0.21(1)$	&	$0.27(2)$	&	$0.26(1)$	&	$0.24(2)$	&	$0.27(2)$	\\ 
		 	&	$\alpha$		&	$1.09(3)$	&	$0.97(3)$	&	$1.21(7)$	&	$1.10(4)$	&	$1.03(4)$	&	$1.10(3)$	\\ 
	\hline
	$(12)$ 	&	$r_0~[\fm]$	&	$0.30(1)$	&	$0.28(1)$	&	$0.29(3)$	&	$0.33(2)$	&	$0.28(1)$	&	$0.31(1)$	\\
		 	&	$\alpha$		&	$1.28(2)$	&	$1.22(3)$	&	$1.35(3)$	&	$1.35(3)$	&	$1.19(2)$	&	$1.36(2)$	\\
	\hline
	\end{tabular}
	\end{center}	
	\caption{Fit parameters $r_0$ and $\alpha$ of Eq.~(\ref{eq:fit_vol}) for each lattice ensemble and each axial distribution $f^{(mn)}_{\gamma_{i}\gamma_5}$ ($mn=11,12$).}
	\label{tab:ro}
\end{table}

\subsection{Summation over $r$: the couplings $g_{11}$, $g_{12}$ and $g_{22}$} 
\label{sec:sumrules}

\begin{figure}[t] 
	\begin{minipage}[c]{0.28\linewidth}
	\centering 
	\includegraphics*[width=\linewidth]{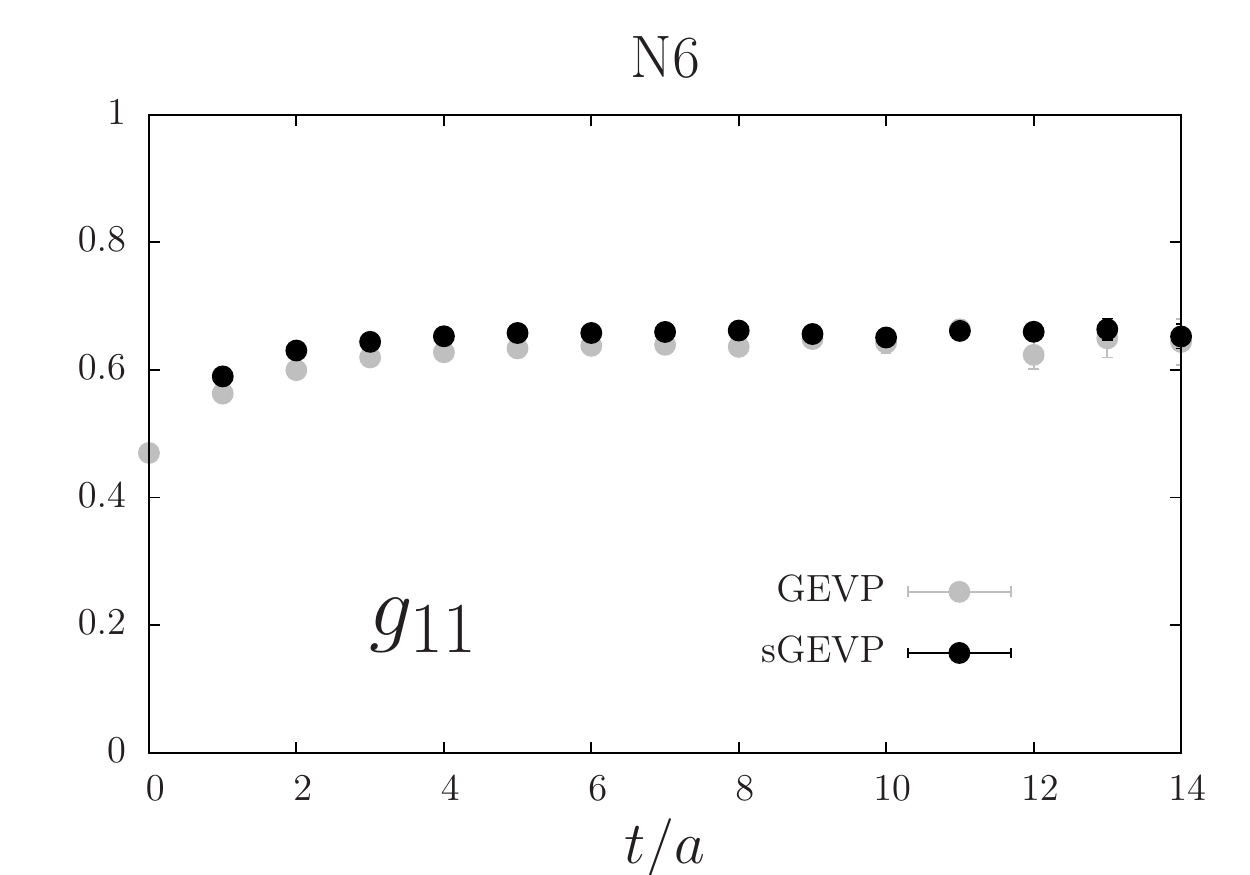}
 	\end{minipage}
	\begin{minipage}[c]{0.28\linewidth}
	\centering 
	\includegraphics*[width=\linewidth]{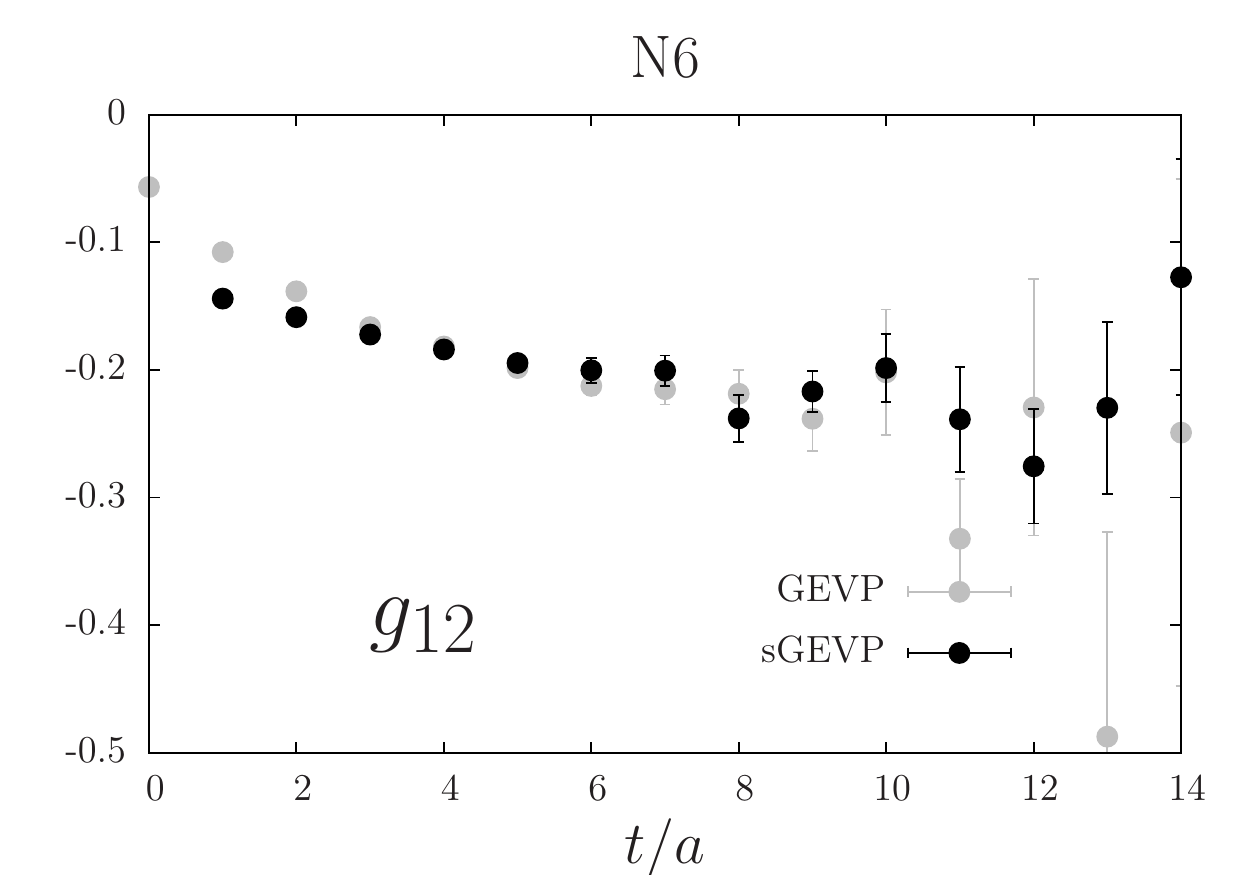}
	\end{minipage}
	\begin{minipage}[c]{0.28\linewidth}
	\centering 
	\includegraphics*[width=\linewidth]{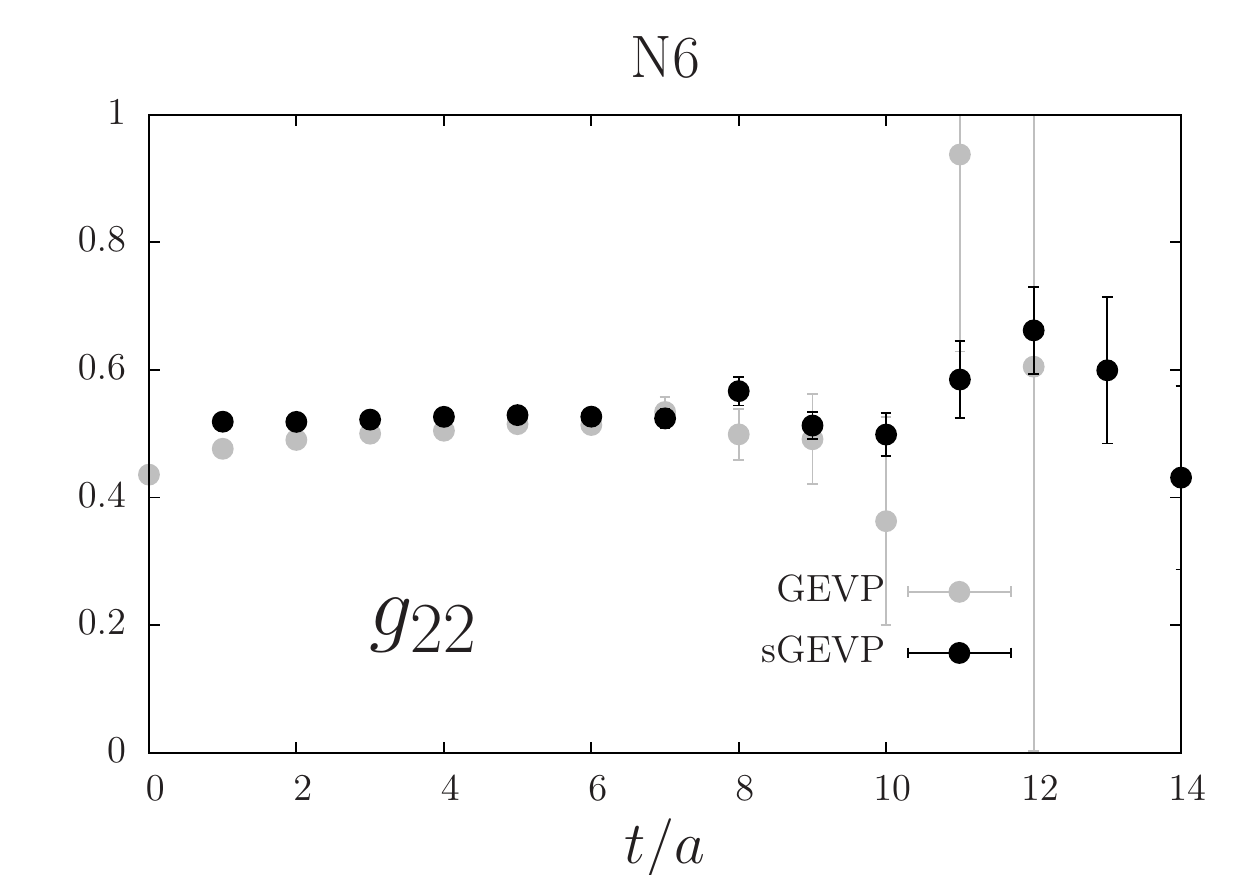}
	\end{minipage}
	
	\caption{Plateaus of the axial density summations over $r$ for the lattice ensemble N6 using the GEVP method (grey points, Eq.~(\ref{eq:Agevp}) and the sGEVP method (black points, Eq.~(\ref{eq:Asgevp})) : $g_{11}$ (\textit{left}), $g_{12}$ (\textit{center}) and $g_{22}$ (\textit{right}). }	
\label{fig:axial_sum_rule}
\end{figure}

\begin{figure}[t] 

	\hspace{-0.6cm}
	\begin{minipage}[c]{0.33\linewidth}
	\centering 
	\includegraphics*[width=1.1\linewidth]{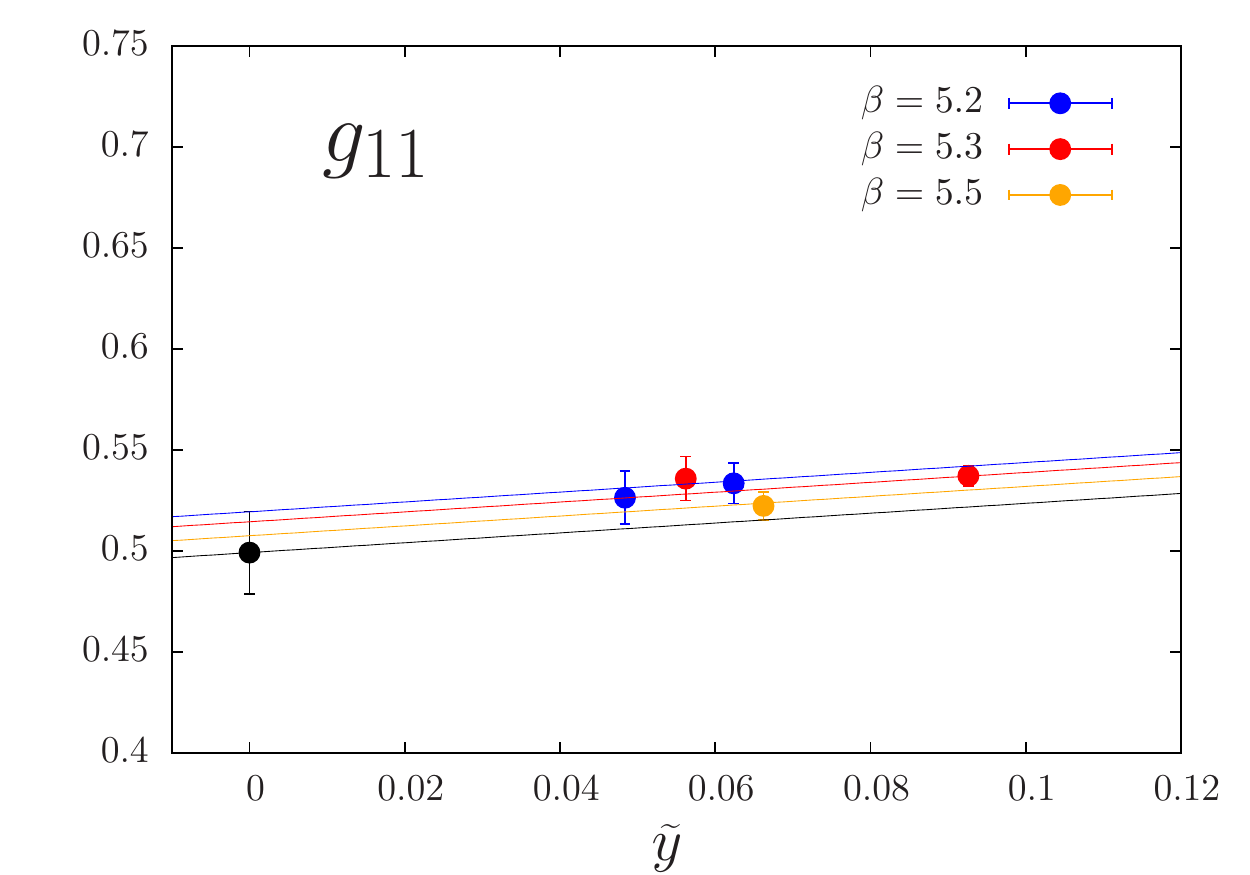}
	\end{minipage}
	\begin{minipage}[c]{0.33\linewidth}
	\centering 
	\includegraphics*[width=1.1\linewidth]{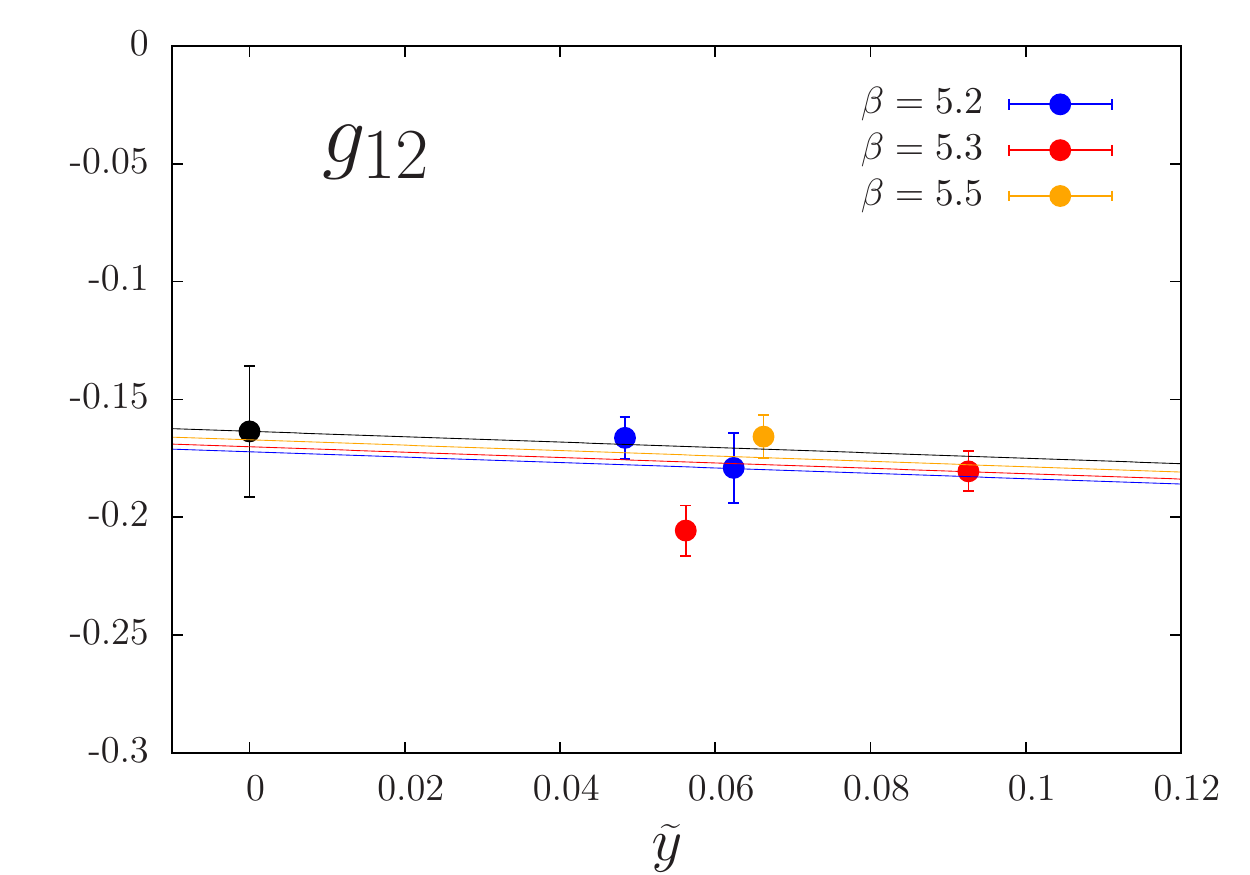}
	\end{minipage}
	\begin{minipage}[c]{0.32\linewidth}
	\centering 
	\includegraphics*[width=1.1\linewidth]{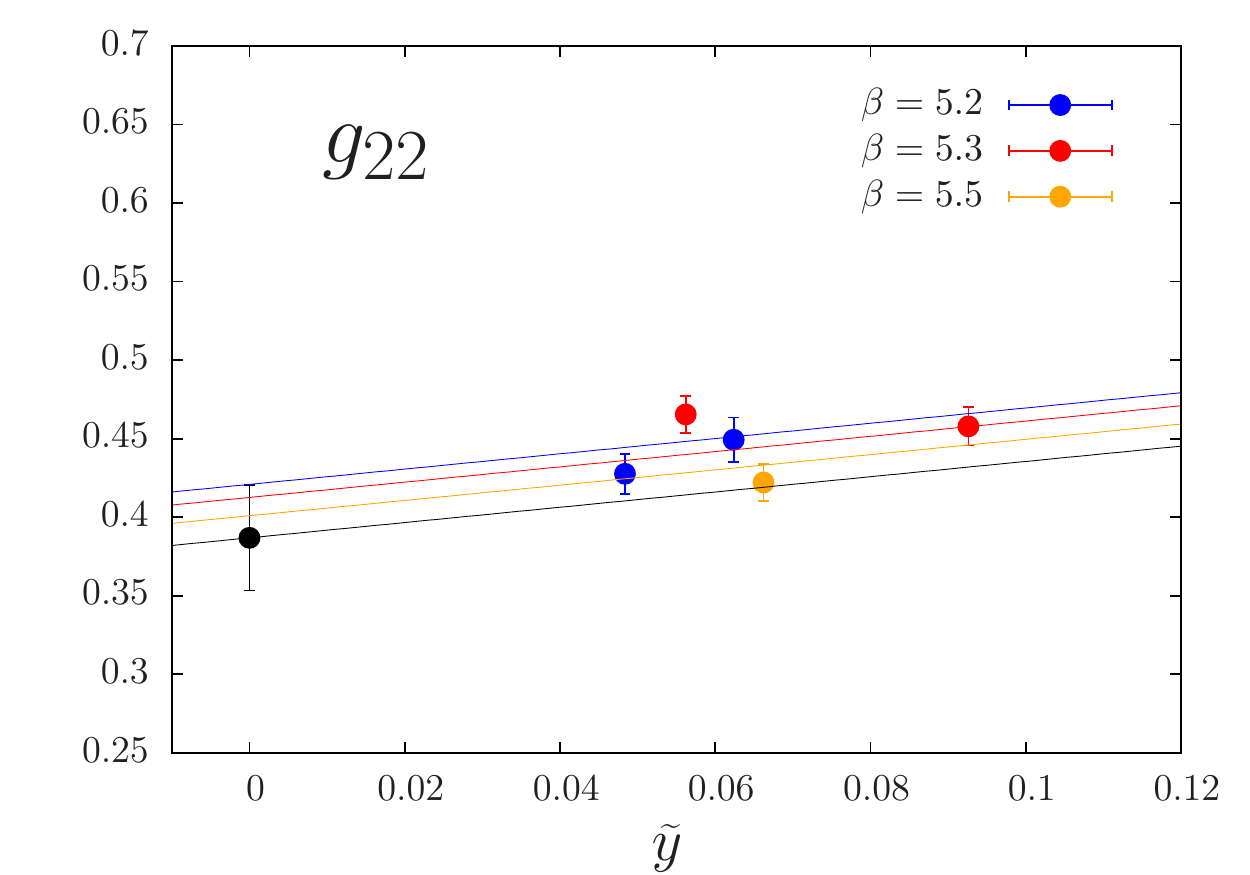}
	\end{minipage}
	
	\caption{Extrapolations of $\overline{g}_{11}$, $\overline{g}_{22}$ and $\overline{g}_{12}$ to the chiral and continuum limits using the fit function~(\ref{eq:fit1}).}	
\label{fig:fit1}
\end{figure}

The bare couplings $g_{mn}$ are computed by summing the densities $f^{mn}_{\gamma_i \gamma_5}(r)$ over radii $r$ and the renormalized couplings in the $\mathcal{O}(a)$-improved theory are given by\footnote{The renormalized $\mathcal{O}(a)$-improved axial current reads $\overline{A}_i = Z_A (1 + b_A a m_q)(A_i + a c_A \partial_i P)$ but the last term does not contribute at vanishing spatial momentum.}  $\overline{g}_{mn} = Z_A (1 + b_A a m_q)\, g_{mn}$ where $Z_A$ is the light axial vector current renormalisation constant computed in \cite{DellaMorteXB, Fritzsch:2012wq} and $b_A$ is an improvement coefficient computed in \cite{SintJX} at one loop order in perturbation theory. Values of $Z_A$ for each $\beta$ used in this work are given in Table~\ref{tab:ZA} and the associated error has been checked to be completely negligible at our level of precision. In the static limit of HQET, the renormalized couplings $\overline{g}_{11}$ and $\overline{g}_{22}$ are related to the $g_{B^{*}B\pi}$ and $g_{B^{*\prime} B^{\prime}\pi}$ couplings. On the other hand, $\overline{g}_{12}$ is related to the form factor $A_1(q^2)$ at $q^2=q^2_{\rm max} \neq 0$ whereas the $g_{B^{*\prime}B\pi}$ coupling is defined at $q^2=0$. Results for the bare couplings, obtained using the sGEVP method with $t_0>t/2$ for the lattice ensemble N6, are depicted in Fig.~\ref{fig:axial_sum_rule} where a comparison with the GEVP method result is also given. In the plateau region, where the contamination by higher excited states is negligible, data are fitted to a constant and results are given in Table~\ref{tab:sum_rule}. We have also checked that the results obtained using the GEVP method (Eq.~(\ref{eq:Agevp})) are in agreement within statistical error bars but with noisier plateaus, especially for $g_{12}$ and $g_{22}$ which involve the radial excitations. As the ensemble B6 has been added since our previous study, we have performed new extrapolations to the physical points using the fit ansatz
\begin{equation}
\overline{g}_{nm}(a,m_{\pi}) = \overline{g}_{nm} + C_1 \, a^2 + C_2 \, \widetilde{y} \,,
\label{eq:fit1}
\end{equation}
where $\widetilde{y} = m^2_{\pi} / \left( 8 \pi^2 f_{\pi}^2 \right)$ and $f_{\pi}$ is the pion decay constant \cite{AgasheKDA} (the physical value is $f_{\pi}=130.4~\MeV$ within our conventions). Since the fit parameter $C_2$ is compatible with zero we also tried the following fit function to estimate the systematic error in the quark mass extrapolation
\begin{equation}
\overline{g}_{nm}(a,m_{\pi}) = \overline{g}_{nm} + \widetilde{C}_1 \, a^2 \,. 
\label{eq:fit2}
\end{equation}
In the case of $\overline{g}_{11}$, the NLO formula of HM$\chi$PT is known \cite{Detmold:2011rb, Fajfer:2006hi}. Therefore we also tried the following fit ansatz which include the contribution from both positive and negative parity states
\begin{equation}
\overline{g}_{11}(a,m_{\pi}) = \overline{g}_{11} \left[ 1- (1+ 2 \, \overline{g}_{11}^{\, 2})\, \widetilde{y} \log \widetilde{y} - \frac{h^2 m_{\pi}^2}{16\,  \delta^2} \left( 3 + \frac{\widetilde{g}}{ \overline{g}_{11} }\right) \widetilde{y} \log \widetilde{y}\right]  + \overline{C}_1 \, a^2  + \overline{C}_2 \, \widetilde{y}\,. 
\label{eq:fit3}
\end{equation}
Here, the coupling $\widetilde{g}$ plays a role similar to $g$ but within the positive parity doublet $(B_0^{*}, B_1^{*})$ and the coupling $h$ is related to the transition between a scalar ($B_0^{*}$) and a pseudoscalar $B$ meson. The values of these couplings and of the mass difference $\delta=m_{B_0^{*}} - m_B$ between the scalar and pseudoscalar $B$ mesons are extracted from \cite{BlossierVEA}.
Extrapolations are shown in Fig.~\ref{fig:fit1} and results read
\begin{equation*} 
\overline{g}_{11} = 0.502(20)_{\stat}{(^{+8}_{-45})}_{\chi} \quad , \quad \overline{g}_{12} = -0.164(28)_{\stat}{(6)}_{\chi} \quad , \quad \overline{g}_{22} = 0.387(33)_{\stat}{(20)}_{\chi} \,,
\end{equation*}
where the first error is statistical and the second error includes the systematic error from the chiral extrapolation, estimated as half the difference between fit results using Eq.~(\ref{eq:fit1}) and Eq.~(\ref{eq:fit2}). For $\overline{g}_{11}$, we also used Eq.~(\ref{eq:fit3}) to estimate the systematic error. Results are in perfect agreement with those obtained in our previous paper \cite{BlossierQMA} (that had been obtained without the large volume set B6, and using time-diluted stochastic sources) and with the study \cite{BernardoniKLA} focused on $g_{11}$ and $g_{22}$ concerning the results obtained at $N_f=2$ and the physical point.

\renewcommand{\arraystretch}{1.1}
\begin{table}[t]
	\begin{center}
	\begin{tabular}{c@{\quad}c@{\quad}c@{\quad}c}
	\hline
	$\beta$	&	$5.2$	&	$5.3$	&	$5.5$	\\ 
	\colrule 
	$Z_A$	&	$0.7703$		&	$0.7784$		&	$0.7932$		\\ 
	\hline
	\end{tabular}
	\end{center}	
	\vspace{-0.3cm}
	\caption{Light axial-vector current renormalisation constant $Z_A$ for each value of $\beta$ \cite{DellaMorteXB, Fritzsch:2012wq}.}
	\label{tab:ZA}
\end{table}

\renewcommand{\arraystretch}{1.3} 
\begin{table}[t]
	\begin{center}
	\begin{tabular}{c@{\quad}c@{\quad}c@{\quad}c@{\quad}c@{\quad}c@{\quad}c@{\quad}c@{\quad}c}
	\hline
	id	&	$a\Sigma_{12}$	&	$\Sigma_{12}~[\GeV]$&	$g_{11}$	&	$g_{12}$	&	$g_{22}$&	$\overline{g}_{11}$	&	$\overline{g}_{12}$	&	$\overline{g}_{22}$	\\ 
	\hline 
	A5	&	$0.253(7)$	&	675(15)	&	$0.692(13)$	&	$-0.232(19)$	&	$0.583(18)$	&	$0.533(10)$	&	$-0.179(15)$	&	$0.449(14)$	\\  
	B6	&	$0.235(8)$	&	632(18)	&	$0.683(17)$	&	$-0.216(11)$	&	$0.555(17)$	&	$0.526(13)$	&	$-0.166(9)$	&	$0.428(13)$	\\  
	E5	&	$0.225(10)$	&	679(26)	&	$0.690(6)$	&	$-0.232(11)$	&	$0.587(16)$	&	$0.537(5)$	&	$-0.181(9)$	&	$0.457(12)$	\\  
	F6	&	$0.213(11)$	&	648(27) 	&	$0.688(14)$	&	$-0.264(14)$	&	$0.598(15)$	&	$0.536(11)$	&	$-0.206(11)$	&	$0.465(12)$	\\  
	N6	&	$0.166(9)$	&	681(33)	&	$0.658(10)$	&	$-0.209(12)$	&	$0.532(15)$	&	$0.522(7)$	&	$-0.166(9)$	&	$0.422(12)$	\\ 
		Q2	&	$0.195(5)$	& 641(16) &	$0.732(10)$	&	$-0.217(14)$	&	$0.598(20)$	&	$0.596(8)$	&	$-0.177(12)$	&	$0.488(16)$	\\ 
	\hline
	\end{tabular}
	\end{center}	
	\caption{Mass splitting $\Sigma_{12}=E_2-E_1$, bare couplings $g_{11}$, $g_{12}$, $g_{22}$ obtained using the sGEVP method and  renormalized couplings $\overline{g}_{11}$, $\overline{g}_{12}$, $\overline{g}_{22}$ for each lattice ensemble.}
	\label{tab:sum_rule}
\end{table}

\subsection{Discussions} 

\renewcommand{\arraystretch}{1.4}
\begin{table}[t!]
\begin{center}
\begin{tabular}{l@{\qquad}cc@{\qquad}cc@{\qquad}cc}
	\hline	
				& 		\multicolumn{2}{c}{$a=0.075~\fm$}	& \multicolumn{2}{c}{$a=0.065~\fm$}	& $a=0.048~\fm$ \\
	\cline{2-3} \cline{4-5} \cline{6-6}
	$m_{\pi}$			&	$330~\MeV$	&	$280~\MeV$	&	$440~\MeV$	&	$310~\MeV$	&	$340~\MeV$  \\
	\colrule 
	$\langle r^2 \rangle_A~[\fm^2]$	&	$0.398(38)$	&	$0.455(49)$	 &  	$0.358(15)$	&	$0.390(26)$	& 	$0.297(14)$  	\\ 
	\colrule 
	$r_n~[\fm]$	&	$0.369(13)$	&	$0.374(12)$	&  	$0.369(11)$	&	$0.379(20)$	& 	$0.365(12)$  	\\  
	\hline
 \end{tabular} 
\end{center}
\caption{Square radius of the ground state radial distribution $f^{(11)}_{\gamma_i\gamma_5}(r)$ and position of the node $r_n$ of the radial distribution $f^{(12)}_{\gamma_i\gamma_5}(r)$ for each lattice ensemble. }
\label{tab:square_radius}
\end{table}

The behavior of the densities helps to understand the small value of the coupling $g_{12}$ compared to the ground state coupling $g_{11}$. In particular, the presence of the node reduces significantly the value of the off-diagonal coupling. In the case of $g_{22}$, the densities fall slower than for $g_{11}$ and this coupling is not significantly suppressed. We have computed the first moment of the spatial component of the ground state radial distribution defined by
\begin{equation*}
\langle r^{2} \rangle_A = \frac{ \displaystyle \int_0^{\infty} \, \mathrm{d}r \, r^4 \, f^{(11)}_{\gamma_i\gamma_5}(r)  }{ \displaystyle  \int_0^{\infty} \, \mathrm{d}r \, r^2 \, f^{(11)}_{\gamma_i\gamma_5}(r) }\,,
\end{equation*}
and results are given in Table \ref{tab:square_radius}. We observe a clear dependence on both the pion mass and lattice spacing. Therefore, we have tried a linear extrapolation to the physical point and obtained $\langle r^2 \rangle_A = 0.251(41)~{\rm fm}^2$. The first moment probes the large $r$ region where the overlap of the tails is significant. In particular, the result are sensitive to the order of the polynomial function used to fit the data, explaining the large error in the determination of $\langle r^{2} \rangle_A$. We have also determined the position of the node $r_n$ of $f^{(12)}_{\gamma_i\gamma_5}(r)$ involving the radial excitation: results are collected in Table~\ref{tab:square_radius}. At our level of precision $r_n$ is remarkably stable and we do not see any dependence on the lattice spacing nor pion mass.

\subsection{Comparison with quenched results} 
\label{Quenched_results}

\begin{figure}[t]
	\begin{minipage}[c]{0.28\linewidth}
	\centering 
	\includegraphics*[width=\linewidth]{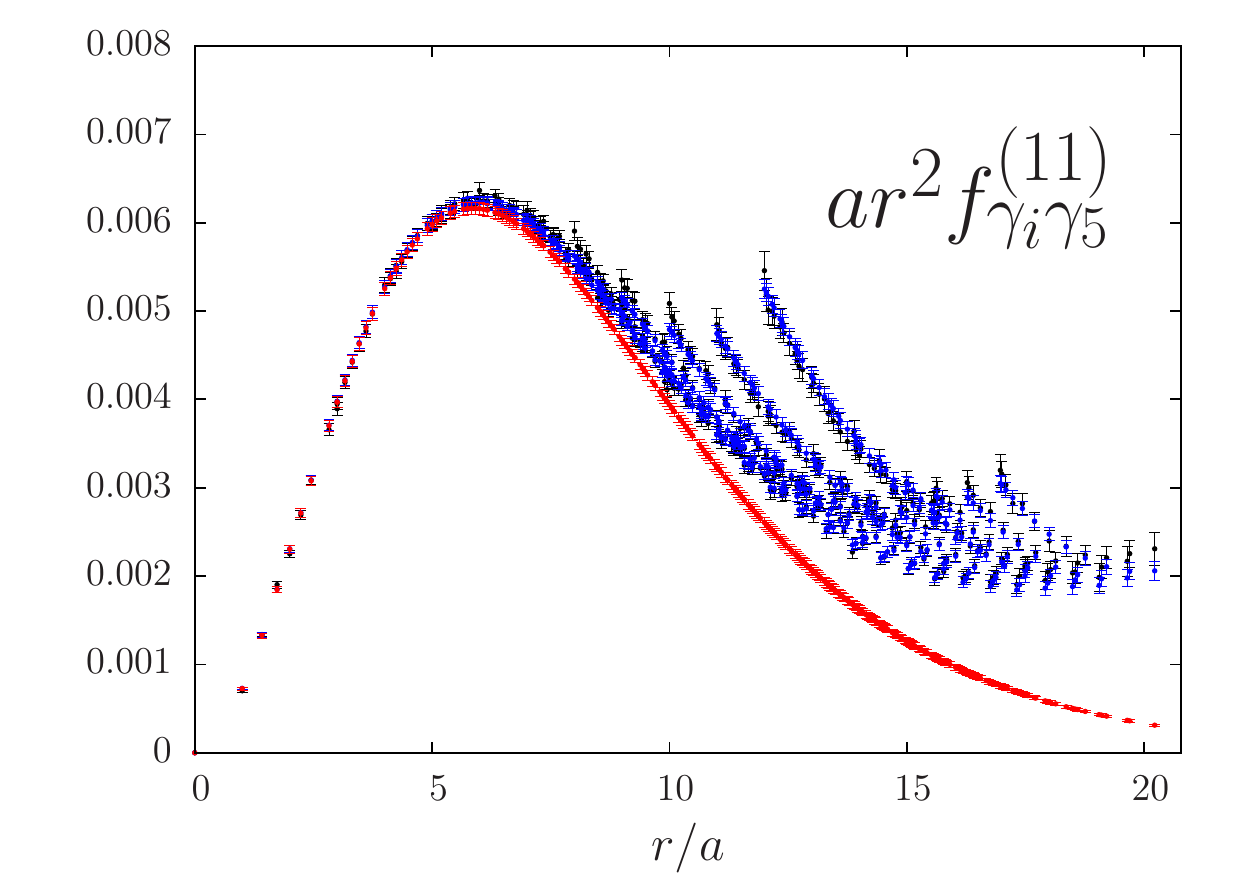}
	\end{minipage}
	\begin{minipage}[c]{0.28\linewidth}
	\centering 
	\includegraphics*[width=\linewidth]{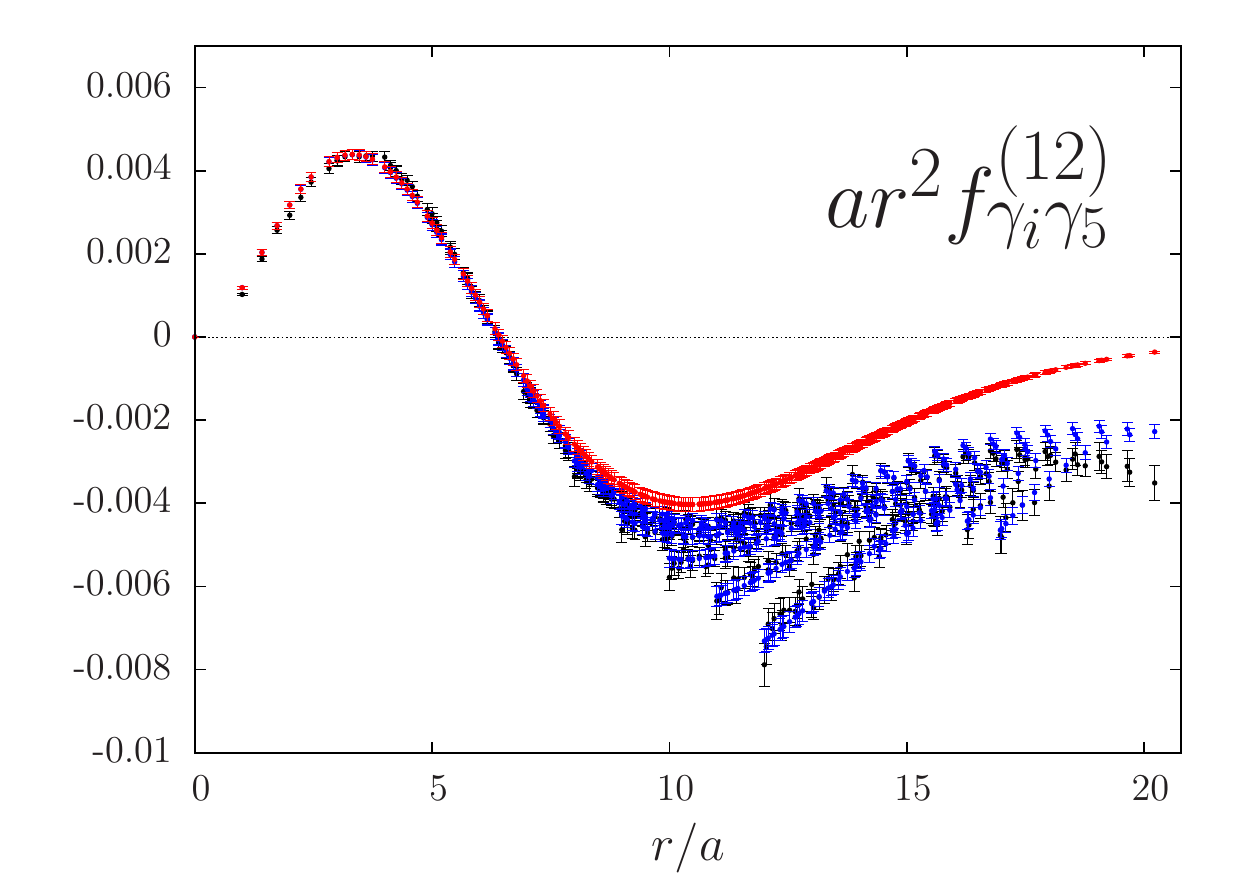}
	\end{minipage}
	\begin{minipage}[c]{0.28\linewidth}
	\centering 
	\includegraphics*[width=\linewidth]{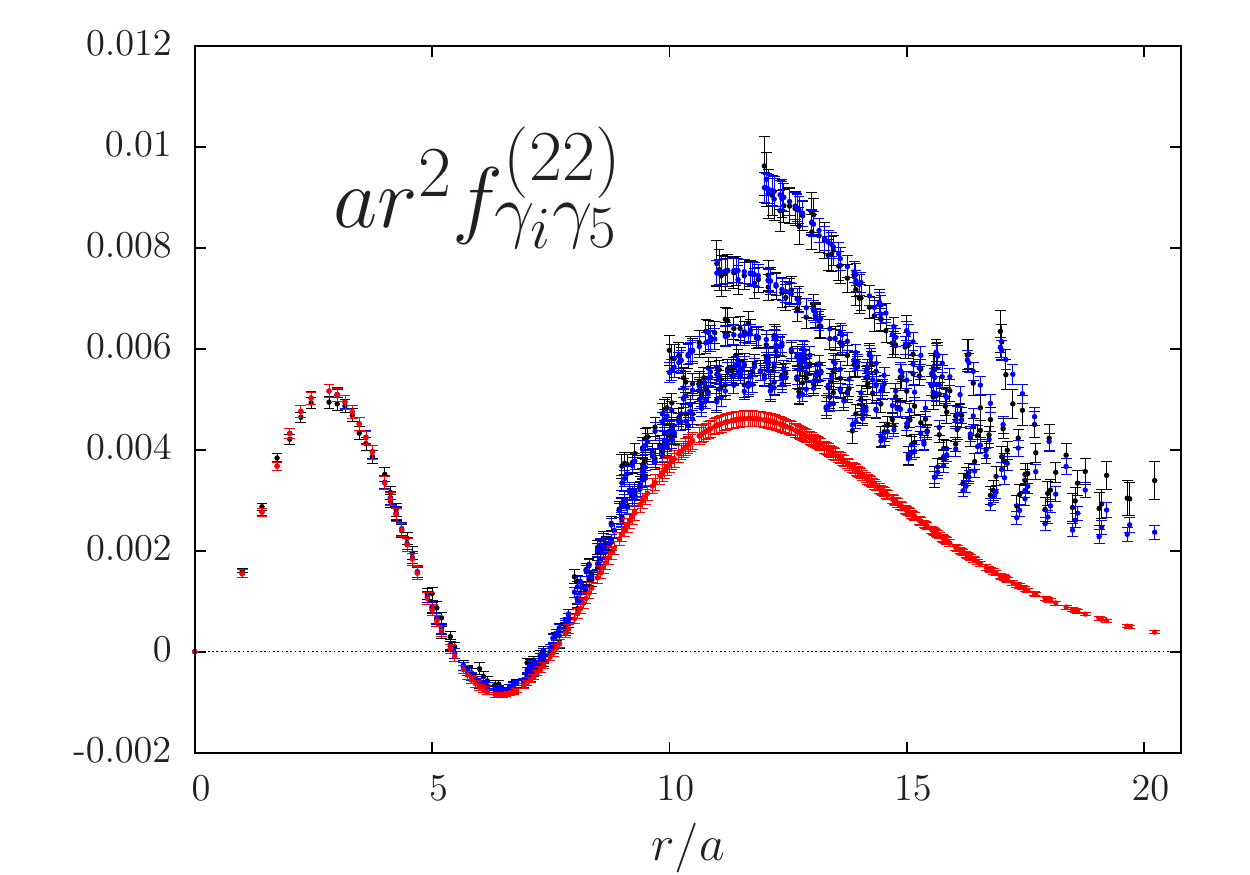}
	\end{minipage}

	\caption{Density distributions $ar^2\, f^{(11)}_{\gamma_i\gamma_5}(r/a)$, $ar^2\, f^{(12)}_{\gamma_i\gamma_5}(r/a)$ and $ar^2\, f^{(22)}_{\gamma_i\gamma_5}(r/a)$ for our quenched lattice ensemble Q2. The blue curves correspond to results after a fit and the red curves to the distributions after removing volume effects.}	

	\label{fig:quenched_distrib}
\end{figure}

\begin{figure}[t]

	\begin{minipage}[c]{0.28\linewidth}
	\centering 
	\includegraphics*[width=\linewidth]{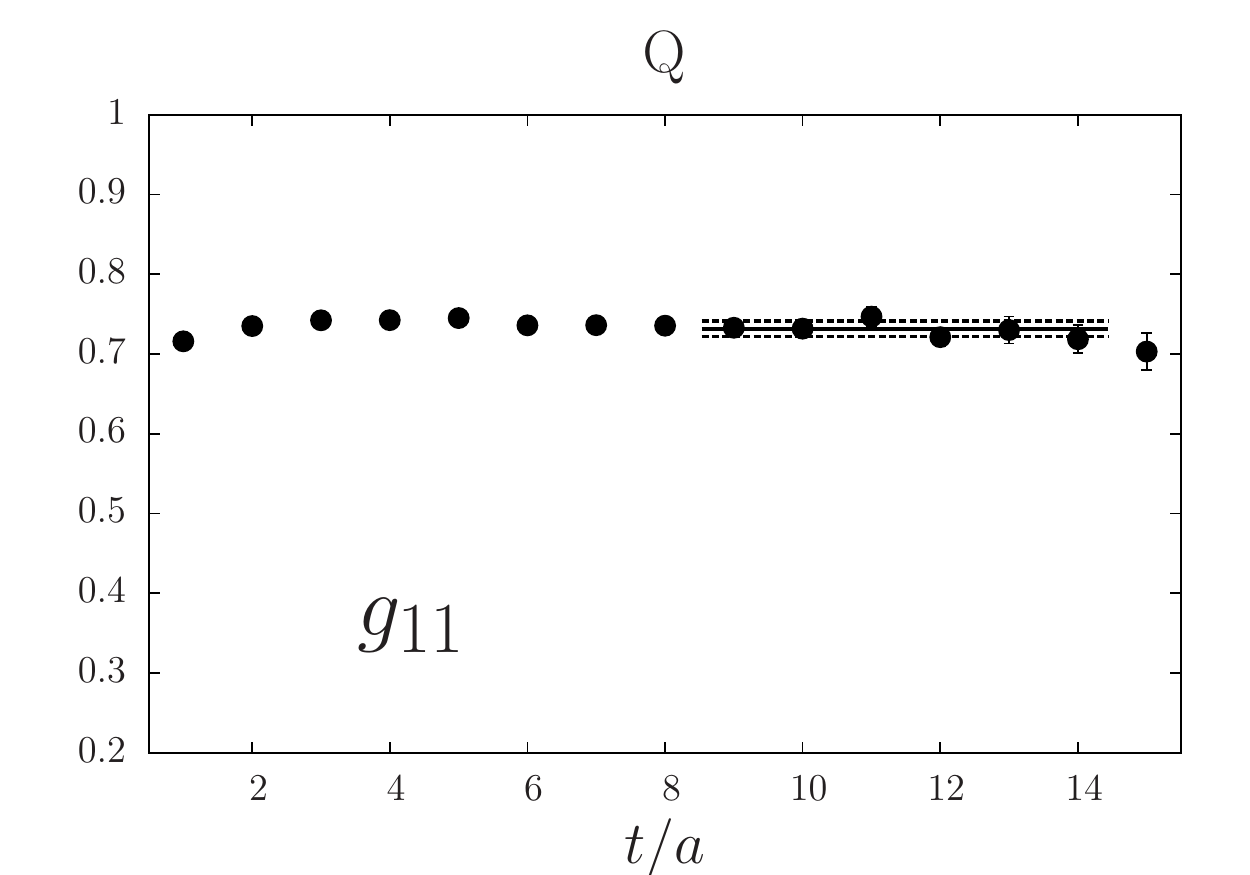}
	\end{minipage}
	\begin{minipage}[c]{0.28\linewidth}
	\centering 
	\includegraphics*[width=\linewidth]{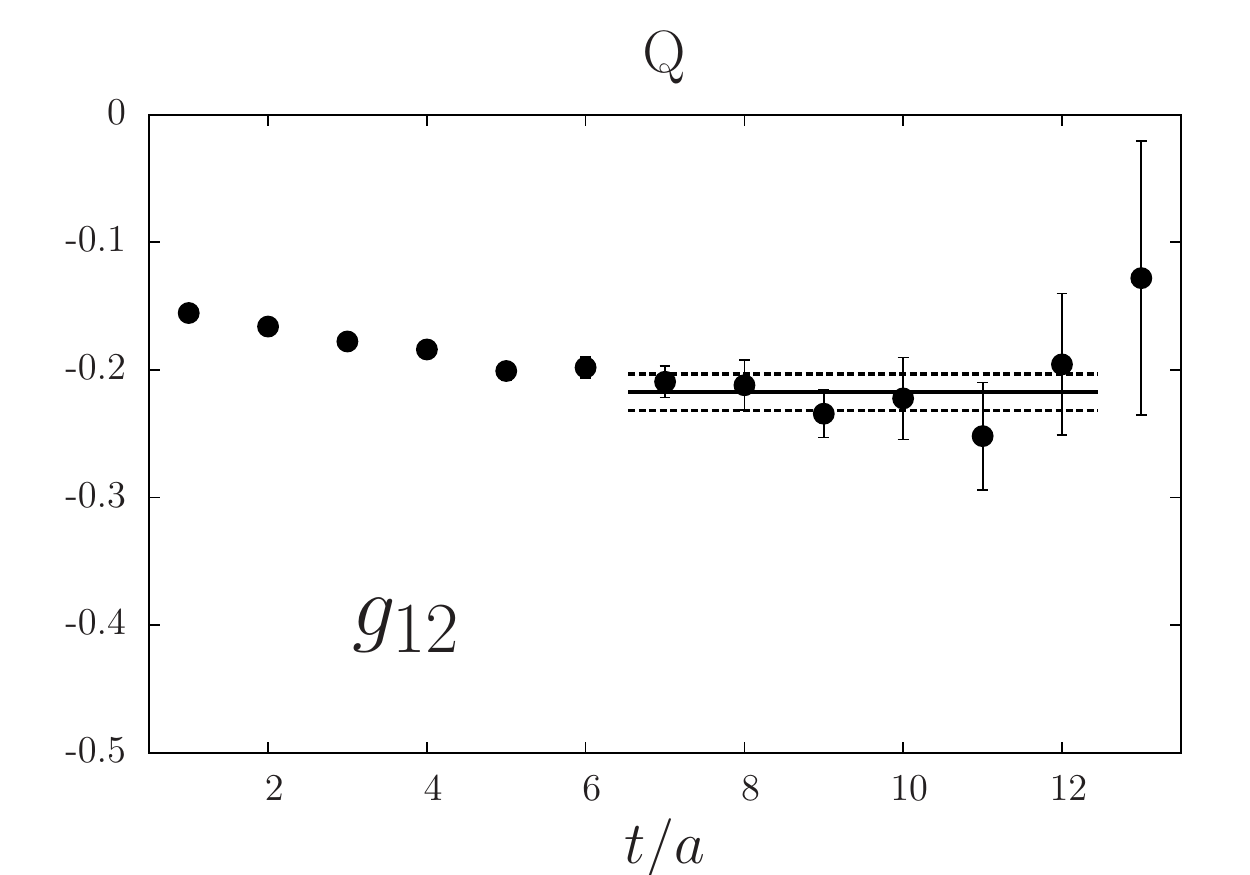}
	\end{minipage}
	\begin{minipage}[c]{0.28\linewidth}
	\centering 
	\includegraphics*[width=\linewidth]{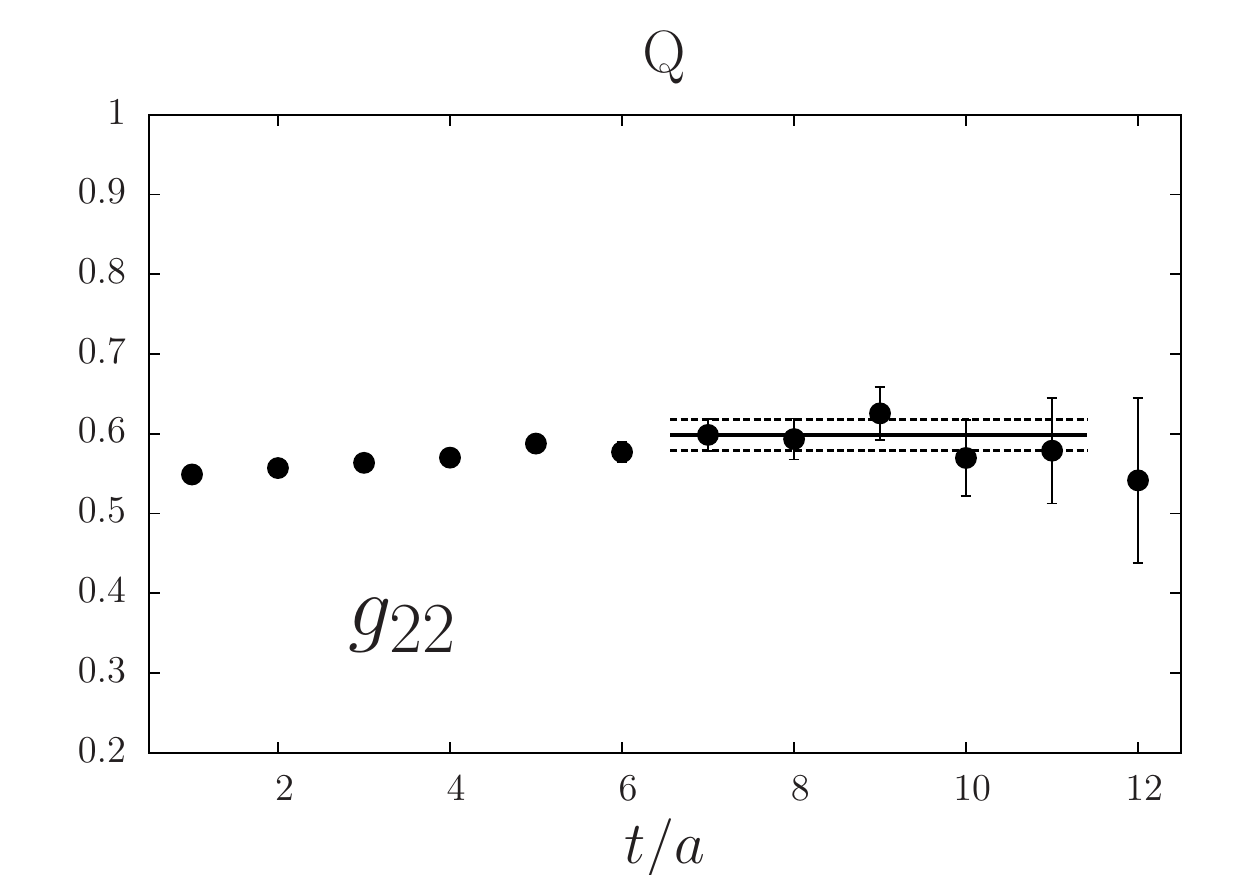}
	\end{minipage}

	\caption{Plateaus for the summation over $r$ of the axial densities computed in section~\ref{sec:sumrules} for our quenched lattice ensemble Q2: $g_{11}$, $g_{12}$ and $g_{22}$.}	
	\label{fig:quenched_couplings}
\end{figure}

We have repeated the same analysis for the quenched lattice ensembles Q1 and Q2, at the strange quark mass determined in \cite{GardenFG}. Results are plotted in Fig.~\ref{fig:quenched_distrib} where the volume effects are taken into account using the method presented in Section \ref{sec:volume_effects}. Our results read $\alpha=1.40(1)$, $r_0 = 0.32(1)~{\rm fm}$ and $\langle r^2 \rangle_A = 0.319(8)~{\rm fm}^2$. The position of the node of $f^{(12)}_{\gamma_i\gamma_5}$ is $r_n=0.390(9)~{\rm fm}$. Concerning the summation over $r$, results for $\Sigma_{12}$ and the couplings $g_{mn}$ are given in Table~\ref{tab:sum_rule}. In particular, the value of the renormalized coupling $\overline{g}_{11}$ is in perfect agreement with \cite{BernardoniKLA}. Here, we used the value $Z_A = 0.81517$ from ref.~\cite{LuscherJN} and the plateaus are depicted in Fig.~\ref{fig:quenched_couplings}. \\

\begin{figure}[t!]

	\begin{minipage}[c]{0.45\linewidth}
	\centering 
	\includegraphics*[width=0.9\linewidth]{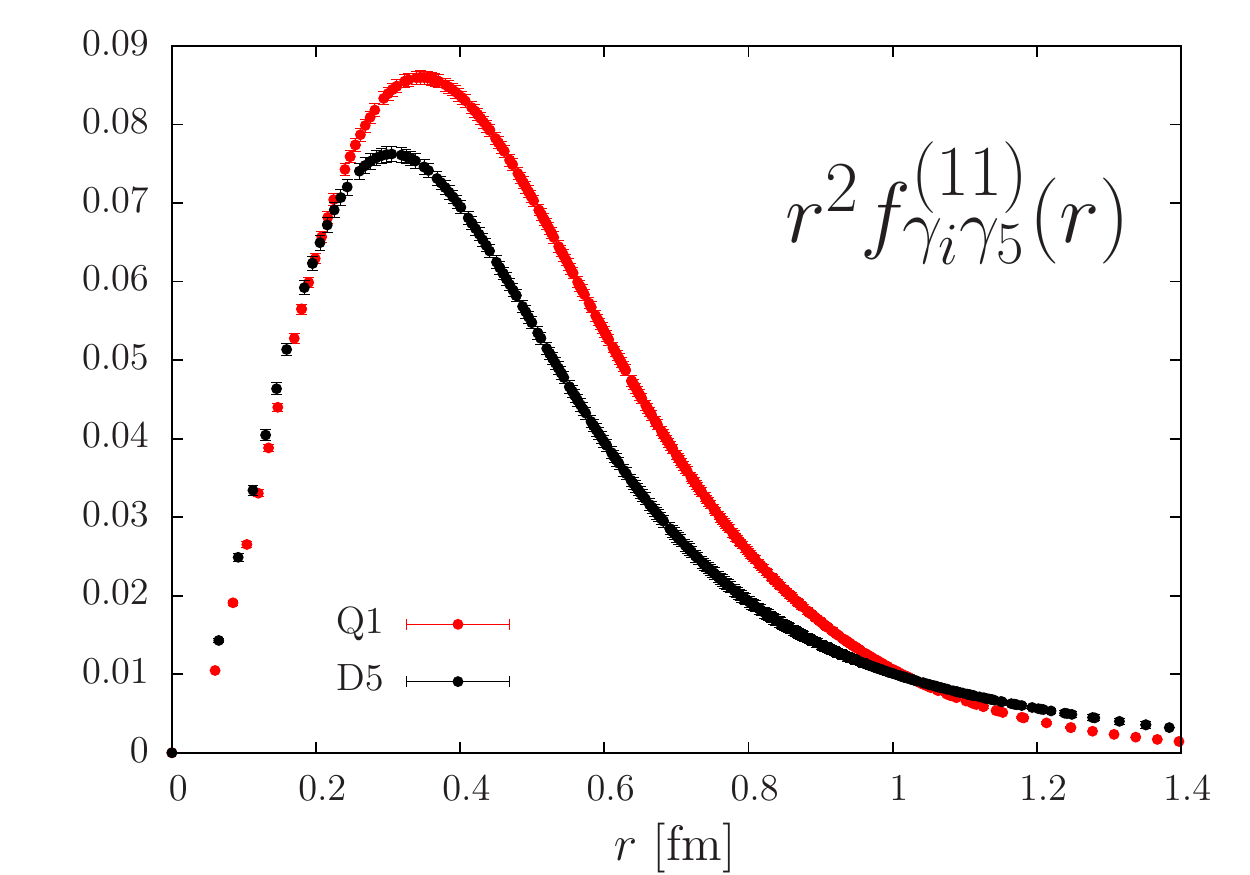}
	\end{minipage}
	\begin{minipage}[c]{0.45\linewidth}
	\centering 
	\includegraphics*[width=0.9\linewidth]{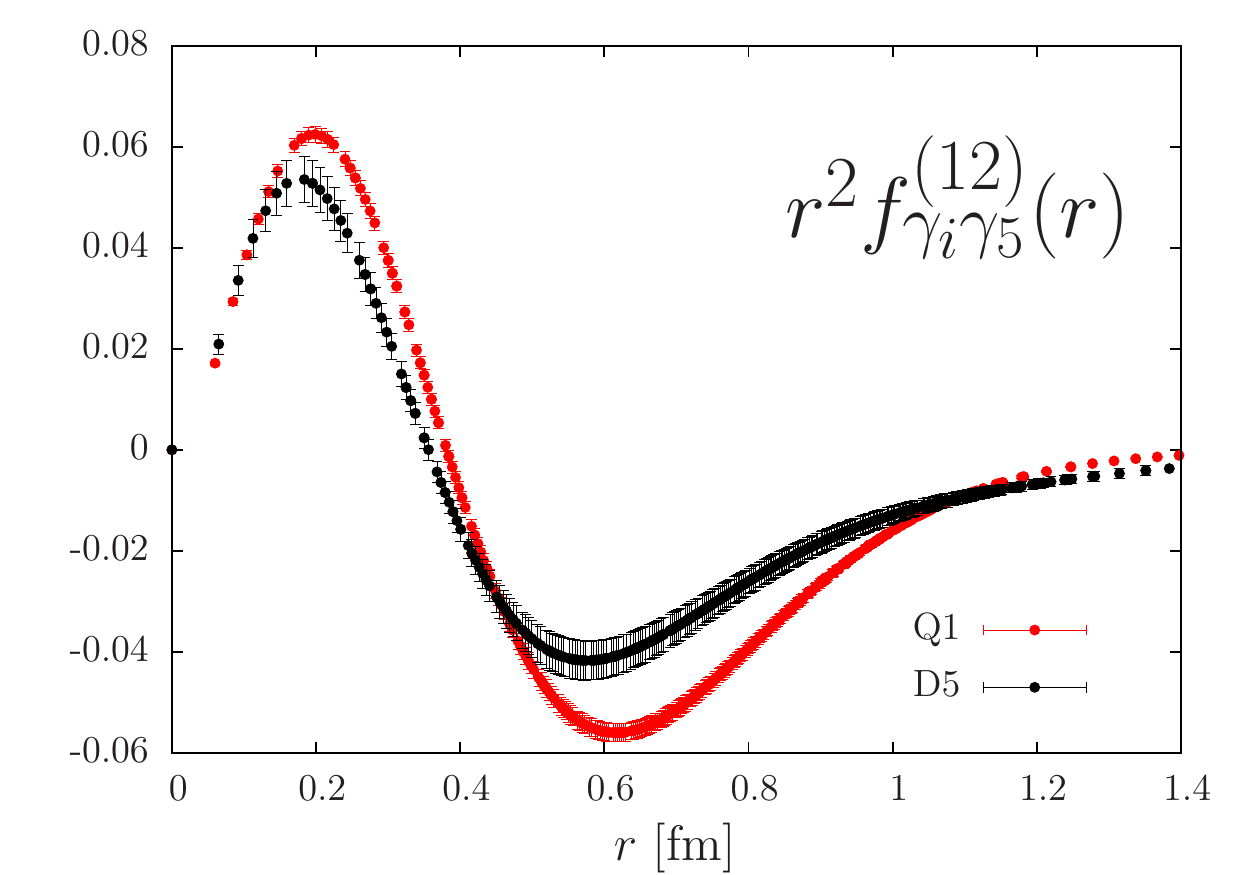}
	\end{minipage}
	
	\caption{Comparison of the renormalized distributions $r^2\, f^{(mn)}_{\gamma_i\gamma_5}(r)$ in physical units. The quenched result (ensemble Q1) is plotted in red and the dynamical case in black (ensemble D5).}	

	\label{fig:quenched_cmp}
\end{figure}

To compare quenched results with the ${\rm N_f}=2$ case, we have plotted in Fig.~\ref{fig:quenched_cmp} the axial distributions for two sets that are close from each other in parameter space, except in $N_f$. For the ground state distribution $f^{(11)}_{\gamma_{i}\gamma_5}$, one notices a faster fall-off in the quenched case ($\alpha_{n_f=0}=1.40 > \alpha_{n_f=2} = 1$). However, in that case, the distribution is more spread, and of a larger magnitude: it explains the larger value of the quenched coupling $g_{11}$. Concerning the axial distribution $f^{(12)}_{\gamma_{i}\gamma_5}$, the position of the node is slightly higher in the quenched case without being able to provide any explanation of that observation.

\subsection{Time component of the axial radial density distributions} 

\begin{figure}[t]

	\begin{minipage}[c]{0.49\linewidth}
	\centering 
	\includegraphics*[width=0.99\linewidth]{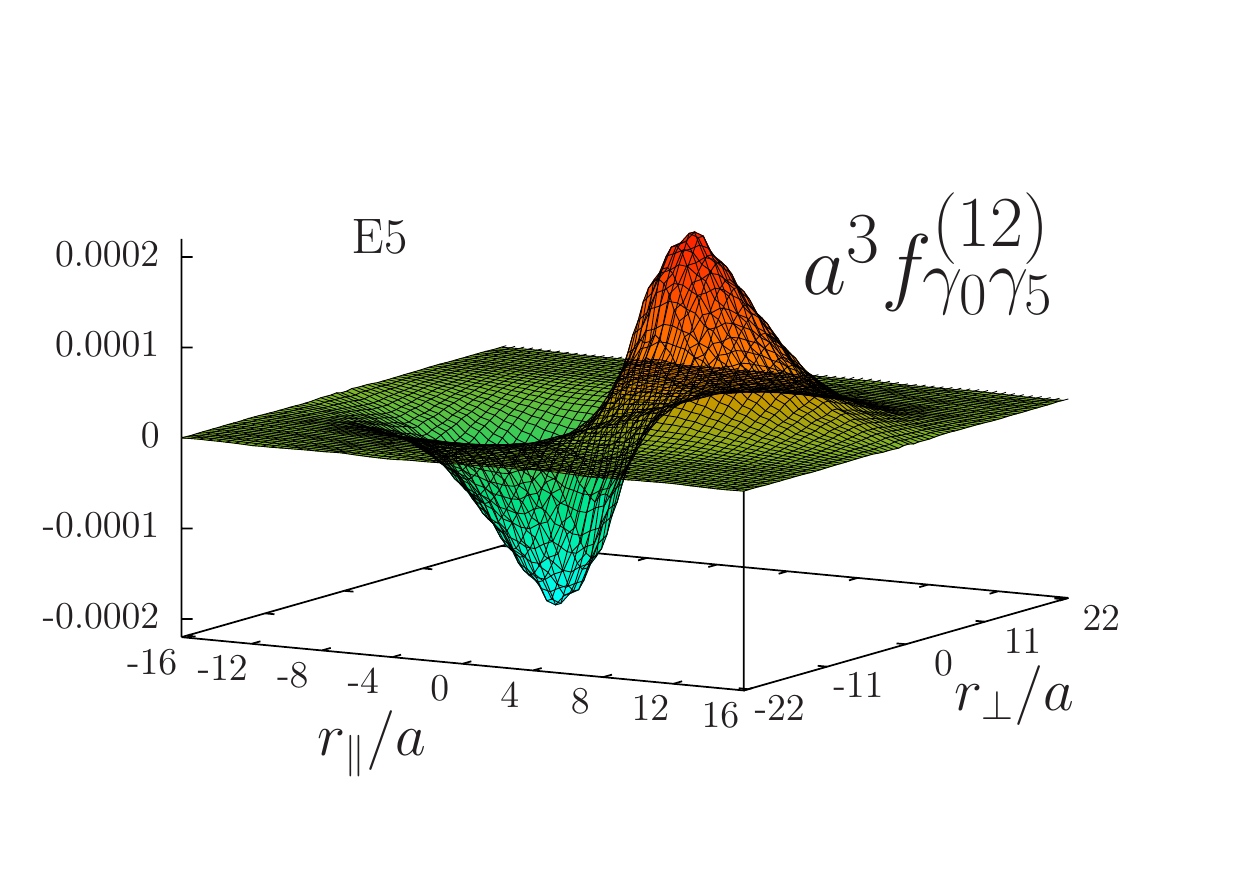}		
	\end{minipage}
	\begin{minipage}[c]{0.49\linewidth}
	\centering 
	\includegraphics*[width=0.99\linewidth]{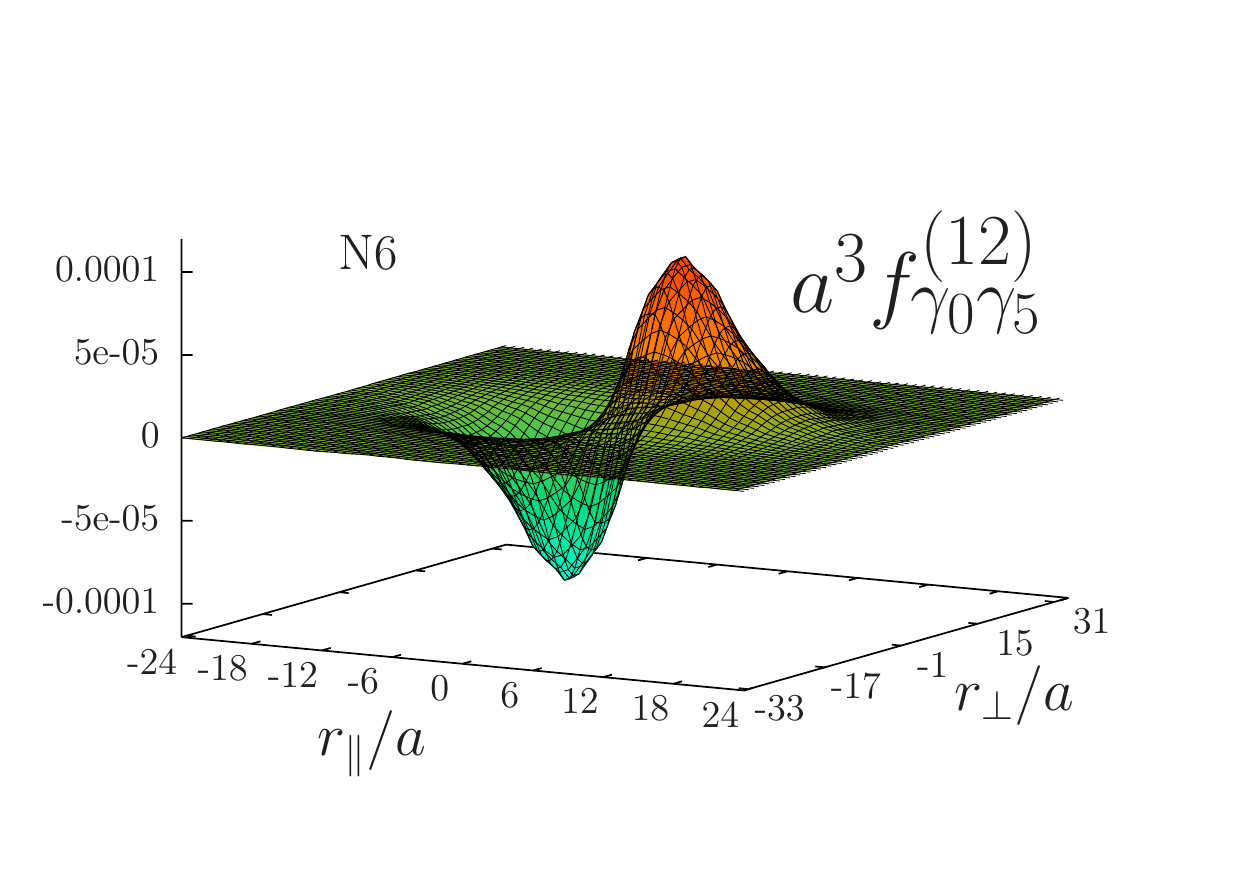}		
	\end{minipage}
	
	\caption{Imaginary part of the time component $f^{(12)}_{\gamma_0\gamma_5}(r)$ of the axial density distribution for the lattice ensembles E5 and N6. Choosing $i=z$ we define $r_{\parallel} = r_z$ and $r_{\perp} = \sqrt{r_x^2+r_y^2}$.
\label{fig:distrib_time} }
\end{figure}

The time component of the axial radial density distribution is odd with respect to the projection $r_i$ along the vector meson polarisation (see Section \ref{sec2}). Therefore, we have only averaged the raw data over equivalent points in the plane orthogonal to direction $i$, which corresponds to $(N+1)^2(N+2)/2$ independent points where $N = L/(2a)$. Results of the axial density distribution $f^{(12)}_{\gamma_0 \gamma_5}$ are depicted in Fig.~\ref{fig:distrib_time} for the lattice ensembles E5 and N6.

\subsection{The $g_{B^{*\prime}B\pi}$ coupling}

\begin{figure}[t]
	\hspace{-0.7cm}
	\begin{minipage}[c]{0.45\linewidth}
	\centering 
	\includegraphics*[width=0.9\linewidth]{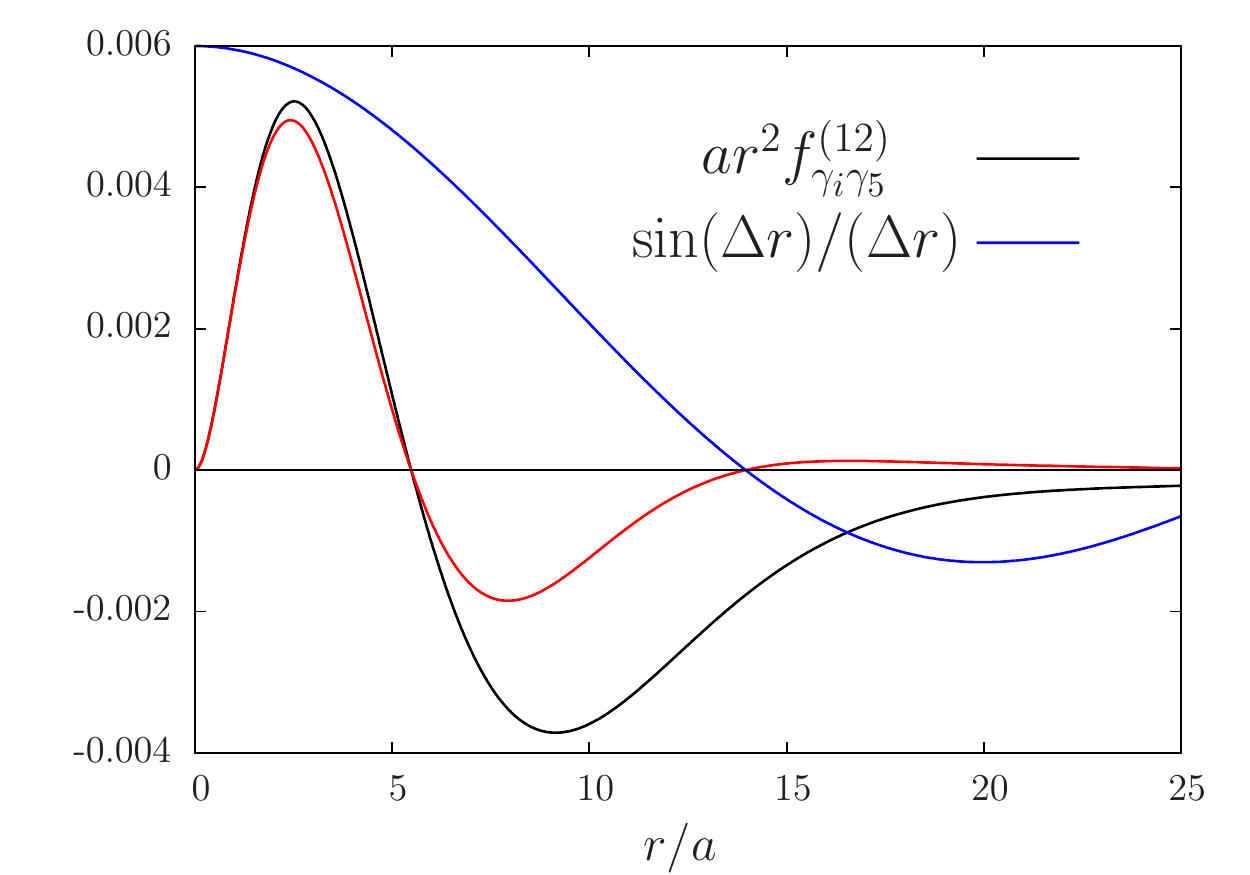}
	\end{minipage}
	\begin{minipage}[c]{0.45\linewidth}
	\centering 
	\includegraphics*[width=0.9\linewidth]{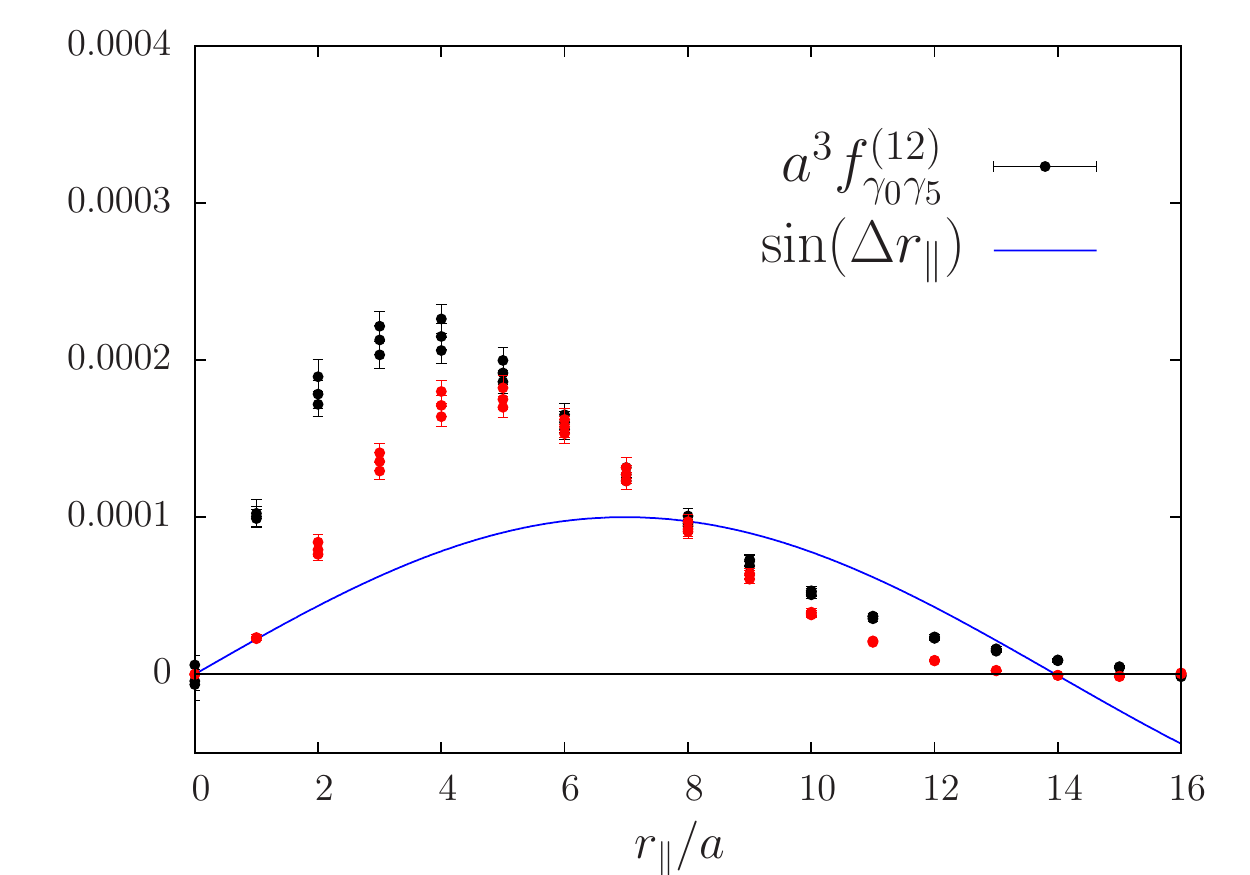}
	\end{minipage}	
	\caption{Integrand in Eqs.~(\ref{eq:extrap_q_spat}) and (\ref{eq:extrap_q_time}) to obtain the form factor at $q^2=0$. The red curve is the product of the blue and dark curves and the normalisation of the blue curve is arbitrary. For the time component (right plot), we plot the imaginary part and only points with $|r_{\perp}|\leq 1$ are shown.}
	\label{fig:numerical_int}		
\end{figure}

To compute the form factors associated to $\langle B|A_\mu|B^{*\prime} \rangle$ at every $q^2$ from $q^2=q^2_{\rm max}$, and especially at the kinematical point $q^2=0$, we consider  the Fourier transform of the radial density distributions at a momentum $\vec{q}$ aligned with the polarisation vector $\vec{\epsilon}$, namely
\begin{equation}
\mathcal{M}_{\mu}(q^2_{\rm max} - \vec{q}\,^2) = \int \mathrm{d} \vec{r} \, f_{\gamma_{\mu} \gamma_5}^{(12)}(\vec{r}) \, e^{i\vec{q}\cdot \vec{r}} \,.
\label{eq:extrap_q}
\end{equation}
For the spatial component of the distribution, using the radial symmetry, one obtains
\begin{equation}
\mathcal{M}_i (q^2_{\rm max} - \vec{q}\,^2) = 4\pi \int_0^{\infty} \mathrm{d}r \ r^2 \, \frac{\sin(|\vec{q}|r)}{|\vec{q}| r}  f_{\gamma_{i} \gamma_5}^{(12)}(\vec{r}) \,,
\label{eq:extrap_q_spat}
\end{equation}
and the special case $q^2=0$ corresponds to $|\vec{q}| = \Delta = m_{B^{*\prime}} - m_B$. 
Concerning the time component of the distribution, the radial symmetry is lost but we still have a cylindrical symmetry with respect to the axis $r_{\parallel}$ ($r_{\parallel}$ refers to the direction given by $\vec{q}$) which leads to\footnote{With a covariant index, $q_i \equiv -|\vec{q}|$.}
\begin{equation}
\frac{q_0}{q_i} \mathcal{M}_0 (q^2_{\rm max} - \vec{q}\,^2) = - q_0 4 i \pi  \int_{0}^{\infty} \mathrm{d} r_{\parallel}  \int_{0}^{\infty} \mathrm{d} r_{\perp} \, r_{\perp} \, f^{(12)}_{\gamma_0\gamma_5}(r_{\parallel},r_{\perp}) \, \frac{\sin(|\vec{q}| \, r_{\parallel})}{|\vec{q}|}  \,.
\label{eq:extrap_q_time}
\end{equation}
Here, we cannot use the method described in previous sections to cure finite volume effects and the sum~(\ref{eq:extrap_q}) is simply replaced by a discrete sum. However, this approximation is expected to be good since Fourier transforms lower the contribution from large radii at $q^2=0$, as can be seen in Fig.~\ref{fig:numerical_int} where we plot the integrand of Eqs.~(\ref{eq:extrap_q_spat}) and (\ref{eq:extrap_q_time}). At $q^2=0$, the small $r$ region is enhanced whereas the large $r$ region contribution is reduced. In particular, results are not affected by the large radii behavior of the function used to fit the data in Eq.~(\ref{eq:extrap_q_spat}).

\begin{figure}[t] 
	\begin{minipage}[c]{0.33\linewidth}
	\centering 
	\includegraphics*[width=\linewidth]{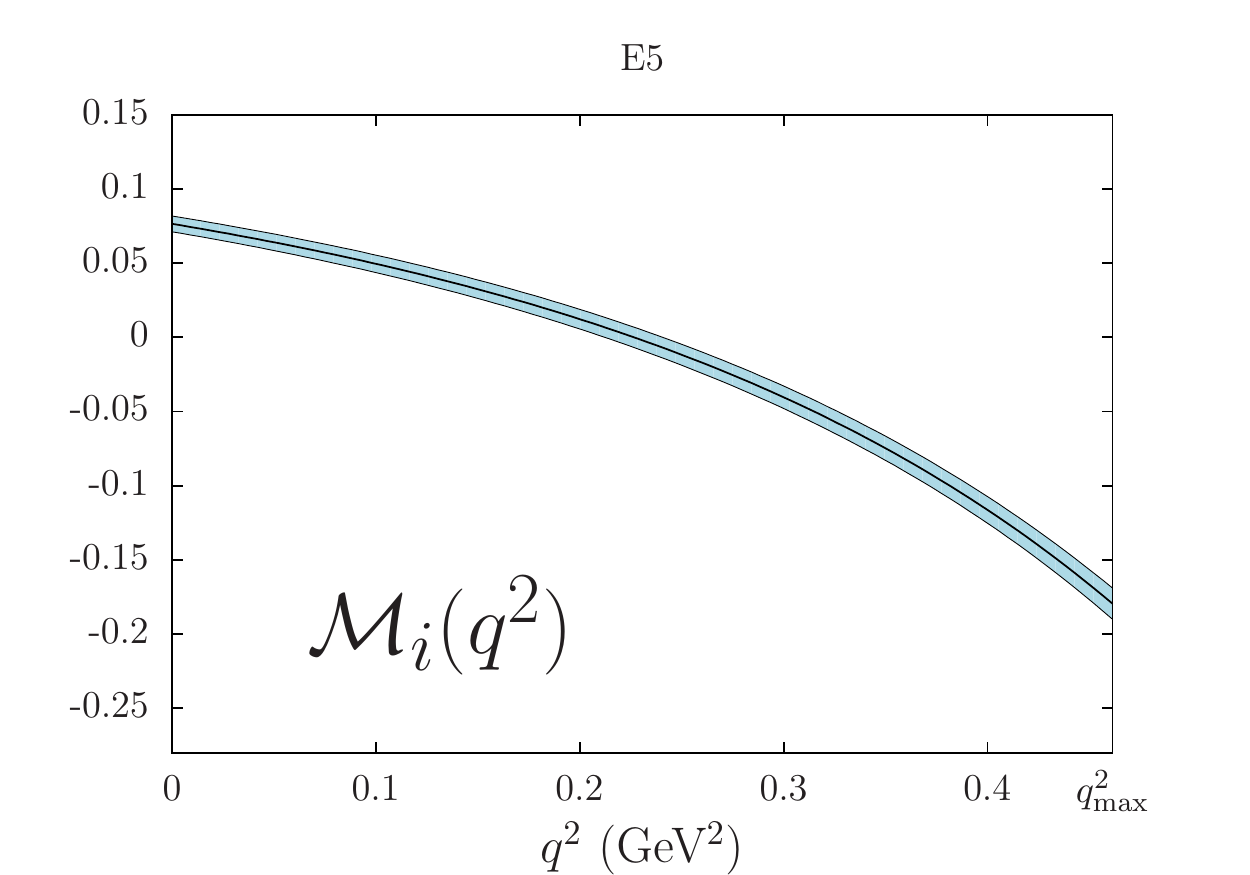}
	\end{minipage}
	\begin{minipage}[c]{0.33\linewidth}
	\centering 
	\includegraphics*[width=\linewidth]{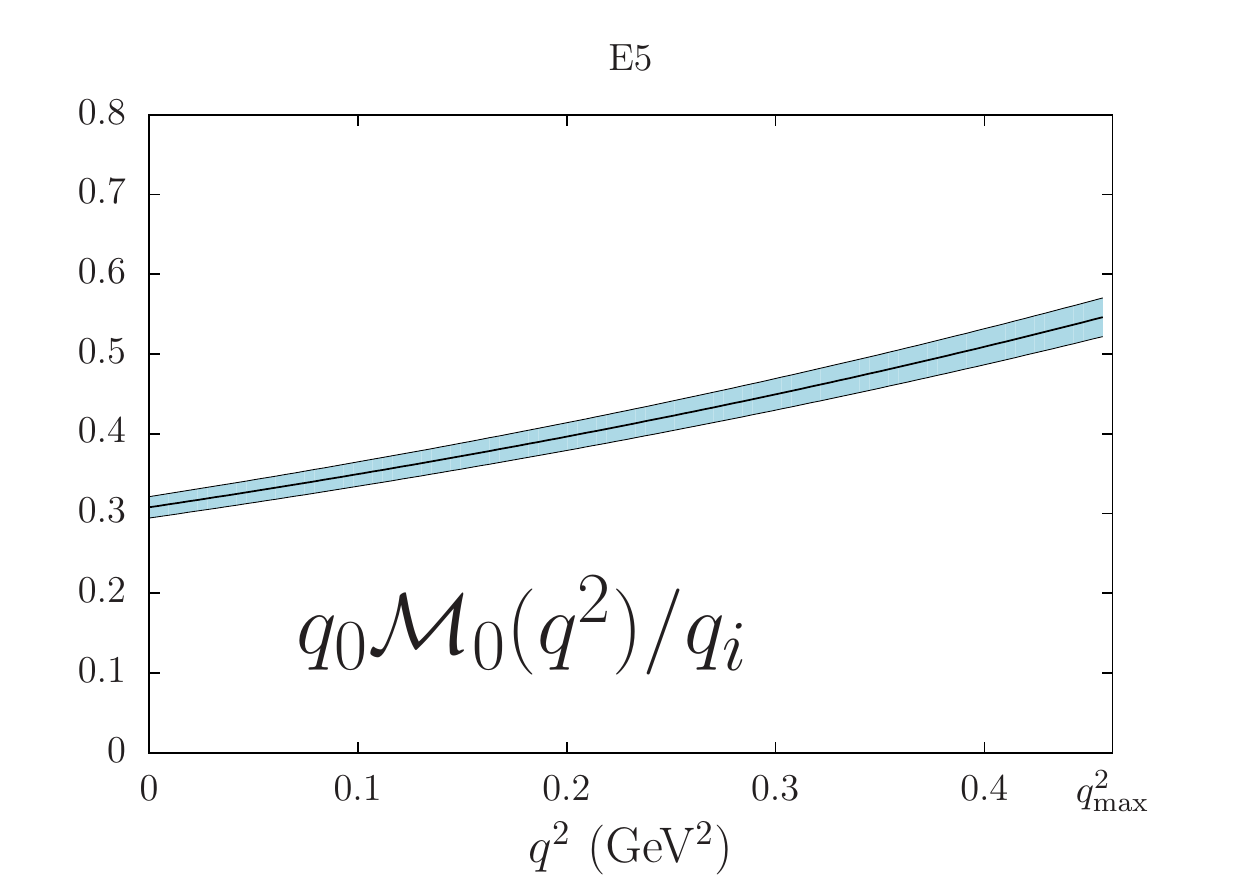}
	\end{minipage}
	\begin{minipage}[c]{0.32\linewidth}
	\centering 
	\includegraphics*[width=\linewidth]{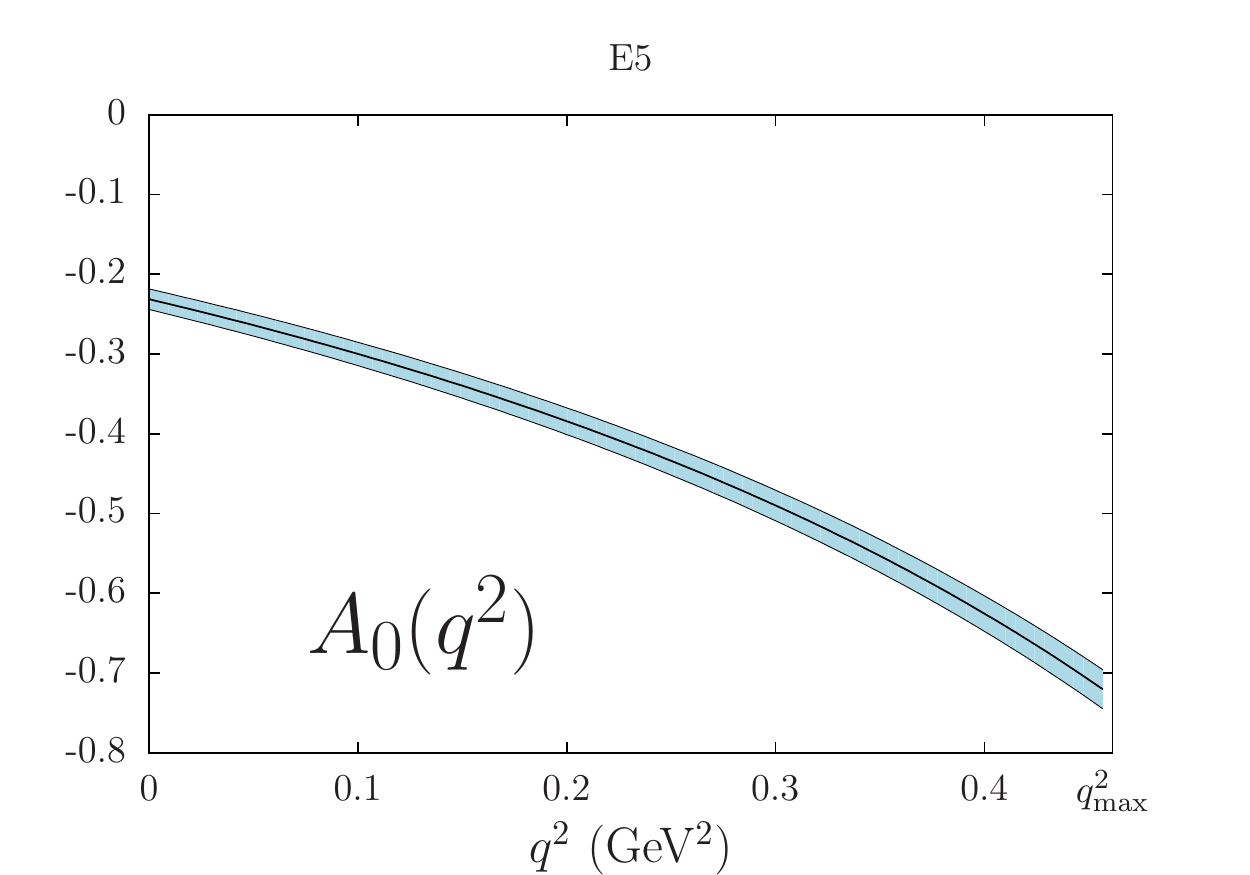}
	\end{minipage}	
	\caption{$q^2$ dependence of the spatial (left) and time(middle) contributions to the form factor $A^{12}_{0}(q^2)$ for the lattice ensemble E5. It should be stressed that the form factor $A_0$ is interesting \emph{per se} only at $q^2=0$, where it is related to the $g_{B^{*\prime} B\pi}$ coupling.}	
\label{fig:RQ}
\end{figure}

\subsubsection{Lattice results} 

\begin{figure}[t]
	\centering 
	\includegraphics*[width=0.45\linewidth]{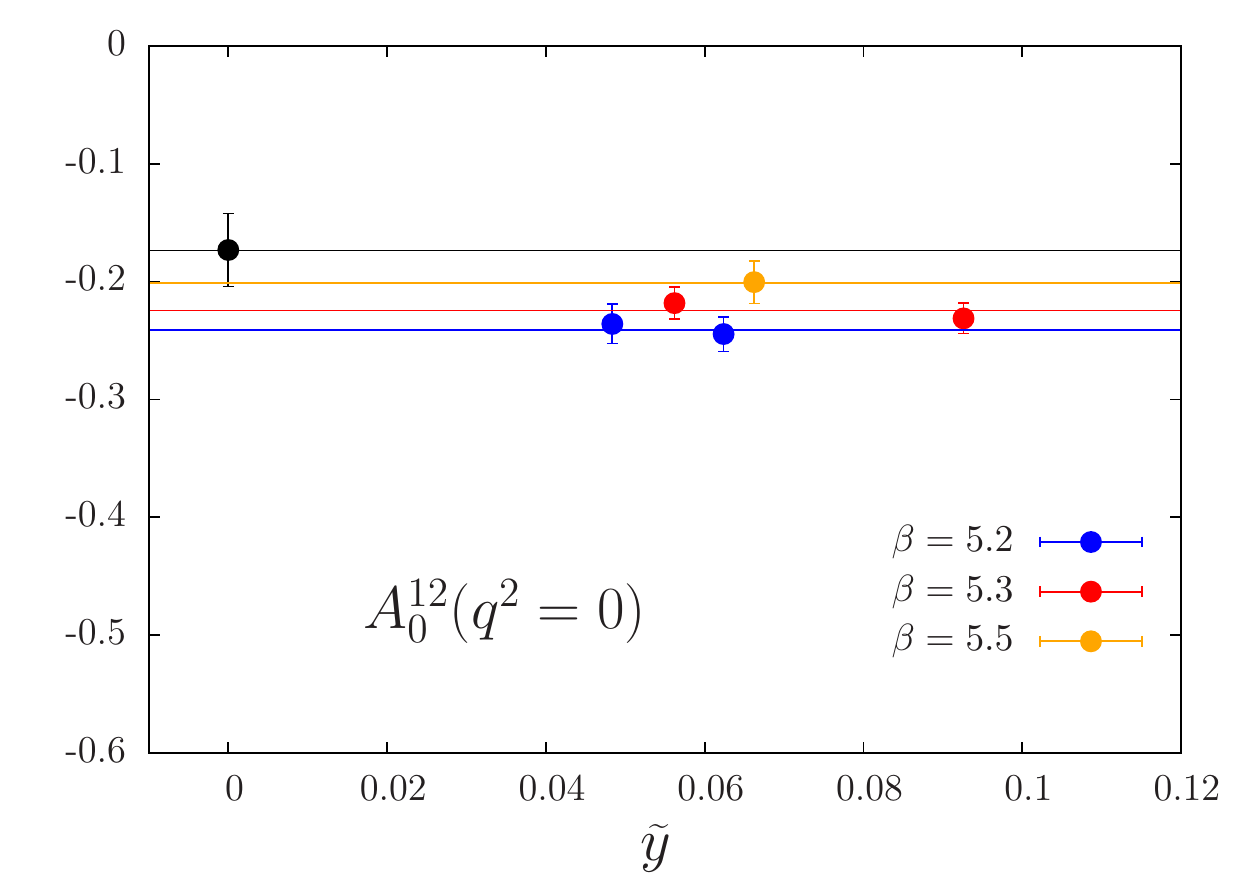}
	\caption{Extrapolations of the form factor $A^{12}_0$ at $q^2=0$ to the chiral and continuum limits.}	
\label{fig:all_extrap}
\end{figure}

In Eq.~(\ref{eq:formfac}), there is no sum over $i$ but we average the data over equivalent directions on the lattice. Results for the spatial and time components of distributions are depicted in Fig.~\ref{fig:RQ} for $q^2$ in the range $[0, q^2_{\rm max}]$ and the lattice set E5.  All results are collected in Table.~\ref{tab:res_ff}: we observe a large variation of the spatial component between $q^2=0$ and $q^2_{\max}$ and even a change of sign whereas the time component is slowly varying: it dominates in magnitude over the spatial component.  Finally, we have performed a chiral and continuum extrapolations using the same fit formulae as $\overline{g}_{11}$. Findings for each component at both $q^2=0$ and $q^2=q^2_{\rm max}$ are given in Table~\ref{tab:ff_comp} and the extrapolations of the form factor $A^{12}_0$ at $q^2=0$ is depicted in Fig.~\ref{fig:all_extrap}. The small difference between $\overline{g}_{12}$ computed in Sec.~\ref{sec:sumrules} and $\mathcal{M}_i(q^2_{\max})$ can be explained by the use of the fitted densities in the latter case. However, both results are perfectly compatible within error bars. Our final results read
\begin{equation*} 
A^{12}_0(0) =  -0.173(31)_{\stat}(16)_{\syst} \quad , \quad g_{B^{*\prime} B\pi} =  -15.9(2.8)_{\stat}(1.4)_{\syst} \,,
\end{equation*}
where the first error is statistical and the second error includes the systematics coming from the chiral extrapolation and the uncertainty associated to $\Delta = 701(65)~\MeV$, the mass difference between the ground state and the first radial excitation of the $B$ meson. It is useful to remember that the sign of the form factor corresponds to the convention where all decay constants are positive. Rigorously, $\mathcal{O}(a)$-improvement is only partially implemented for off-shell form factors. Therefore, we also tried a linear fit in the lattice spacing but it failed to reproduce the data, which strongly suggests that $\mathcal{O}(a)$ artefacts are small.

\renewcommand{\arraystretch}{1.3}  
\begin{table}[t]
	\begin{center}
	\begin{tabular}{l@{\quad}l@{\quad}c@{\quad}c@{\quad}c@{\quad}c@{\quad}c@{\quad}c@{\quad}c}
	\hline
				&	& 	A5 			&	B6			&	E5			&	F6			&	N6 &	Q2		\\ 
	\hline 
	\multirow{2}{*}{$q_0 \mathcal{M}_0(q^2)/q_i$} 	&	$q^2=q^2_{\rm max}$ &	$0.669(33)$			&	$0.675(45)$	&	$0.546(24)$ 	&	$0.559(30)$		&	$0.487(31)$	& $0.253(18)$\\ 
	 	&	$q^2=0$	&	$0.347(17)$			&	$0.301(15)$	&	$0.308(13)$	&	$0.282(13)$		&	$0.266(16)$	& $0.196(14)$\\ 
	\hline
	\multirow{2}{*}{$\mathcal{M}_i(q^2)$}	 	&	$q^2=q^2_{\rm max}$	&	$-0.172(11)$			&	$-0.161(12)$		&	$-0.180(8)$		&	$-0.184(12)$		&	$-0.166(9)$	& $-0.154(8)$ \\
	 	&	$q^2=0$	&	$0.072(4)$		&	$0.065(8)$		&	$0.076(4)$		&	$0.063(6)$		&	$0.065(5)$ & $0.053(4)$\\
	\hline
	\end{tabular}
	\end{center}	
	\caption{Time and spatial contributions to the form factor $A^{12}_0$, at two different values of $q^2$, for each lattice ensemble.}
	\label{tab:res_ff}
\end{table}

\subsubsection{Comparison with quenched data} 

\begin{figure}[t]

	\begin{minipage}[c]{0.49\linewidth}
	\centering 
	\includegraphics*[width=0.9\linewidth]{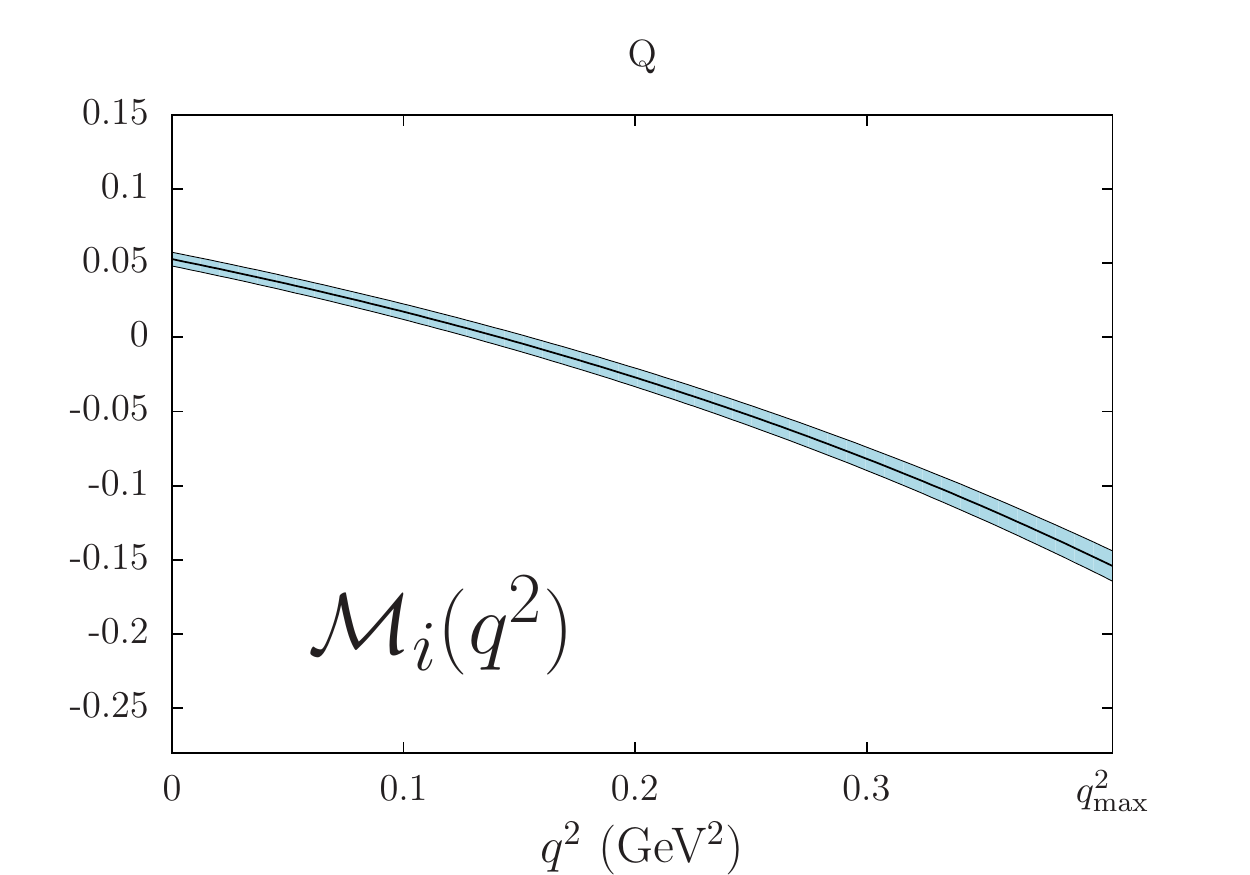}
	\end{minipage}
	\begin{minipage}[c]{0.49\linewidth}
	\centering 
	\includegraphics*[width=0.9\linewidth]{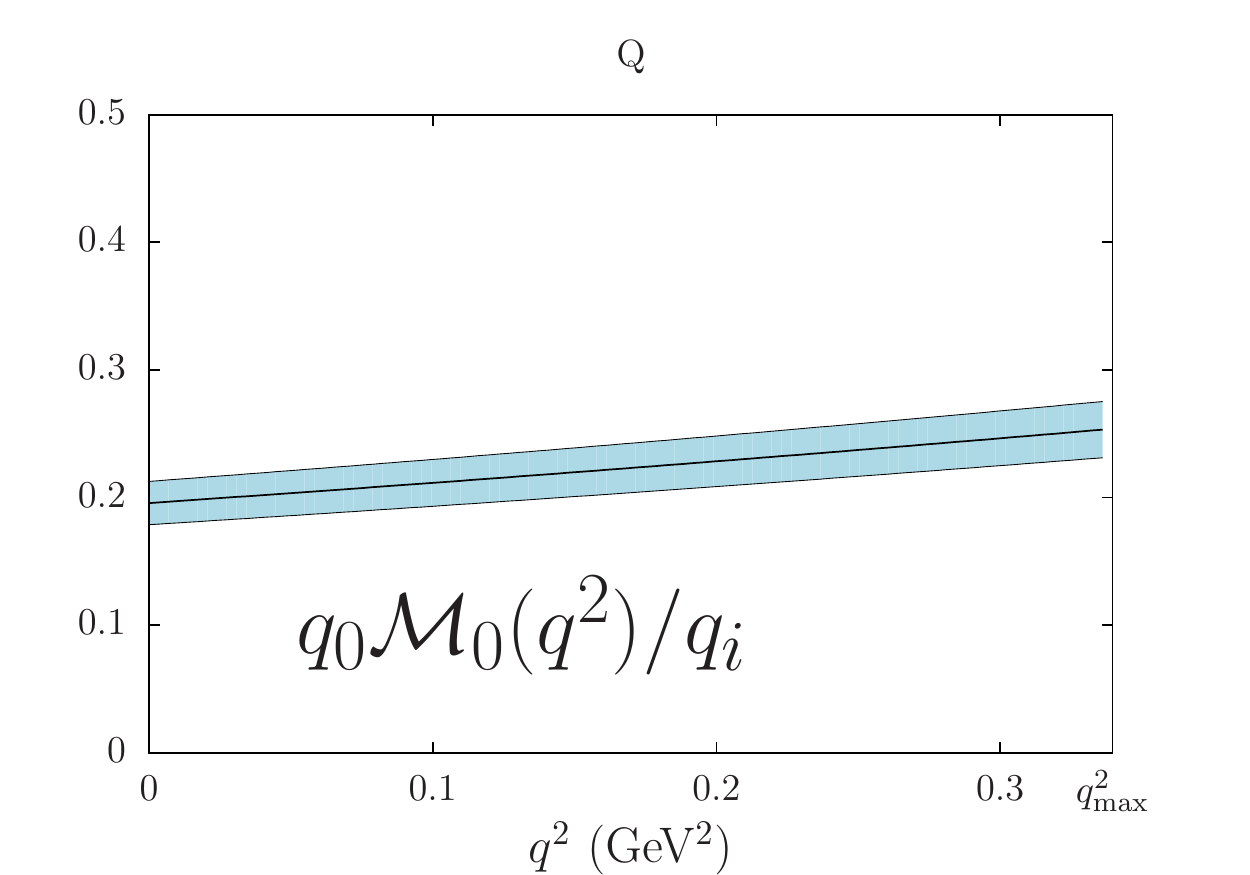}
	\end{minipage}
	
	\caption{$q^2$ dependence of the spatial and time contributions to the form factor $A^{12}_{0}(q^2)$ for the quenched lattice ensemble Q2.}	
\label{fig:form_factor_quenched}
\end{figure}

We have repeated the same analysis for the quenched ensemble Q2. The spatial and time components of the form factor are depicted in Fig.~\ref{fig:form_factor_quenched} and the results are summarized in Table~\ref{tab:res_ff}. Finally, we obtain for the form factor at $q^2=0$
\begin{equation*} 
A^{12}_0(0) = -0.143(14) \,,
\end{equation*}
where the error is only statistical since only a single pion mass and lattice spacing has been studied.

\begin{table}[t] 
\begin{center}
\begin{tabular}{|c|@{\quad}cc@{\quad}|@{\quad}cc@{\quad}|@{\quad}cc@{\quad}|}
	\hline
	&	\multicolumn{2}{c@{\quad}|@{\quad}}{Lattice}	&	\multicolumn{2}{c@{\quad}|@{\quad}}{Bakamjian-Thomas}	&	\multicolumn{2}{c@{\quad}|}{Dirac} \\
	\hline 
	$q^2$& 	$q^2_{\rm max}$ & $0$ & 	$q^2_{\rm max}$ & $0$ & 	$q^2_{\rm max}$ & 	$0$ \\
	\hline 
	$q_{0} \mathcal{M}_{0}(q^2) / q_i$ & $0.402(54)_{\stat}(27)_{\chi}$ & $0.237(27)_{\stat}(28)_{\chi}$ &0.252&0.173&0.219&0.164\\  
	$\mathcal{M}_{i}(q^2)$ & $-0.172(16)_{\stat}(6)_{\chi}$ & $0.064(9)_{\stat}(13)_{\chi}$&-0.103&0.05&-0.223&-0.056  \\
	\hline
 \end{tabular} 
\end{center}
\caption{Lattice and quark models results for the spatial and time contributions to $A^{12}_0(q^2)$ at the kinematical points $q^2_{\rm max}$ and 0 \cite{LeYaouancprivate}
Left panel: Extrapolated lattice results using the fit formula (\ref{eq:fit2}): the first error is statistical and the second error include the systematics from the chiral extrapolation.
Middle panel: Bakamjian-Thomas with Godfrey-Isgur potential, obtaining $q_0=0.538~\GeV$; 
right panel: Dirac, obtaining $q_0=0.576~\GeV$. In the case of Dirac quark model, the global sign of hadronic matrix elements can not be known independently of the states phases: the convention is such that the discrepancy between Dirac and BT is minimal, $f_B>0$ and $f_{B^{*\prime}}>0$. }
\label{tab:ff_comp}
\end{table}

\subsubsection{Comparison with quark models} 

Heavy-light mesons are systems for which quark models are particularly well suited, especially in the infinite mass limit of the heavy quark, to make predictions or to confront with experimental data in order to better understand the dynamics at work in the non perturbative regime of strong interaction. As a comparison with lattice data, density distributions have been computed with two different quark models: the first one, called \emph{\`a la} Bakamjian-Thomas \cite{BakamjianKH, OsbornGM, LeYaouancWV}, is a relativistic quark model, with a fixed number of constituents, that has for benefits that wave functions are representations of the Poincar\'e group, currents are covariant in the heavy quark limit and the rest-frame Hamiltonian contains the interaction potential, here the very fruitful Godrey-Isgur potential \cite{GodfreyXJ}. The second quark model is based on solving the Dirac equation with a central potential having a confining term, with a scalar Lorentz structure, and a Coulombian part \cite{BogolioubovQM, BecirevicFR}. We collect in Table \ref{tab:ff_comp} the values of $A^{12}_0(q^2)$ obtained with the two models \cite{LeYaouancprivate}. Quite interestingly, quark models and our lattice study agree in the fact that the time contribution to $A^{12}_0$ dominates over the spatial one and explains why the form factor is negative at $q^2=0$.

\section{Multihadrons thresholds and excited states
\label{sec4}}

\renewcommand{\arraystretch}{1.3}
\begin{table}[t]
	\begin{center}
	\begin{tabular}{c@{\quad}c@{\quad}c@{\quad}c@{\quad}c@{\quad}}
\hline
	\toprule
	\ id\	&	$a\Sigma_{12}$	&	$a\delta$		&	$am_{\pi}$	&	$a\delta+am_{\pi}$		\\ 
\hline
	\midrule 
	A5		&	$0.253(7)$		&	$0.155(4)$	&	$0.12625$	&	$0.281(4)$\\ 
	\midrule 
	B6		&	$0.235(8)$		&	$0.141(4)$	&	$0.10732$	&	$0.248(4)$\\ 
	\midrule 
	E5		&	$0.225(10)$		&	$0.133(6)$	&	$0.14543$	&	$0.278(6)$\\ 
	\midrule 
	F6		&	$0.213(11)$		&	$0.129(3)$	&	$0.10362$	&	$0.233(3)$\\ 
	\midrule 
	N6		&	$0.166(9)$		&	$0.092(3)$	&	$0.08371$	&	$0.176(3)$\\ 
	\bottomrule
\hline
	\end{tabular}
	\end{center}	
	\caption{Mass splittings $\Sigma_{12}=m_{B^{*\prime}}-m_B$ and $\delta = m_{B_1^{*}}-m_B$ \cite{BlossierVEA} for each lattice ensemble.}
	\label{tab:threshold}
\end{table}

In many lattice studies, the radial or orbital excitations of mesons lie near the multihadron threshold, making the extraction of excited states properties more challenging. Usually, interpolating operators having a large overlap with a two-body system \cite{MichaelKW} are used but they require more computer time and it is argued that bilinear interpolating operators are coupled only weakly with those states \cite{BarCE}. Here we propose to study this problem from our results on radial distributions.\\

Within our lattice setup, the radial excitation of the vector meson ($B^{*\prime}$) lies near the multiparticle threshold $B_1^* \pi$ in $S$ wave where $B_1^*$ represents the axial $B$ meson (see Table~\ref{tab:threshold}). Its mass, in the static limit of HQET, is extracted from \cite{BlossierVEA}. Assuming that the energy of the two-particle state is simply given by $E=m_{B_1^*}+m_{\pi}$, we conclude that for all lattice ensembles we are below (but near) threshold. Since our interpolating operators are coupled, in principle, to all states with the same quantum numbers, it means that we could be sensitive to the $B^*_1$ state. However, if the coupling were not small, it would be difficult to interpret our GEVP results: we extract a clear signal for the third excitation and it is far above the second energy level. We do not see near-degenerate states. Moreover, the position of the node of the density distribution $f^{(12)}_{\gamma_i \gamma_5}$ is remarkably stable and does not depend on the pion mass, contrary to what would be expected in the case of a mixing with multiparticle states.  Also, the qualitative agreement with quark models makes us confident that our measurement of the density distributions $f^{(12)}_\Gamma(r)$ probes transition amplitudes among $\bar{q}b$ bound states: in the quark model language they correspond to overlaps between wave functions.

\subsection{Multihadron analysis} 

\begin{figure}[t]
	\begin{minipage}[c]{0.49\linewidth}
	\centering 
	\includegraphics*[width=0.9\linewidth]{plots/energy_E5.pdf}
	\end{minipage}
	\begin{minipage}[c]{0.49\linewidth}
	\centering 
	\includegraphics*[width=0.9\linewidth]{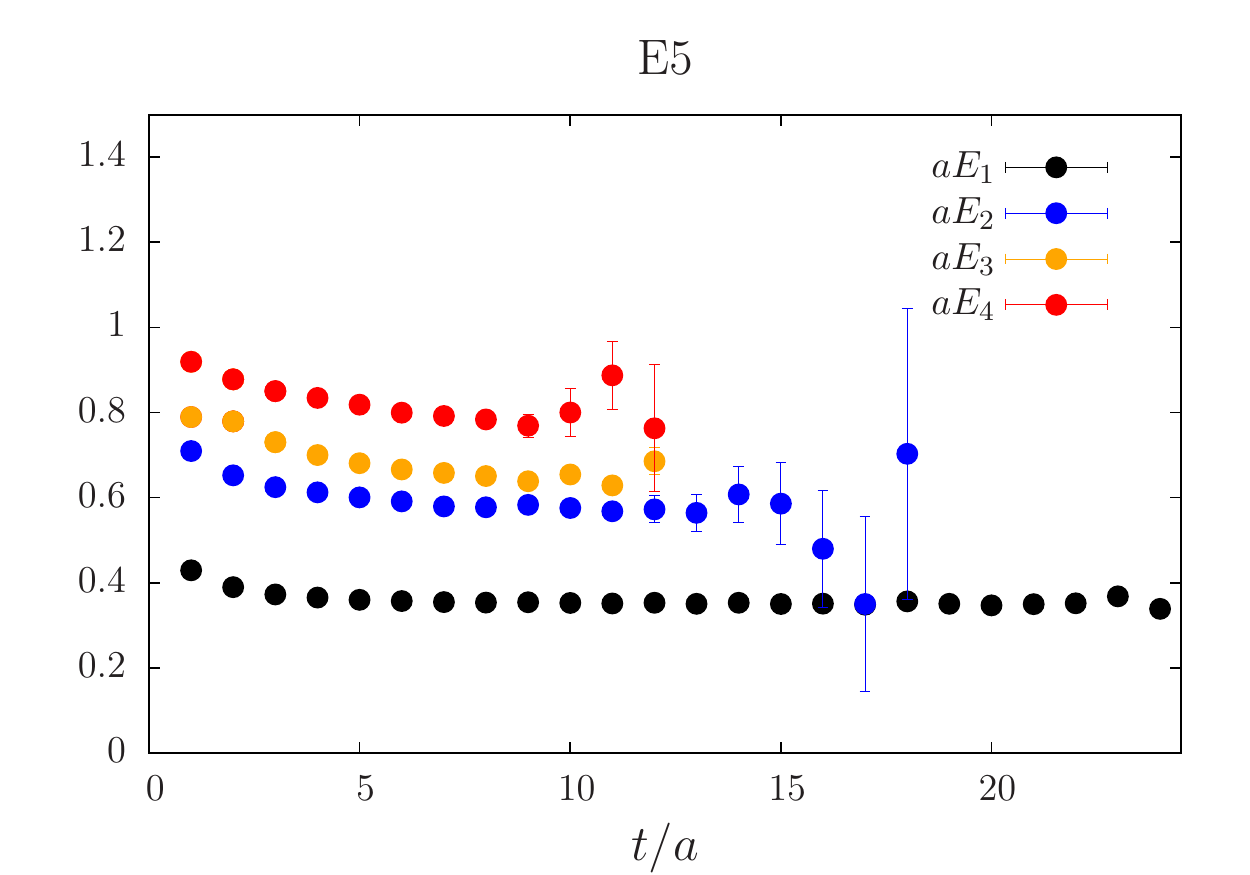}
	\end{minipage}
	\caption{(\textit{left}) Effective mass plot extracted from a $3\times 3$ GEVP for the lattice ensemble E5 using $\bar{q} \gamma_k b$ interpolating operators. (\textit{right}) Effective mass plot extracted from a $4\times 4$ GEVP for the lattice ensemble E5 using $\bar{q}\gamma_k b$ and $\bar{q} \nabla_k b$ interpolating operators.}	
\label{fig:GEVP_threshold}
\end{figure}

In addition to the Gaussian smearing operators $\mathcal{V}^{(i)}_{k}(x) =  \overline{u}^{(i)}(x) \gamma_k h(x)$ used in the previous analysis, we have inserted a second kind of interpolating operators which could couple to the two-particle state: $\mathcal{V}^{(i)}_{k}(x) =  \overline{u}^{(i)}(x) \overleftarrow{\nabla}_k h(x)$. As can been seen in Fig.~\ref{fig:GEVP_threshold}, the GEVP indeed isolates a new state, slightly above the radial excitation of the vector meson, whose interpretation can be guessed from Table~\ref{tab:threshold}. The effective mass of the ground state and first excited state remain unchanged, as we indicate in Table~\ref{tab:eff_mass_multihadron}. \\

\renewcommand{\arraystretch}{1.3}
\begin{table}[t]
\begin{center}
\begin{tabular}{l@{\quad}l@{\quad}c@{\quad}c@{\quad}c}
\hline
\multicolumn{1}{c}{}&\multicolumn{1}{c}{}	&	$a\Sigma_{12}$		&	$a\Sigma_{13}$		&	$a\Sigma_{14}$  \\
\hline 
	\multirow{2}{*}{E5}	&	$\gamma_k$			&	$0.225(8)$		&	$0.417(21)$	&  	$\times$	\\  
\cline{2-5}
	\cmidrule(lr){2-5} 
		&	$\gamma_k, \nabla_k$	&	$0.218(12)$	&	$0.278(17)$	&  	$0.422(12)$	 	\\  
\hline
	\midrule
	\multirow{2}{*}{A5}	&	$\gamma_k$			&	$0.257(6)$		&	$0.467(23)$	&  	$\times$	\\  
\cline{2-5}
	\cmidrule(lr){2-5} 
		&	$\gamma_k, \nabla_k$	&	$0.254(7)$	&	$0.315(11)$	&  	$0.459(24)$	 	\\  
	\bottomrule
\hline
 \end{tabular} 
\end{center}
\caption{Energy levels extracted from the GEVP (ensembles E5 and A5). In the first raw only Gaussian smeared operators $\mathcal{V}^{(i)}_{k}(x) =  \overline{u}^{(i)}(x) \gamma_k h(x)$ are used. In the second raw, both interpolating operators of the form $\mathcal{V}^{(i)}_{k}(x) =  \overline{u}^{(i)}(x) \gamma_k h(x)$ and $\mathcal{V}^{(i)}_{k}(x) =  \overline{u}^{(i)}(x) \protect\overleftarrow{\nabla}_k h(x)$.} 
\label{tab:eff_mass_multihadron}
\end{table}

\begin{figure}[t]

 	\hspace{-3.4cm}
	\begin{minipage}[c]{0.24\linewidth}
	\centering 
	\includegraphics*[width=1.8\linewidth]{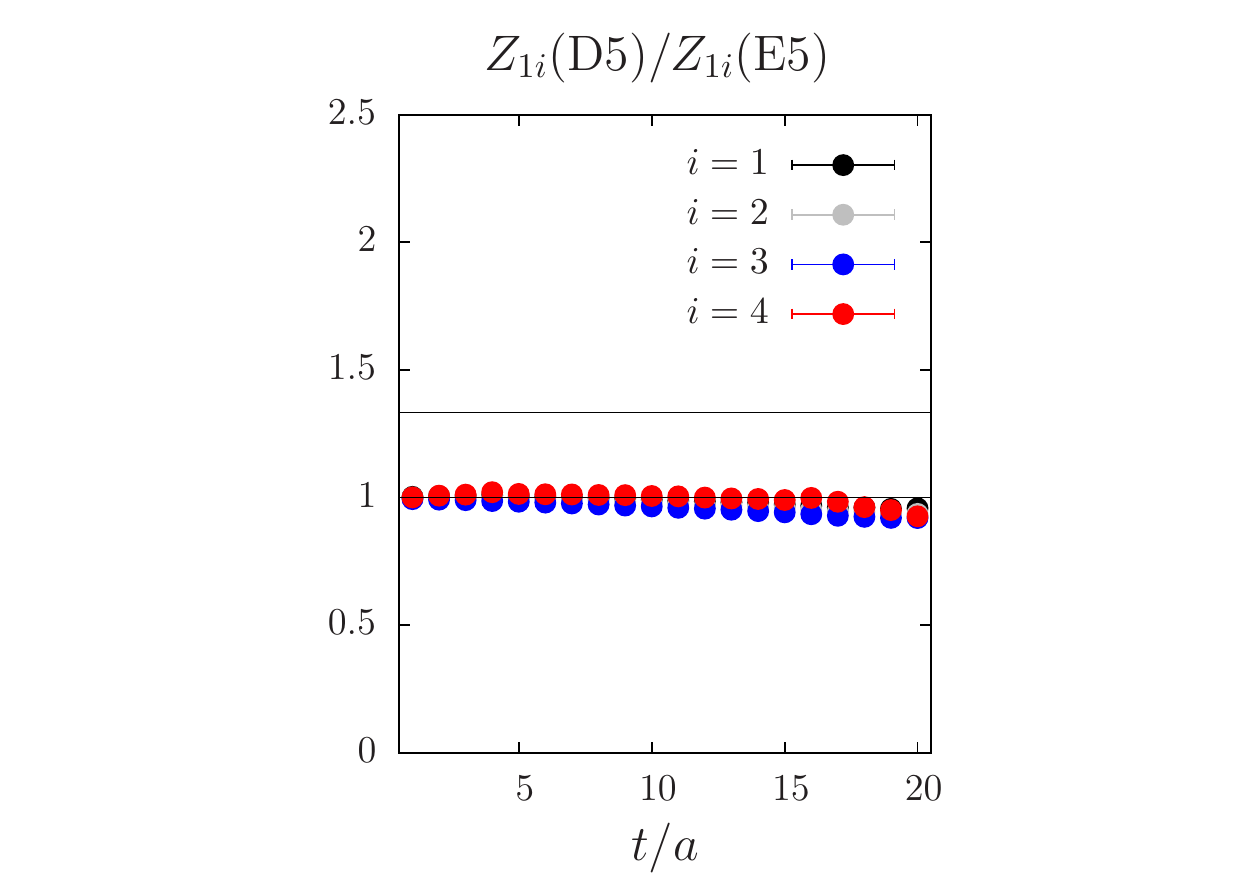}
	\end{minipage}
	\begin{minipage}[c]{0.24\linewidth}
	\centering 
	\includegraphics*[width=1.8\linewidth]{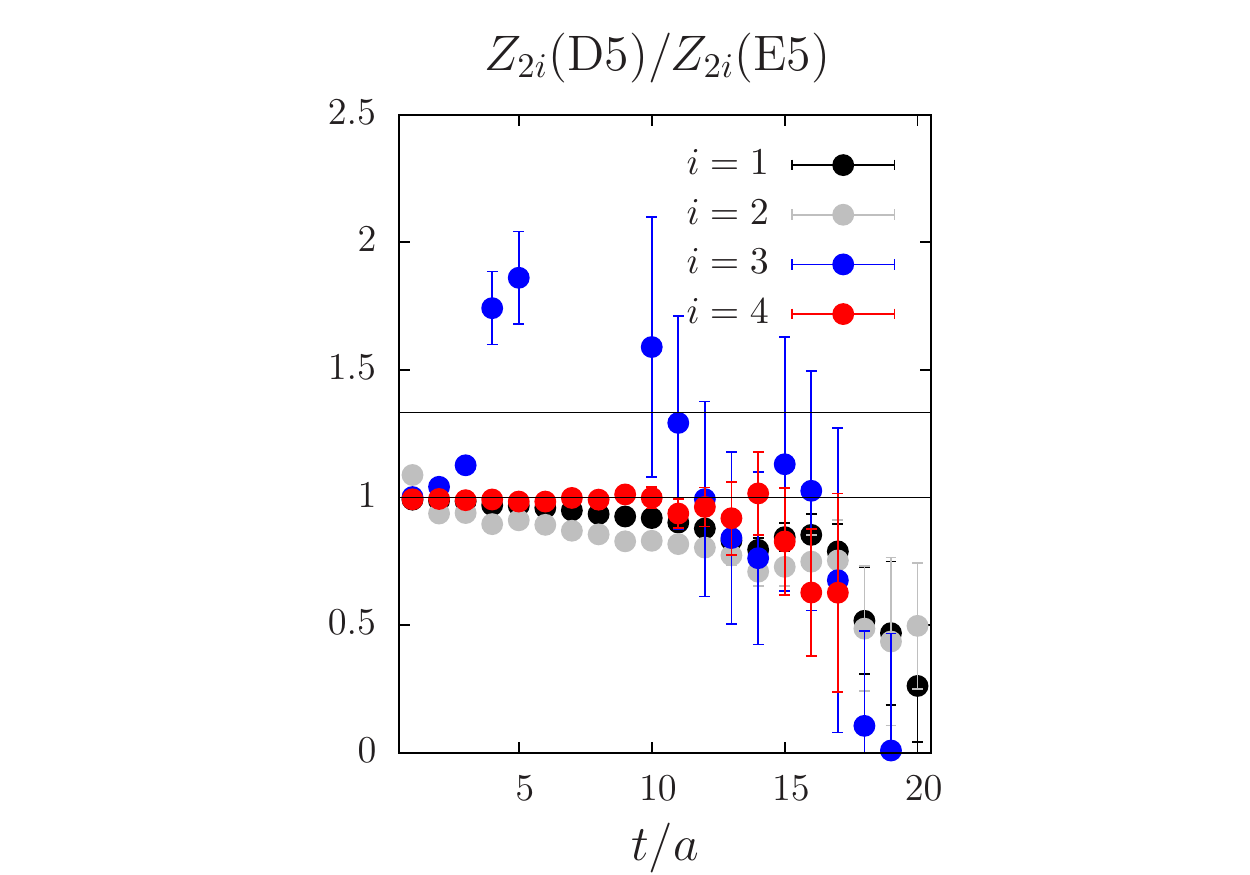}
	\end{minipage}
	\begin{minipage}[c]{0.24\linewidth}
	\centering 
	\includegraphics*[width=1.8\linewidth]{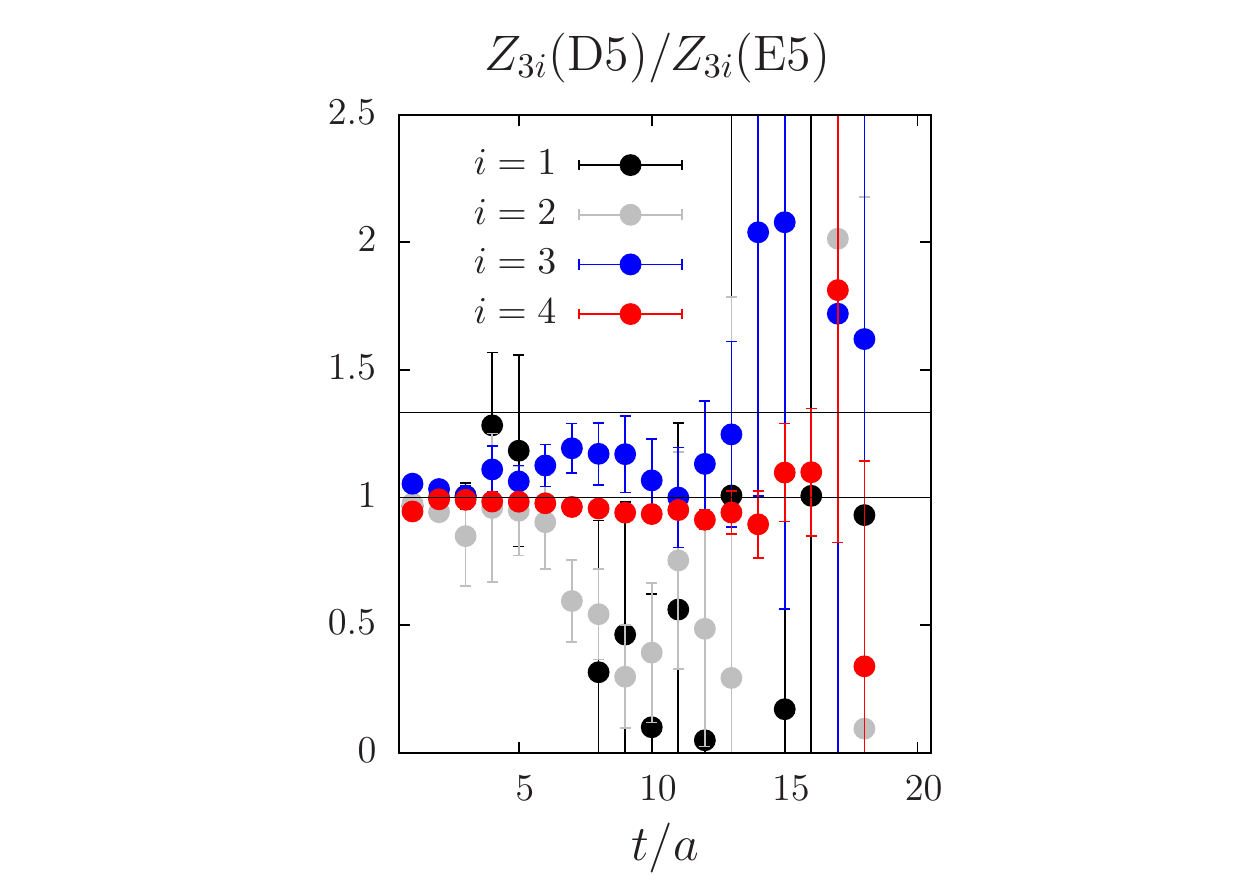}
	\end{minipage}
	\begin{minipage}[c]{0.24\linewidth}
	\centering 
	\includegraphics*[width=1.8\linewidth]{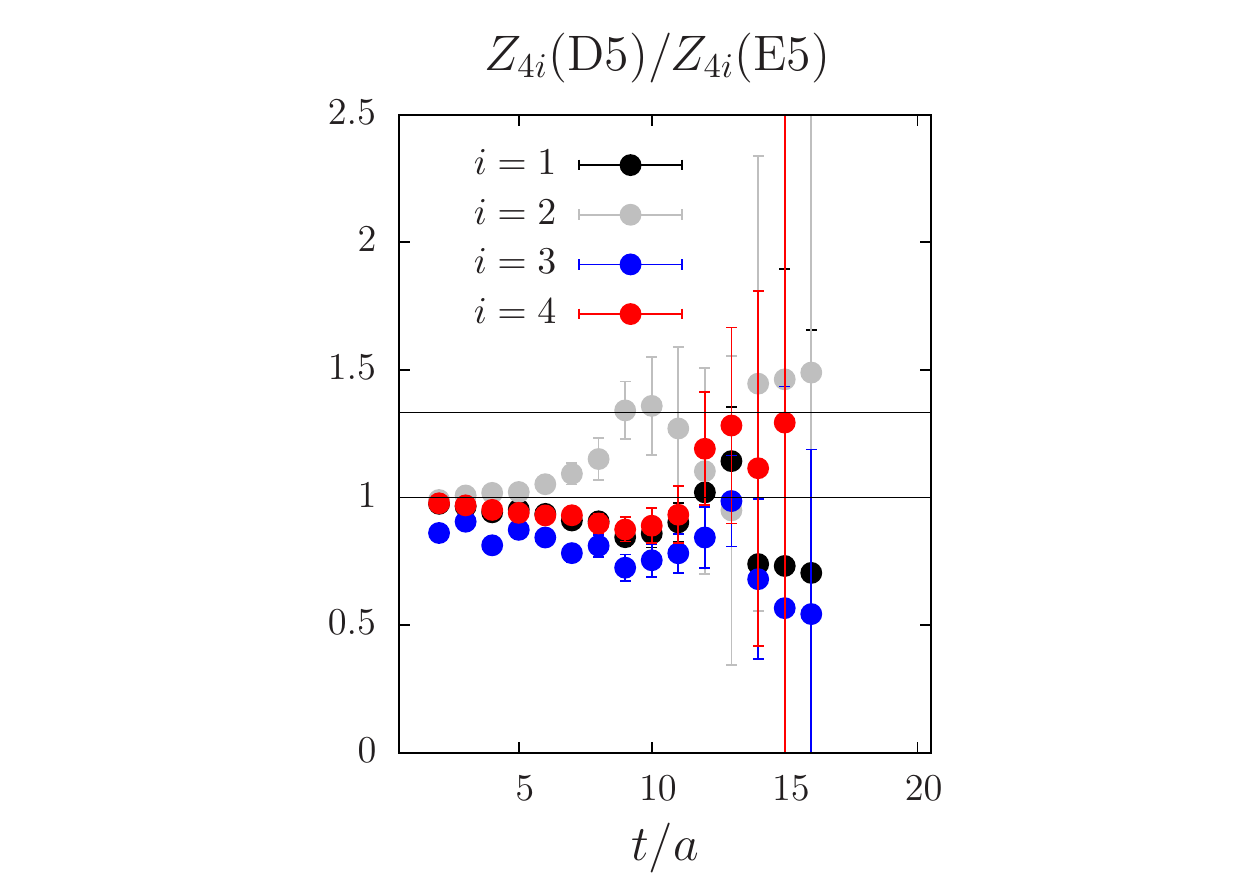}
	\end{minipage}
	
	\caption{Ratio $Z_{ni}(\mathrm{E}5)/Z_{ni}(\mathrm{D}5)$ for the first three levels. The two plain dark lines correspond to a ratio of 1 and~$4/3$.}
	\label{fig:Z}		
\end{figure}

Eigenvectors for single-particle and multi-particle states are expected to have different volume dependence: the former are expected to be almost volume independent whereas the latter should not \cite{MathurJR,LuscherCF}. Then, if any excited state was interpreted as a multi-hadron state, one would expect that the overlap $Z$ to a given interpolating field depends on the volume. We have performed the check on lattice ensembles E5 and D5, which have two different volumes. Using the notations of ref.~\cite{BlossierKD}, the two-point correlation function can be written as
\begin{equation*}
C_{ij}(t) = \frac{a^6}{V} \sum_{\vec{x}, \vec{y}}\langle \mathcal{O}_i(\vec{x},t) \mathcal{O}_j^{\dag}(\vec{y}, 0) \rangle = \sum_{n=1}^{\infty} Z_{ni} \, Z^{*}_{mj}\, e^{-E_n t} \quad , \quad i,j=1,\cdots,N \,,
\end{equation*}
where $Z_{ni}$ corresponds to the overlap between the interpolating field $\mathcal{O}_i$ and the $n^{\text{th}}$ excited state. An estimator for the overlaps $Z_{ni}$ is given by
\begin{equation}
Z_{ni} = \frac{ C_{ij}(t) v_{nj}(t,t_0) }{ \left( v_n(t,t_0), C(t)v_n(t,t_0) \right)^{1/(2} } \left( \frac{\lambda_n(t)}{\lambda_n(t+a)} \right)^{t/(2a)} \,.
\label{eq:GEVP_overlap}
\end{equation}
Results are depicted in Fig.~\ref{fig:Z}. For the ground state, the overlaps are compatible for the two lattice ensembles as expected for a single hadron state. We do not observe neither any volume dependence for both the first and the second excited state whereas, for a multi-hadron states, a volume dependence $Z_{ni}(\mathrm{D}5)/Z_{ni}(\mathrm{E}5) = (32/24)^{3/2} \approx 1.33$ is expected. Therefore, our analysis suggests that this criterion is not satisfied in our case, at least in the time interval considered here; similar conclusions were drawn in~\cite{PrelovsekKG}.

\subsection{A toy model}

\begin{figure}[t]
 
 	\hspace{-3cm}
	\begin{minipage}[c]{0.2\linewidth}
	\centering 
	\includegraphics*[width=1.8\linewidth]{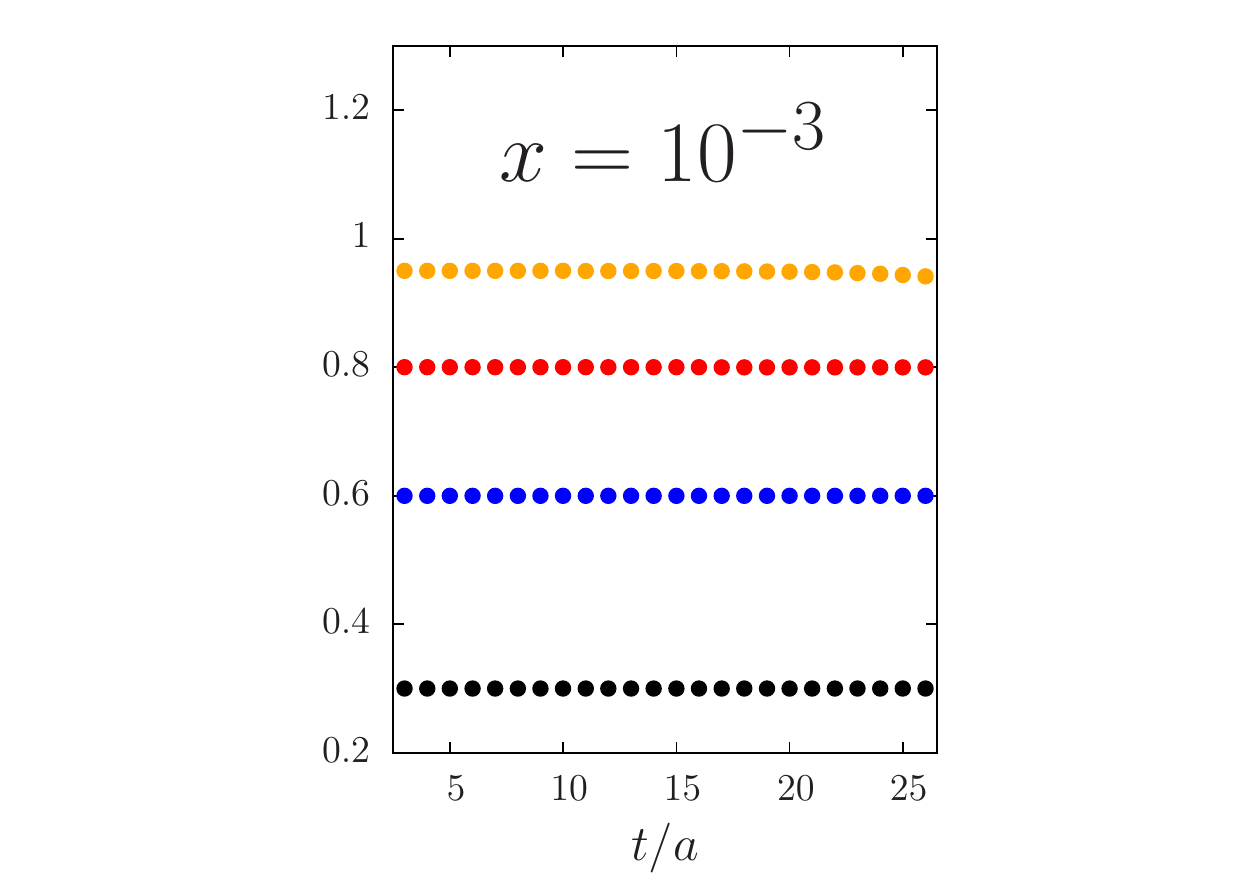}
	\end{minipage}
	\begin{minipage}[c]{0.2\linewidth}
	\centering 
	\includegraphics*[width=1.8\linewidth]{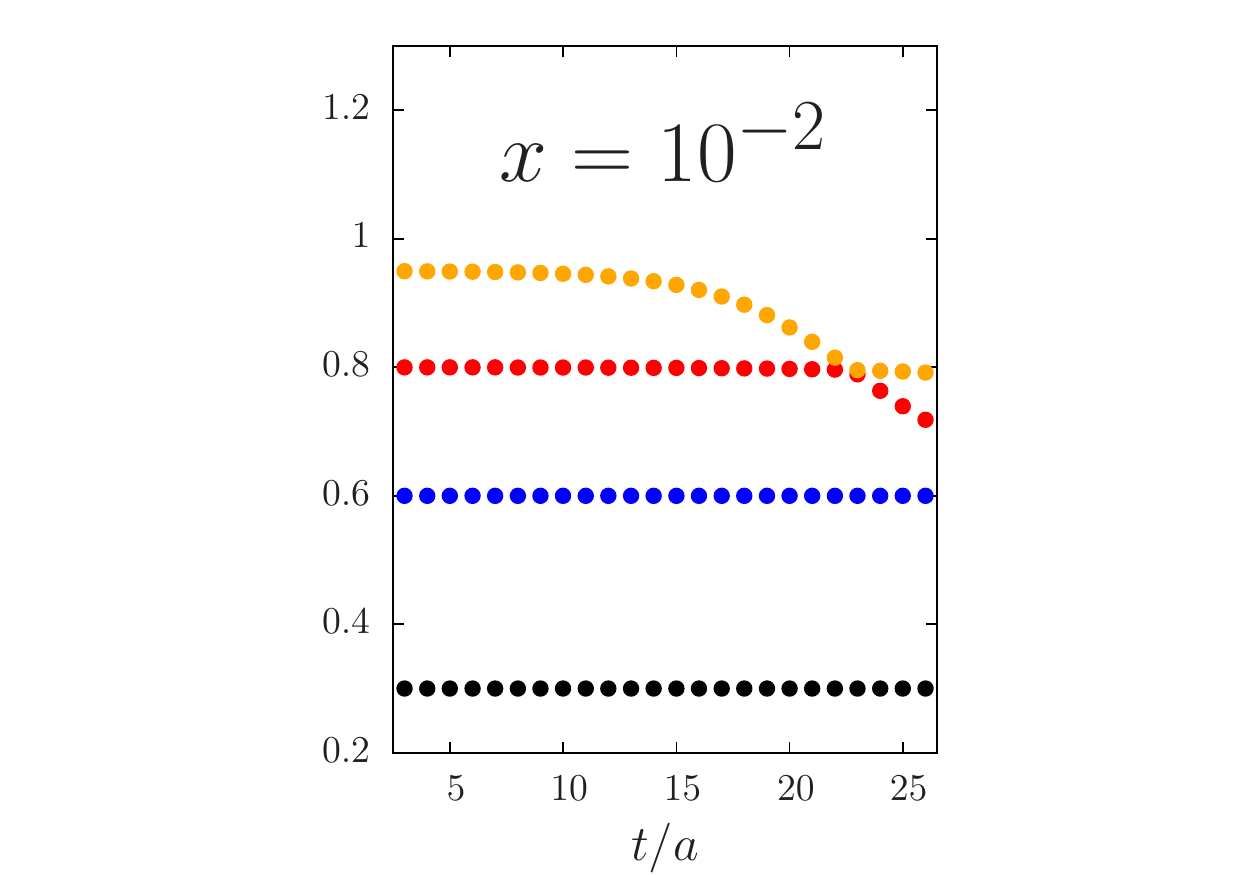}
	\end{minipage}
	\begin{minipage}[c]{0.2\linewidth}
	\centering 
	\includegraphics*[width=1.8\linewidth]{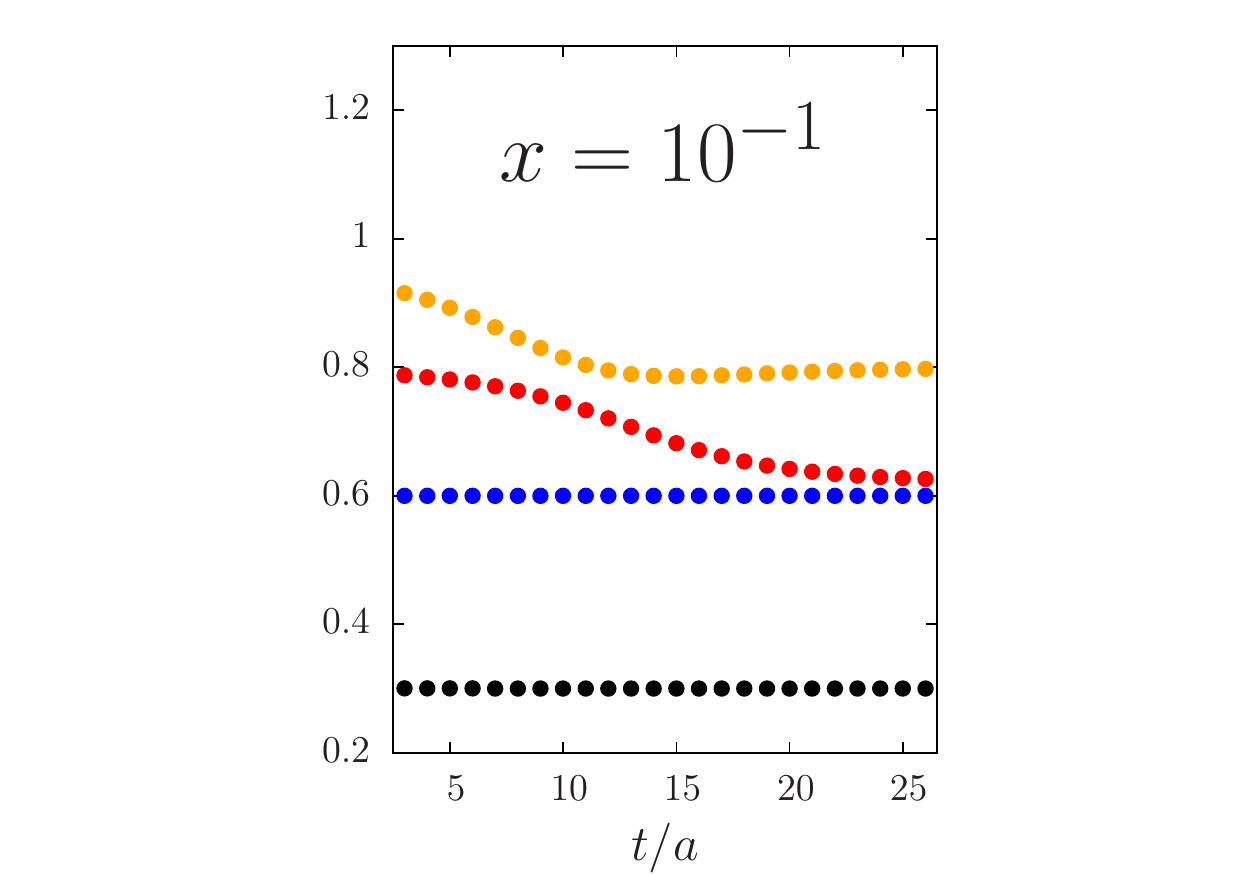}
	\end{minipage}
	\begin{minipage}[c]{0.2\linewidth}
	\centering 
	\includegraphics*[width=1.8\linewidth]{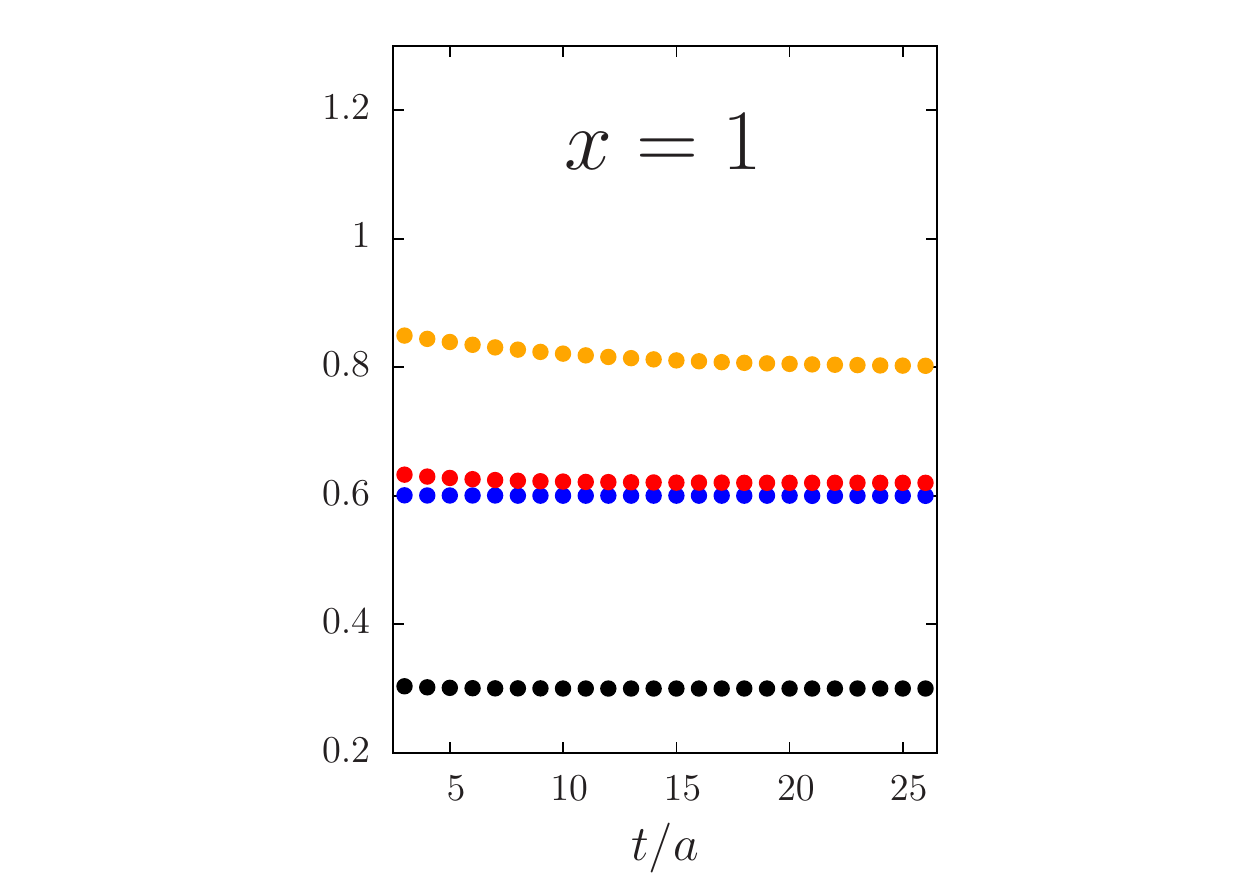}
	\end{minipage}
	
	\caption{\label{fig:plotgevpx} Effective energies of the two-point correlation function (\ref{eq:toy_model_gevp}) obtained by solving a $4\times4$ GEVP for different values of $x$.}
\end{figure}

To understand this fact further, we have performed a test on a toy model. The spectrum contains five states, with energies $E^{(i)}=\{0.3, 0.6, 0.63, 0.8, 0.95\}$. The $1^{\rm st}$ and $2^{\rm nd}$ excited states are almost degenerate. Taking a basis of five interpolating fields, the matrix of couplings has the following structure: 
\begin{equation}
M^{x}=\left[ \begin{array}{ccccc}
0.60&0.25&x \times 0.40&0.10&0.50\\
0.61&0.27&x \times 0.39&0.11&0.51\\
0.58&0.24&x \times 0.42&0.12&0.52\\
0.57&0.25&x \times 0.41&0.10&0.49\\
0.56&0.26&x \times 0.36&0.08&0.48\\
\end{array}
\right],
\end{equation}
where $x$ can be varied from $10^{-3}$ (third interpolating field almost not coupled to the spectrum under investigation) to $1$ (third interpolating field as strongly coupled to the spectrum as the other operators). We solve a GEVP on the $4 \times 4$ matrix of correlators $C^x_{ij}$ defined by
\begin{equation}
C^x_{ij}(t)=\sum_{n=1}^5 M^x_{ni}M^x_{nj} e^{-E_n t}.
\label{eq:toy_model_gevp}
\end{equation}
In Fig. \ref{fig:plotgevpx} we show the effective masses obtained from the generalized eigenvalues, when $x$ is growing. We see clearly the transition: the GEVP isolates the states 1, 2, 4 and 5 at very small $x$ and then, as $x$ is made larger, the states 1, 2, 3 and 4. \emph{Conclusion: a GEVP can ``miss" an intermediate state of the spectrum if, by accident, the coupling of the interpolating fields to that state is suppressed}. Our claim is that, using interpolating fields $\bar{q} \gamma_i h$, we have no chance to couple to multi-hadron states while inserting an operator $\bar{q} \nabla_i h$ could isolate the $B^*_1 \pi$ two-particle state.

\subsection{Density distributions with an enlarged basis of interpolating fields} 

\begin{figure}[t]

 	\hspace{-3.5cm}
	\begin{minipage}[c]{0.2\linewidth}
	\centering 
\includegraphics*[width=2\linewidth]{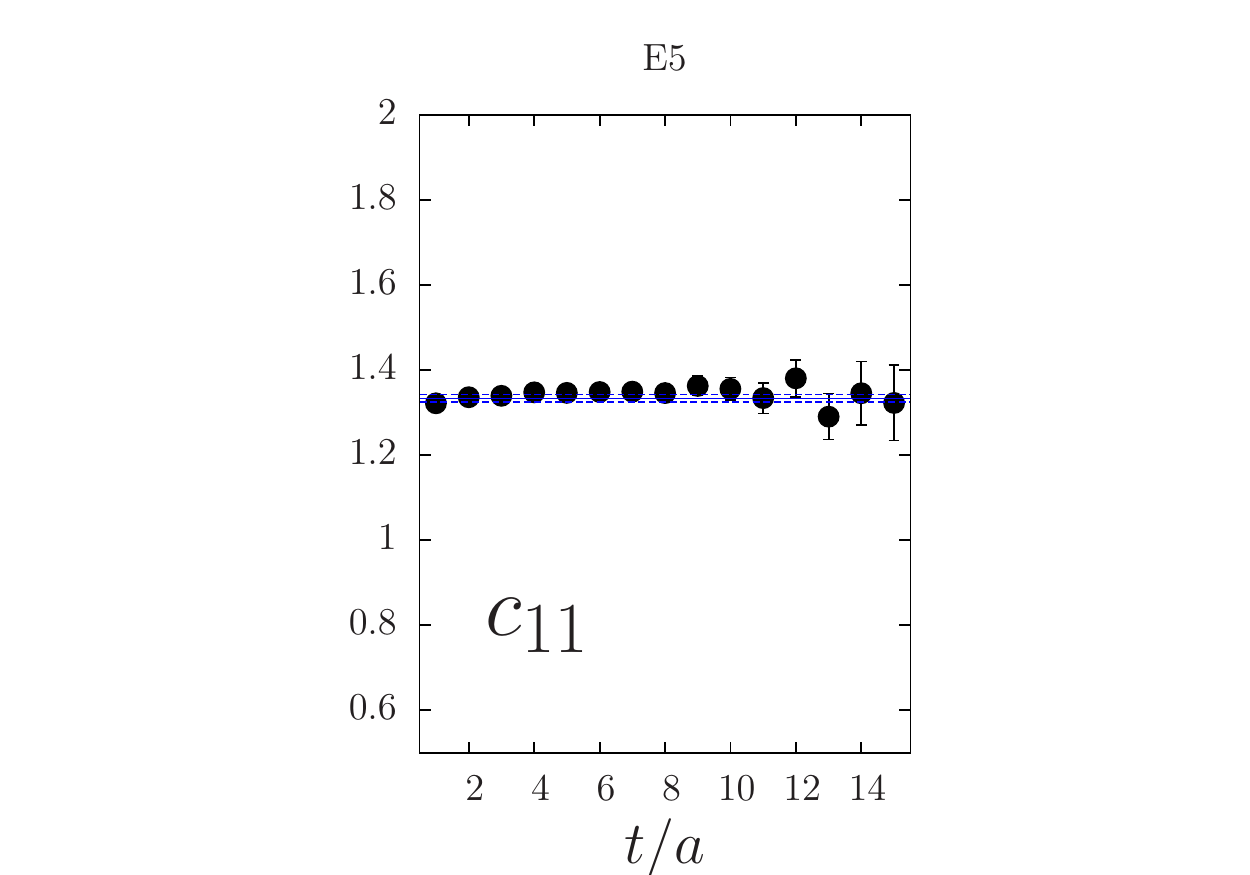}
	\end{minipage}
	\begin{minipage}[c]{0.2\linewidth}
	\centering 
	\includegraphics*[width=2\linewidth]{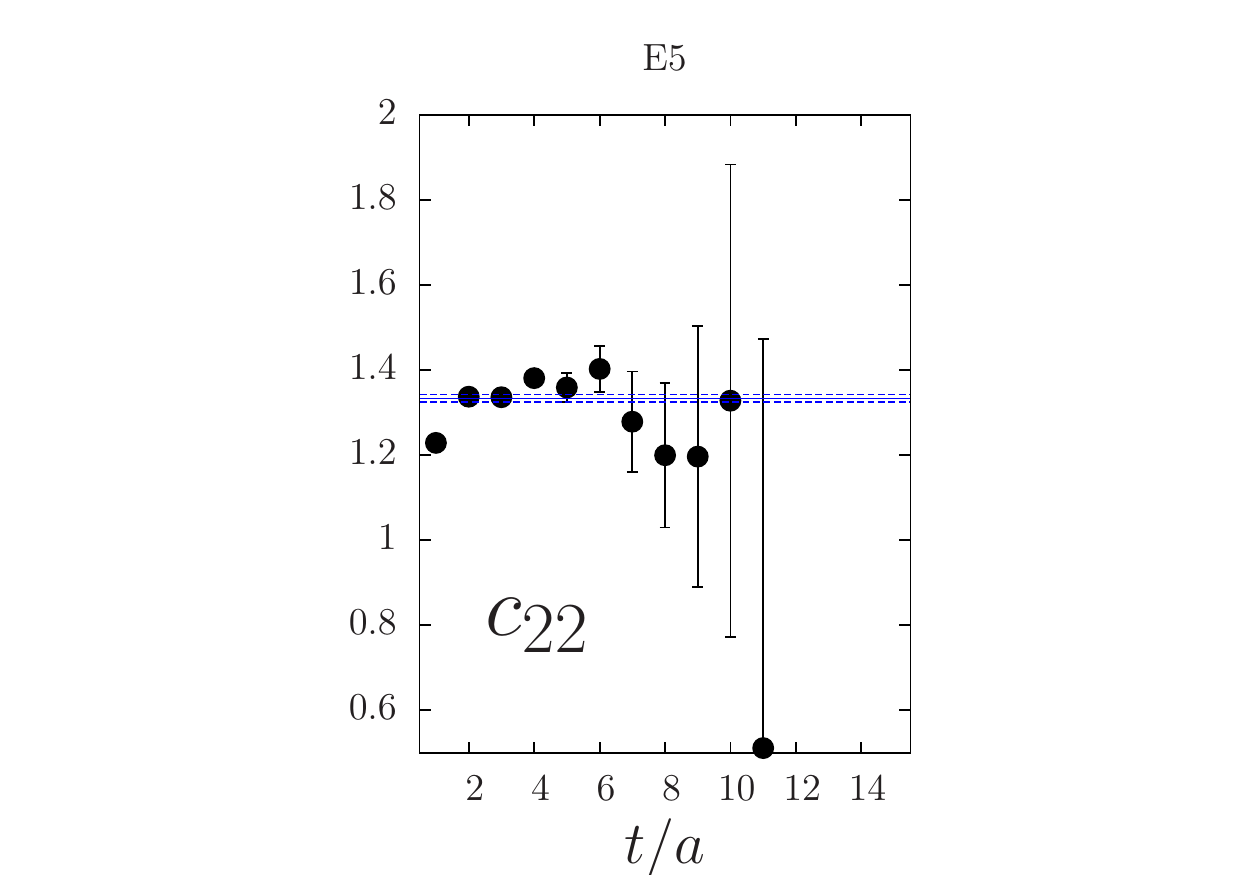}
	\end{minipage}
	\begin{minipage}[c]{0.2\linewidth}
	\centering 
	\includegraphics*[width=2\linewidth]{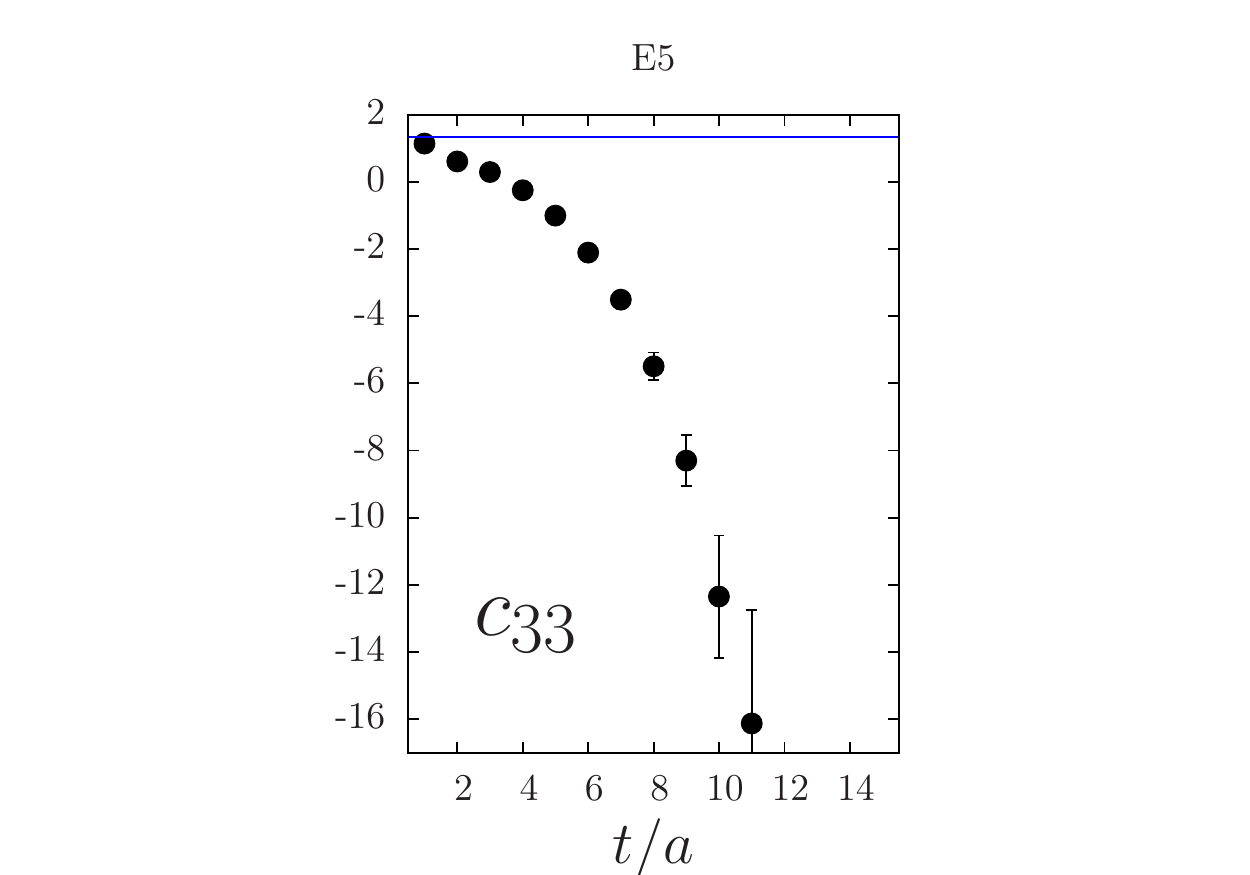}
	\end{minipage}	
	\begin{minipage}[c]{0.2\linewidth}
	\centering 
	\includegraphics*[width=2\linewidth]{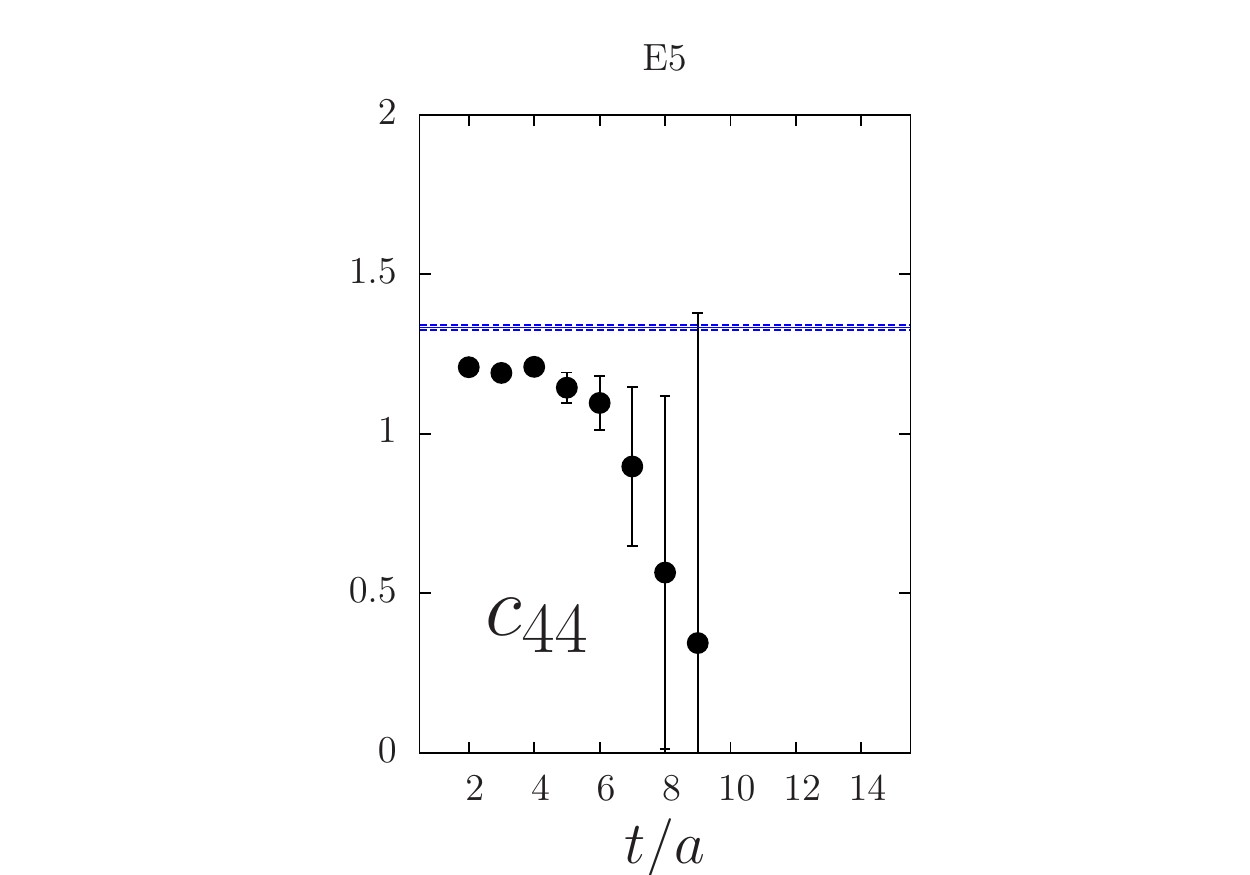}
	\end{minipage}	
\caption{\label{fig:chargesumnabla} Unrenormalized vector charge got from $f^{(nn)}_{\gamma_0}(r)$ on the lattice ensemble E5, using $\bar{q} \gamma_k h$ and $\bar{q} \nabla_k h$ interpolating operators. The blue line corresponds to the expected plateau using the nonperturbative estimate $Z_V = 0.750(5)$ extracted from~\cite{Fritzsch:2012wq}. }
\end{figure}

\begin{figure}[t]
	\begin{minipage}[c]{0.28\linewidth}
	\centering 
	\includegraphics*[width=\linewidth]{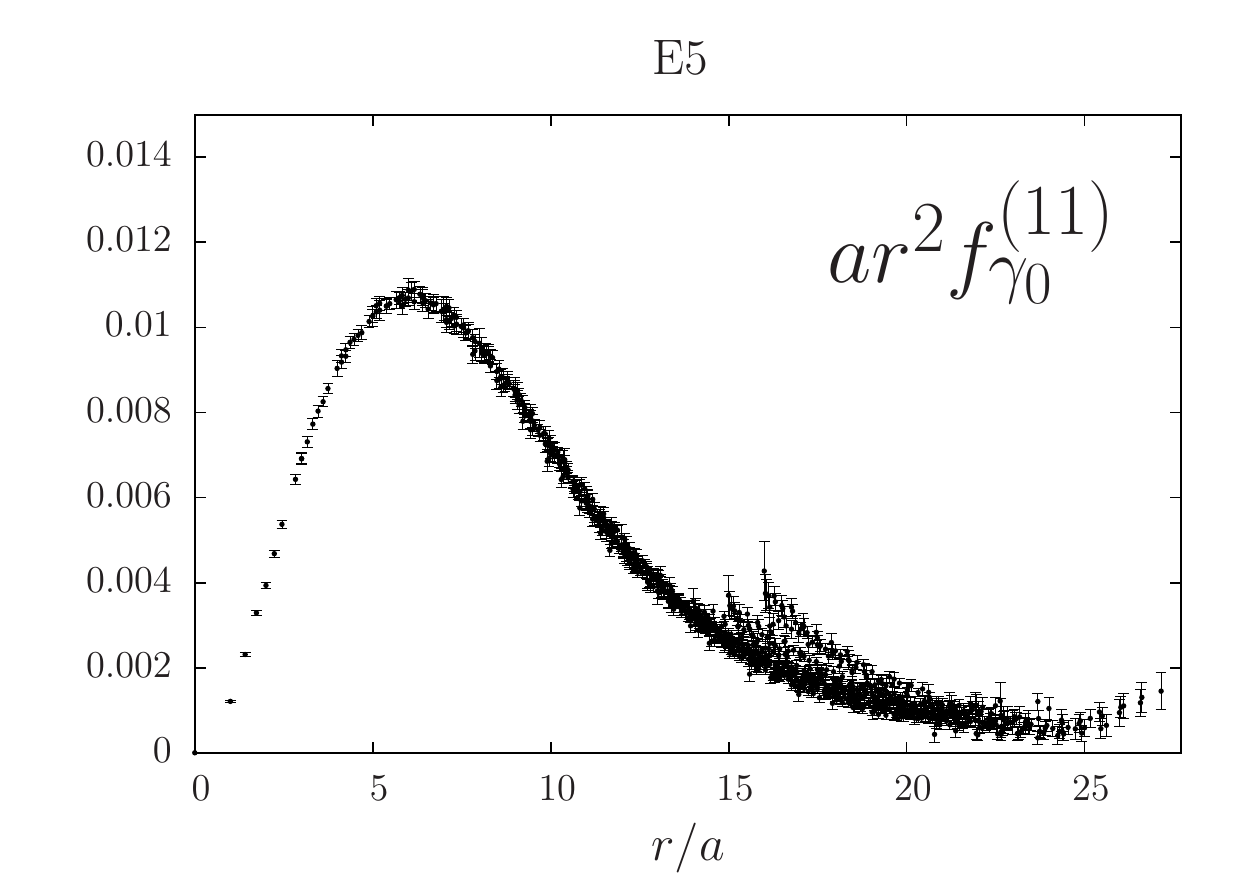}
	\end{minipage}
	\begin{minipage}[c]{0.28\linewidth}
	\centering 
	\includegraphics*[width=\linewidth]{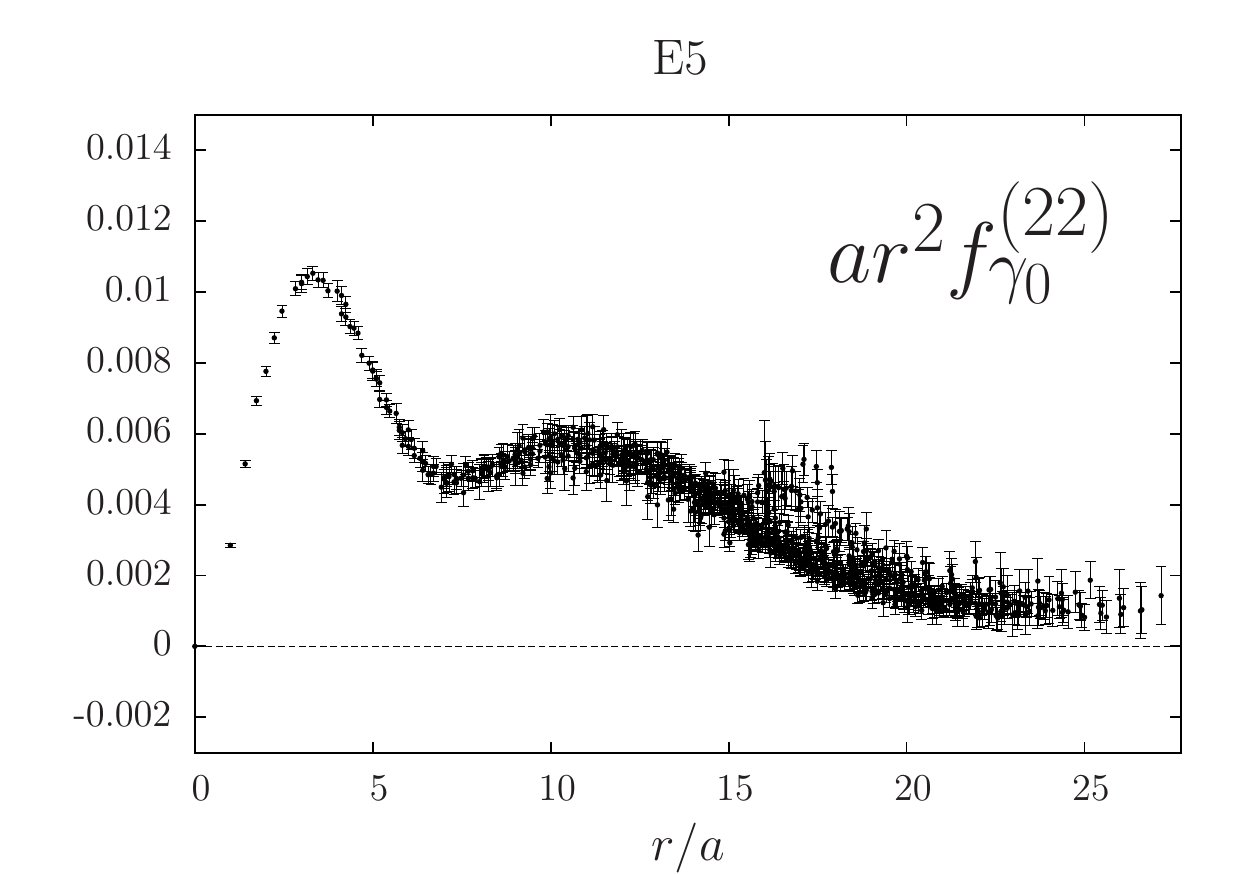}
	\end{minipage}
	\begin{minipage}[c]{0.28\linewidth}
	\centering 
	\includegraphics*[width=\linewidth]{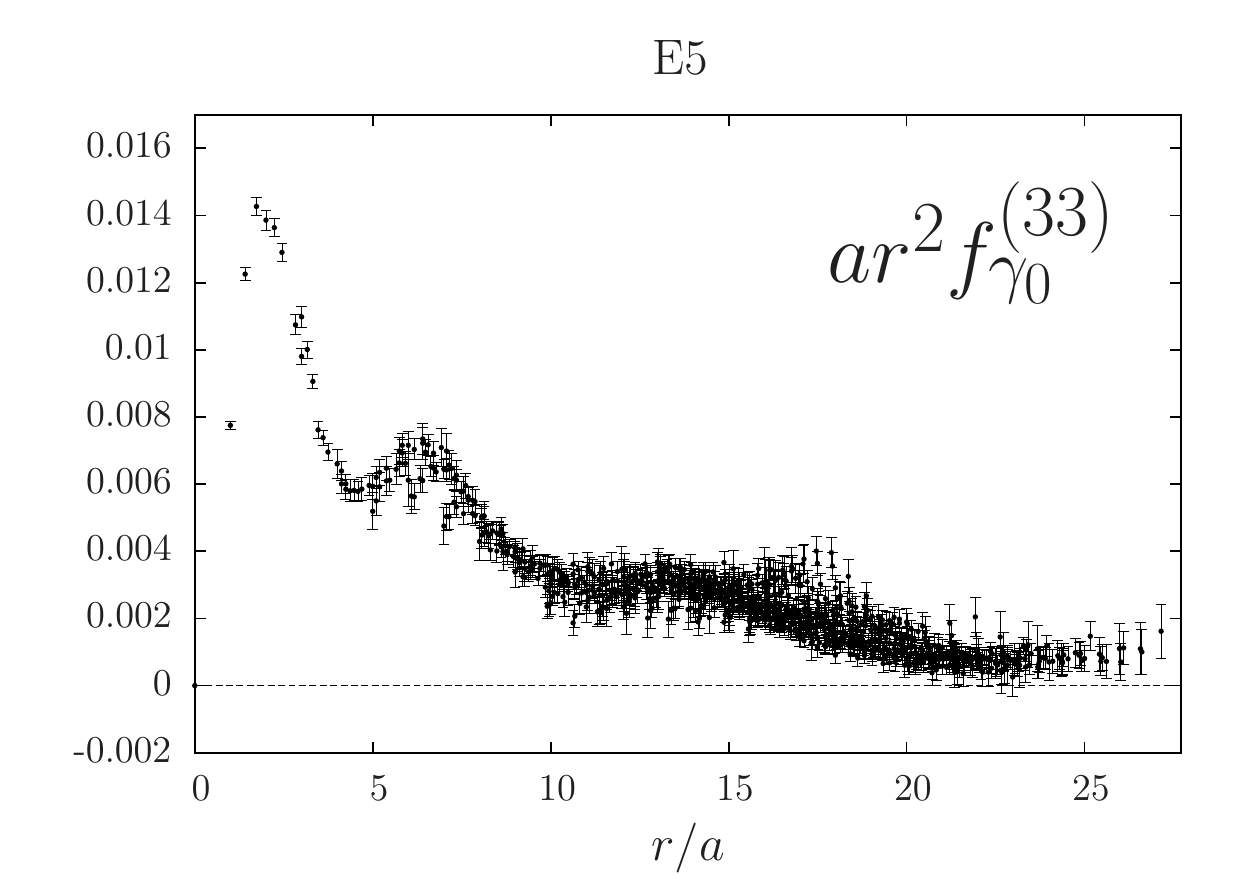}
	\end{minipage}	
	\\
	\begin{minipage}[c]{0.28\linewidth}
	\centering 
	\includegraphics*[width=\linewidth]{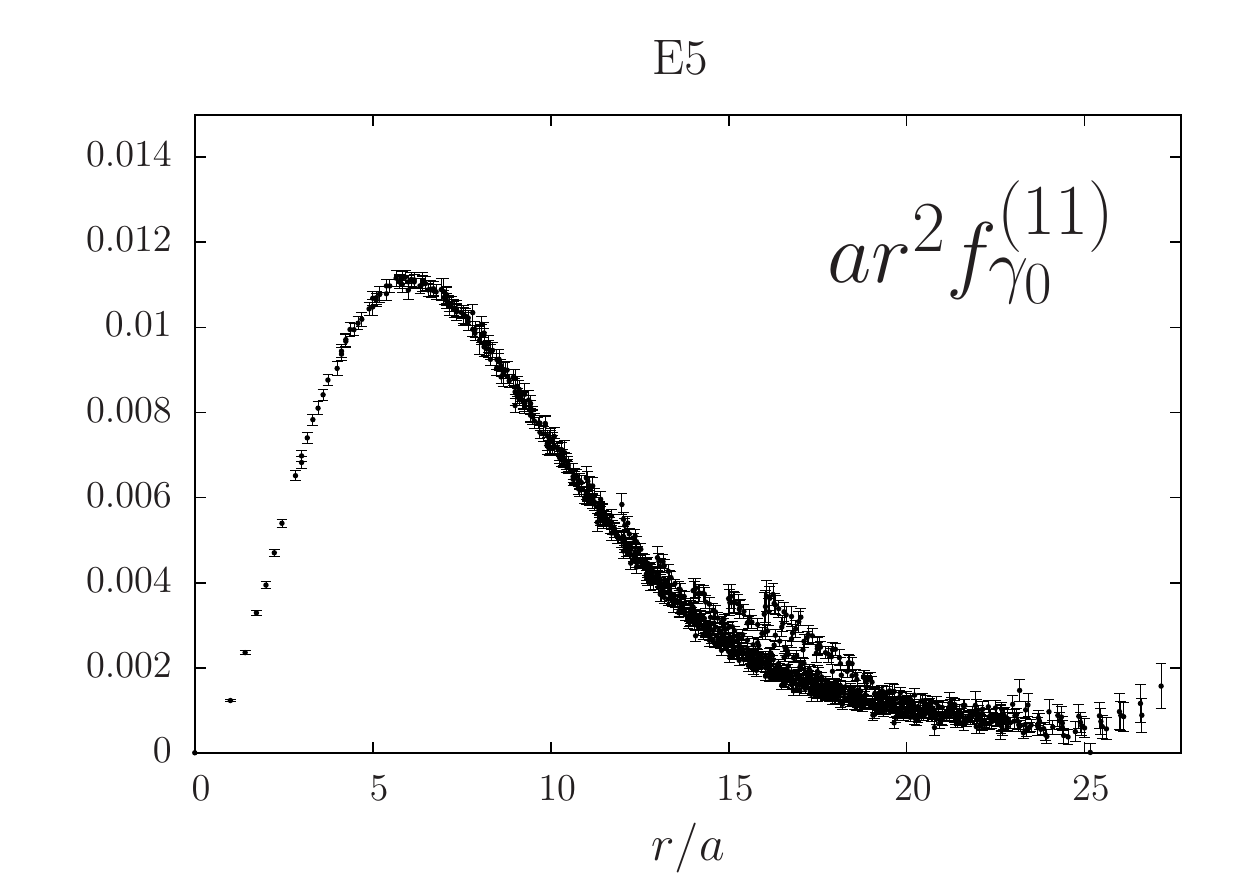}
	\end{minipage}
	\begin{minipage}[c]{0.28\linewidth}
	\centering 
	\includegraphics*[width=\linewidth]{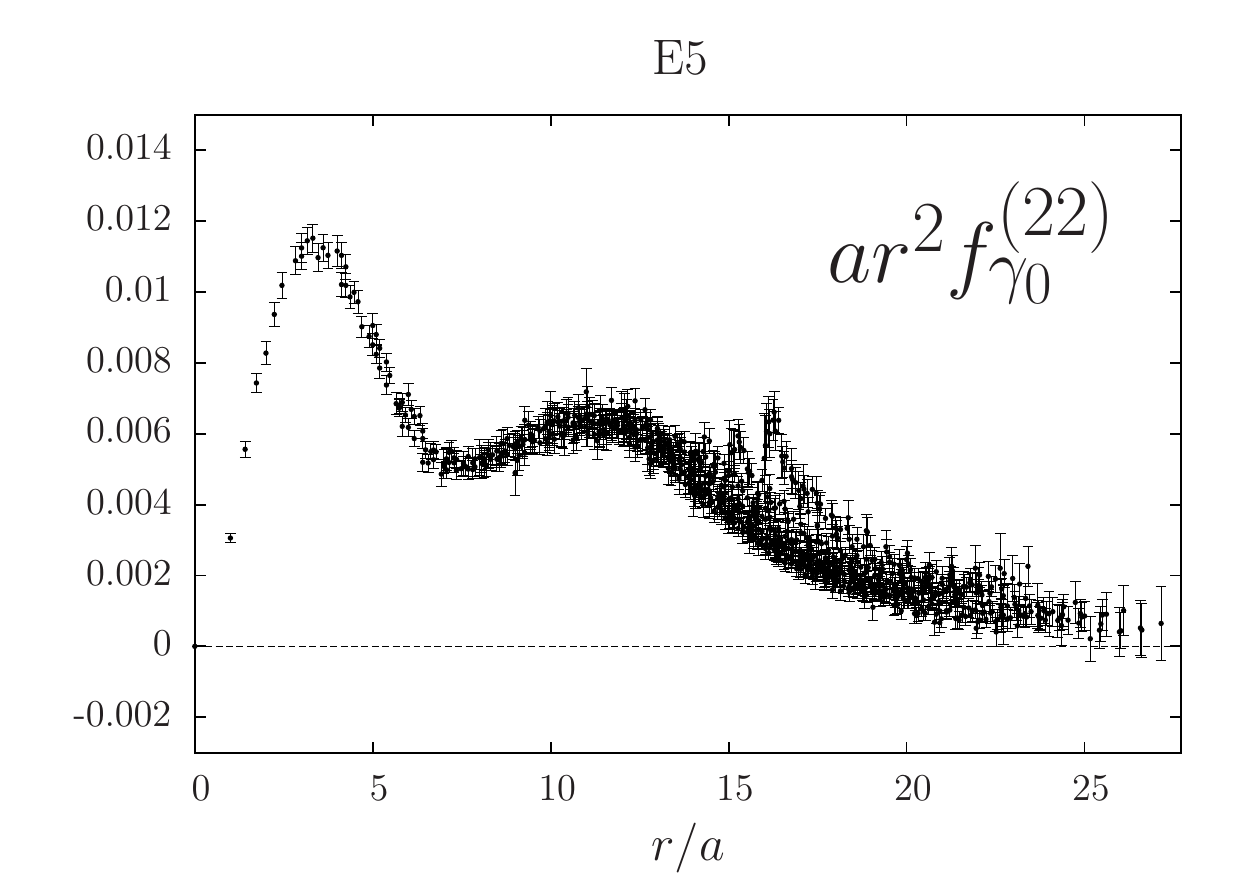} 
	\end{minipage}
	\begin{minipage}[c]{0.28\linewidth}
	\centering 
	\includegraphics*[width=\linewidth]{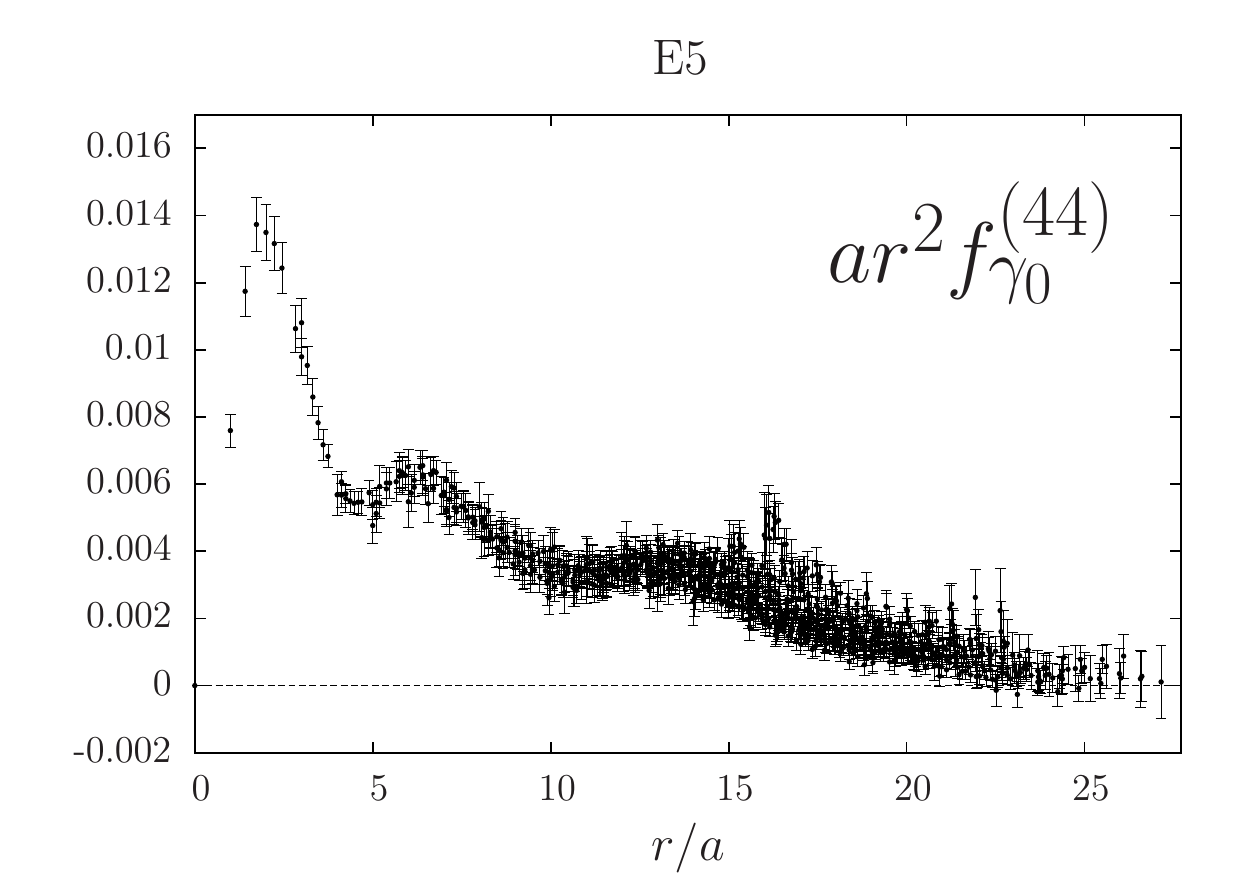}
	\end{minipage}	
	\caption{\label{fig:densitycharge} Density distributions $ar^2f^{(nn)}_{\gamma_0}(r/a)$, $n=1,2,3$ (top) and $n=1,2,4$ (bottom) on the lattice ensemble E5, using only $\bar{q} \gamma_k h$ (top) and including $\bar{q} \nabla_k h$ interpolating fields (bottom) in the analysis.}
\end{figure}

To further investigate this issue, we have computed the radial distribution of the vector density, because the conservation of the vector charge is a precious indicator of a possible source of uncontrolled systematics if it is strongly violated. It is defined similarly to the axial density distribution by replacing the axial density with $\mathcal{O}_{\Gamma}  = \overline{\psi}_l  \gamma_0 \psi_l$. With the interpolating field $\bar{q}\nabla_k h$ included in the basis, together with $\bar{q}\gamma_k h$, we show in Fig. \ref{fig:chargesumnabla} the ``effective" charge density distributions $f^{(nn)}_{\gamma_0}(r)$ integrated over $r$, in function of the time $t$ entering the sGEVP. In the cases of $f^{(11)}_{\gamma_0}(r)$ and $f^{(22)}_{\gamma_0}(r)$, plateaus are clearly compatible with $1/Z_V$ while, for $f^{(33)}_{\gamma_0}(r)$, we observe a divergence with time. Concerning $f^{(44)}_{\gamma_0}(r)$, a (very short) plateau shows up again around $1/Z_V$. Once more, the main lesson is that the second excited state isolated by the GEVP is hard to interpret as a $\bar{q}b$ bound state \emph{whereas the first excited state is}. Density distributions themselves are showed in Fig.~\ref{fig:densitycharge}. Plots on the top correspond to the basis with only $\bar{q}\gamma_i h$-kind interpolating fields of the $B^*$ meson and those on the bottom are obtained after incorporating $\bar{q}\nabla_k h$-kind in the analysis. We note similar facts as for the spectrum: $f^{(11)}_{\gamma_0}(r)$ and $f^{22}_{\gamma_0}(r)$ are almost the same, $f^{(33)}_{\gamma_0}(r)$ of the top looks like $f^{(44)}_{\gamma_0}(r)$ on the bottom. Finally it revealed impossible to obtain a stable density for $f^{(33)}_{\gamma_0}(r)$ when we include $\bar{q} \nabla_k h$ operators in the analysis. Actually, it is just a rephrasing of the observation made at the beginning of the subsection. 

\section{Conclusion}

In that paper we have reported on a lattice estimate of the form factor $A^{12}_0(q^2=0)$ associated with the matrix element $\langle B|{\cal A}^\mu |B^{*\prime}\rangle$ and $g_{B^{*\prime}B\pi}$ coupling. We have measured axial density distributions whose Fourier transforms are used to extrapolate at $q^2=0$ and we obtain $A^{12}_0(0) =  -0.173(31)(16)$ and $g_{B^{*\prime} B\pi} =  -15.9(2.8)(1.4)$. We have confirmed a phenomenological finding that $g_{B^{*\prime}B\pi}$ is negative, with a magnitude accidentally similar to what was computed for the form factor $A^{12}_1(q^2_{\rm max})$, the coupling $\overline{g}_{12}$. We have checked several sources of systematics: cut-off effects, finite-size effects, a possible mixing between radial excitations and multihadron states. In particular, we have clues that interpolating fields of the kind $\bar{q} \nabla_k h$ have a strong coupling to a state that is difficult to interpret as a $\bar{q} b$ bound state because, in that case, density distributions are impossible to extract from our data.

\section*{Acknowledgements}

We thank Damir Becirevic, Philippe Boucaud and Alain Le Yaouanc for valuable discussions and Olivier P\`ene for his careful reading of the manuscript.
We are grateful to CLS for making the gauge configurations used in this work available to us. B.B. has been partly supported by the project USP/COFECUB Uc Ph 157-15 during the work. Computations of the relevant correlation functions are made on GENCI/CINES, under the Grant 2015-056806.

\appendix

\section{Charge and matter distributions}
\label{app:charge_matter_distrib}

In this appendix, we discuss the charge (vector) and matter (scalar) radial distributions. They are defined similarly to the axial density distributions by replacing the axial density $\mathcal{O}_{\Gamma} = \overline{\psi}_l  \gamma_{\mu} \gamma_5 \psi_l$ with $\mathcal{O}_{\Gamma} = \overline{\psi}_l  \mathbb{1} \psi_l$ and $\mathcal{O}_{\Gamma} = \overline{\psi}_l  \gamma_0 \psi_l$ respectively. They have been computed for lattice ensembles E5 and D5. 

\subsection{Correlation functions} 

\begin{figure}[t]
	\vspace{0.5cm}
	\centering
	\unitlength = 1mm
	\begin{fmffile}{feyn/c3pts_disc}
	\begin{fmfgraph*}(60,40)
	\fmfleft{i}
	\fmfright{o}
	\fmftop{t}
	\fmf{phantom,tension=5}{i,v1}
	\fmf{phantom,tension=5}{v2,o}
	\fmf{dbl_plain,width=1}{v1,v2}
	\fmf{phantom,tension=0.8,width=1}{v2,t}
	\fmf{phantom,tension=0.8,width=1}{t,v1}
	\fmf{plain,left=0.45,tension=0.1,width=1}{v1,v2}
	\fmf{plain,left=0.45,tension=1.5,width=1}{t,t}
	\fmfv{decor.shape=circle, decor.filled=full, decor.size=2thick}{v1}
	\fmfv{decor.shape=circle, decor.filled=full, decor.size=2thick}{v2}
	\fmfv{decor.shape=circle, decor.filled=empty, decor.size=4thick, label.angle=0}{t}
	\fmflabel{$(\vec{x},0) \,,\ \gamma_5$}{v1}
	\fmflabel{$ \gamma_i \,, \ (\vec{x},t)$}{v2}
	\fmflabel{$\quad \mathcal{O}^{\mathbb{1}} \,, \ (\vec{x}+\vec{r}, t_1)$}{t}
	\end{fmfgraph*}
	\end{fmffile}
	\vspace{-1.8cm}
	\caption{Disconnected contribution to the three-point correlation function in the case $\Gamma = \mathbb{1}$.}
	\label{fig:disc_contribution}
\end{figure}
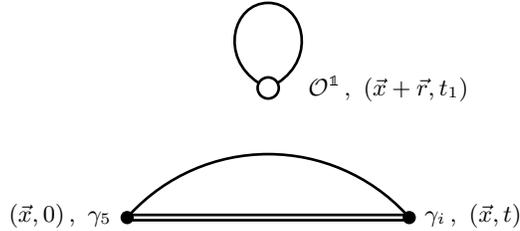

Using the notation $\Gamma=\gamma_0,\mathbb{1}$, the three-point correlation functions associated to the charge and matter distributions are
\begin{align*}
C_{\Gamma, ij}^{(3)}(t, t_1;\vec{r}) &=  \langle \, \mathcal{P}^{(j)}(\vec{x},t) \mathcal{O}_{\Gamma}(\vec{x}+\vec{r},t_1)  \mathcal{P}^{(i)\dag}(\vec{x},0) \, \rangle \,,\\
C_{\Gamma, ij}^{(3)}(t, t_1;\vec{r}) &= \frac{1}{3} \sum_{k=1}^3  \langle \, \mathcal{V}^{(j)}_{k}(\vec{x},t) \mathcal{O}_{\Gamma}(\vec{x}+\vec{r},t_1)  \mathcal{V}^{(i)\dag}_{k}(\vec{x},0) \, \rangle \,.
\end{align*}
In Section \ref{sec3}, we have chosen the isospin combination which excludes the neutral pion, in order to avoid the computation of disconnected diagrams. For the charge distribution, where $\Gamma=\gamma_0$, one can show that disconnected contributions vanish exactly \cite{DraperBP} but this is not true for the matter distribution where $\Gamma = \mathbb{1}$. In the latter case, one should also consider the disconnected contribution (see Fig.~\ref{fig:disc_contribution}) which are more difficult to estimate numerically and have not been computed in this study.

\subsection{Summation over $r$} 

\begin{figure}[t!]

	\begin{minipage}[c]{0.28\linewidth}
	\centering 
	\includegraphics*[width=\linewidth]{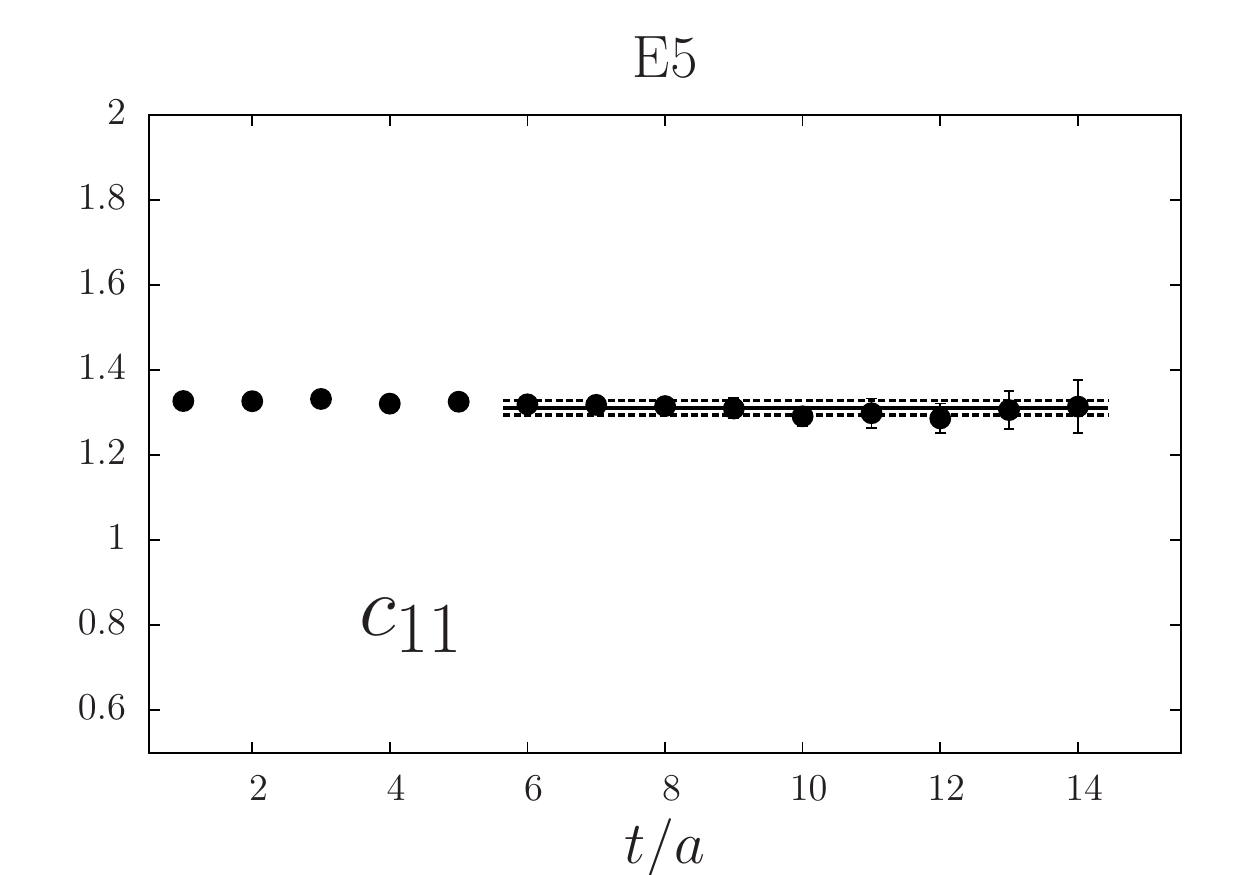}
	\end{minipage}
	\begin{minipage}[c]{0.28\linewidth}
	\centering 
	\includegraphics*[width=\linewidth]{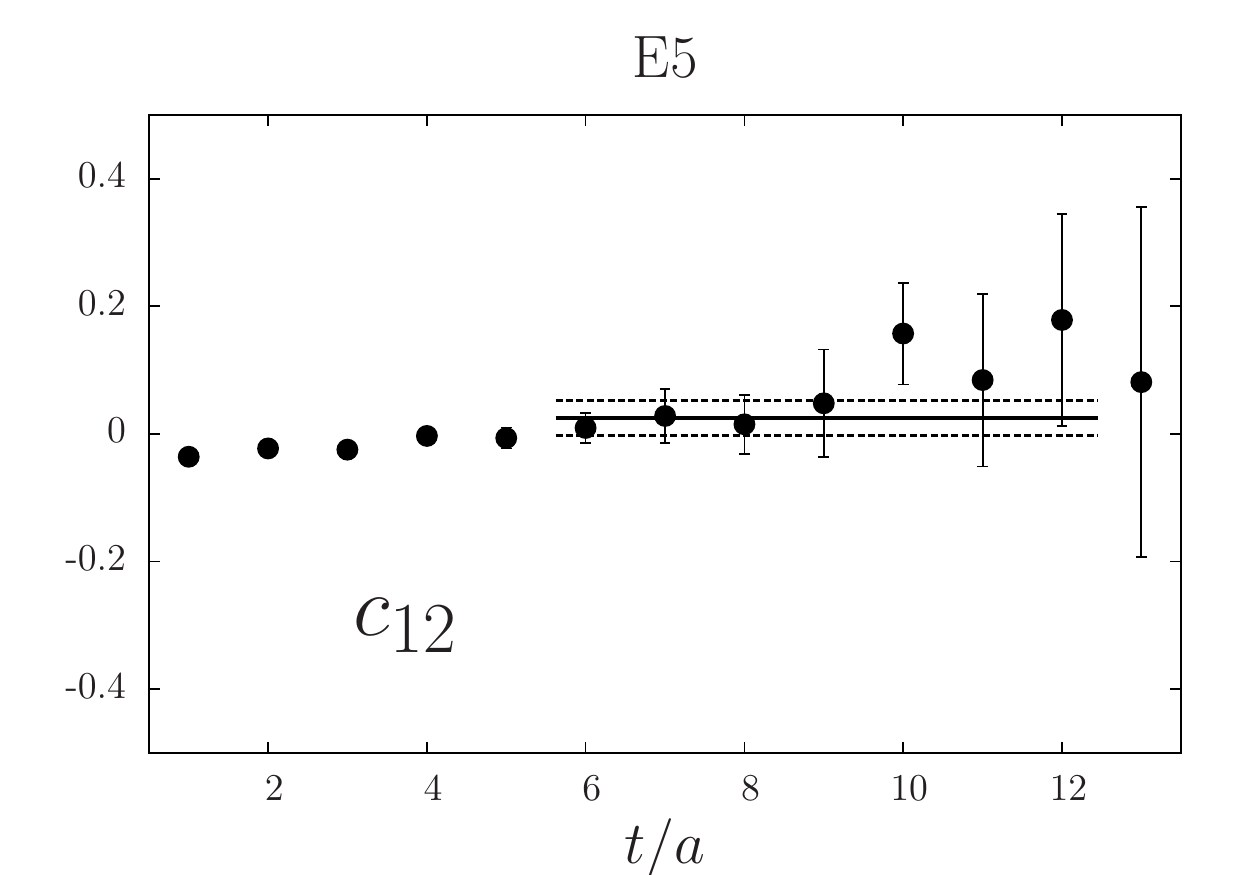}
	\end{minipage}
	\begin{minipage}[c]{0.28\linewidth}
	\centering 
	\includegraphics*[width=\linewidth]{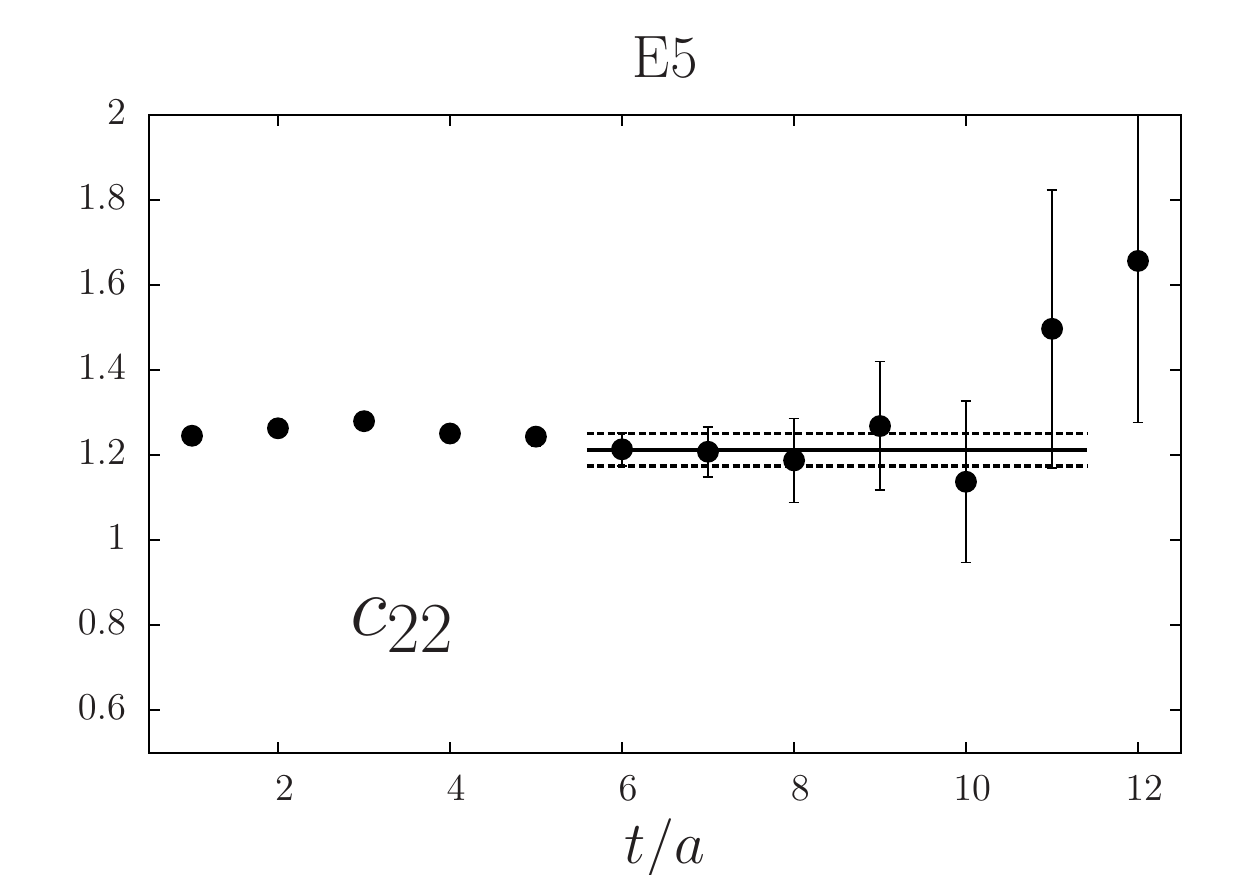}
	\end{minipage}
	
	\begin{minipage}[c]{0.28\linewidth}
	\centering 
	\includegraphics*[width=\linewidth]{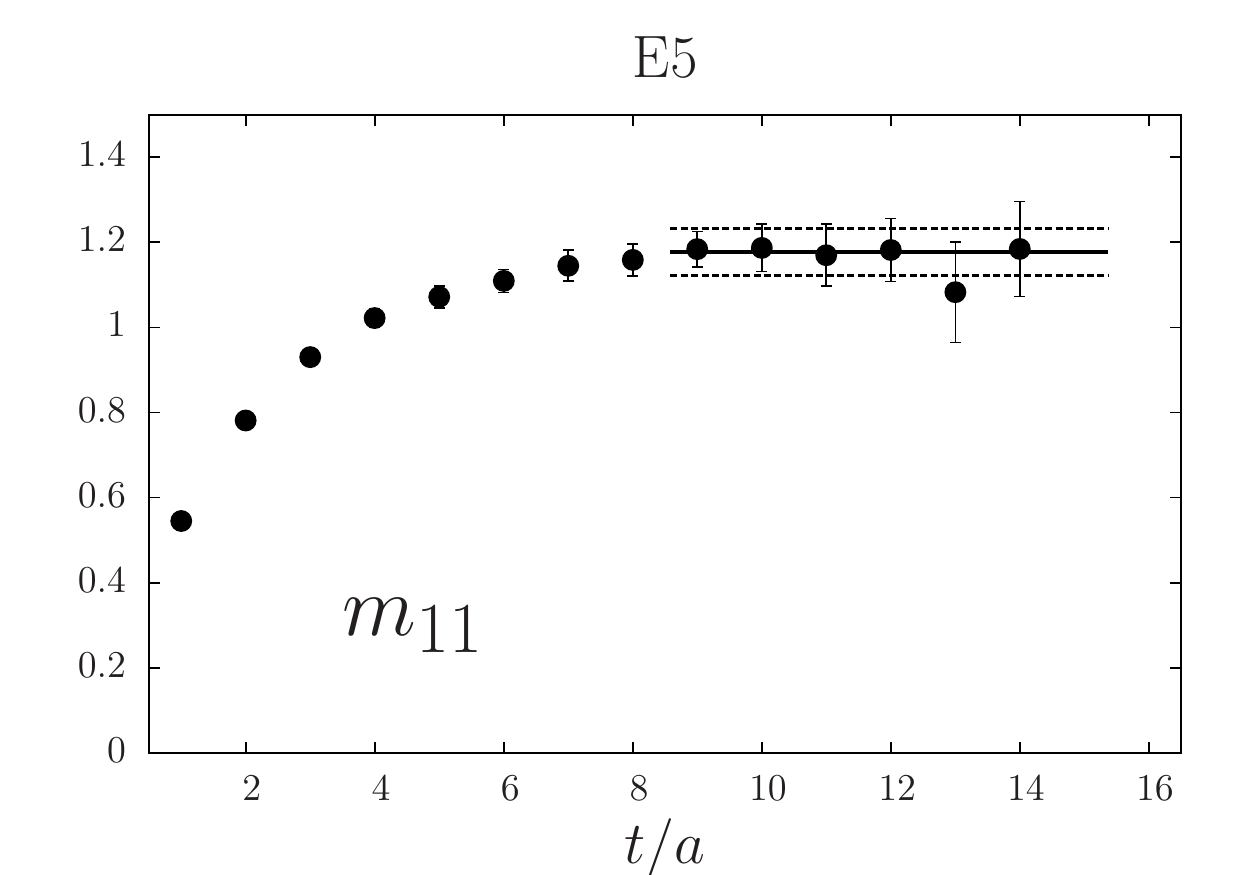}
	\end{minipage}
	\begin{minipage}[c]{0.28\linewidth}
	\centering 
	\includegraphics*[width=\linewidth]{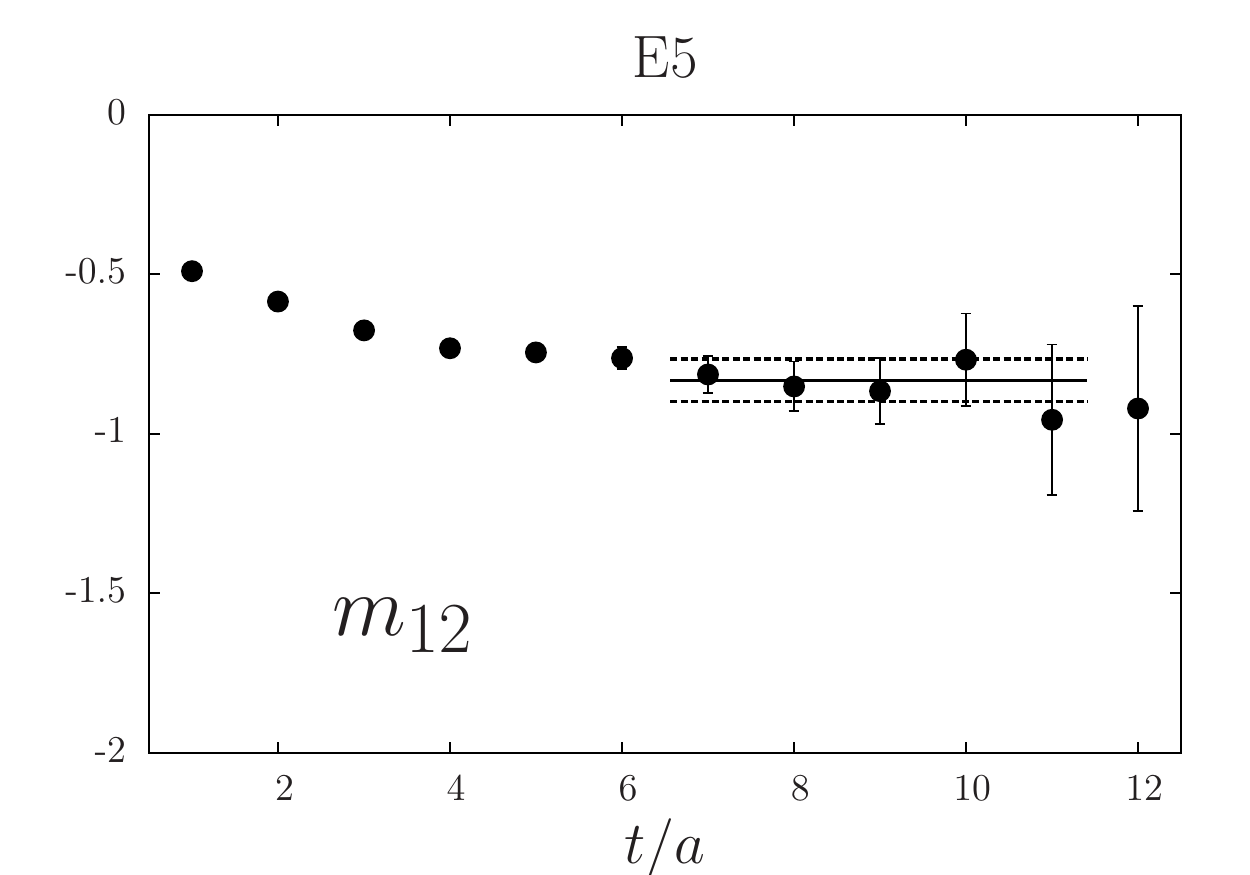}
	\end{minipage}
	\begin{minipage}[c]{0.28\linewidth}
	\centering 
	\includegraphics*[width=\linewidth]{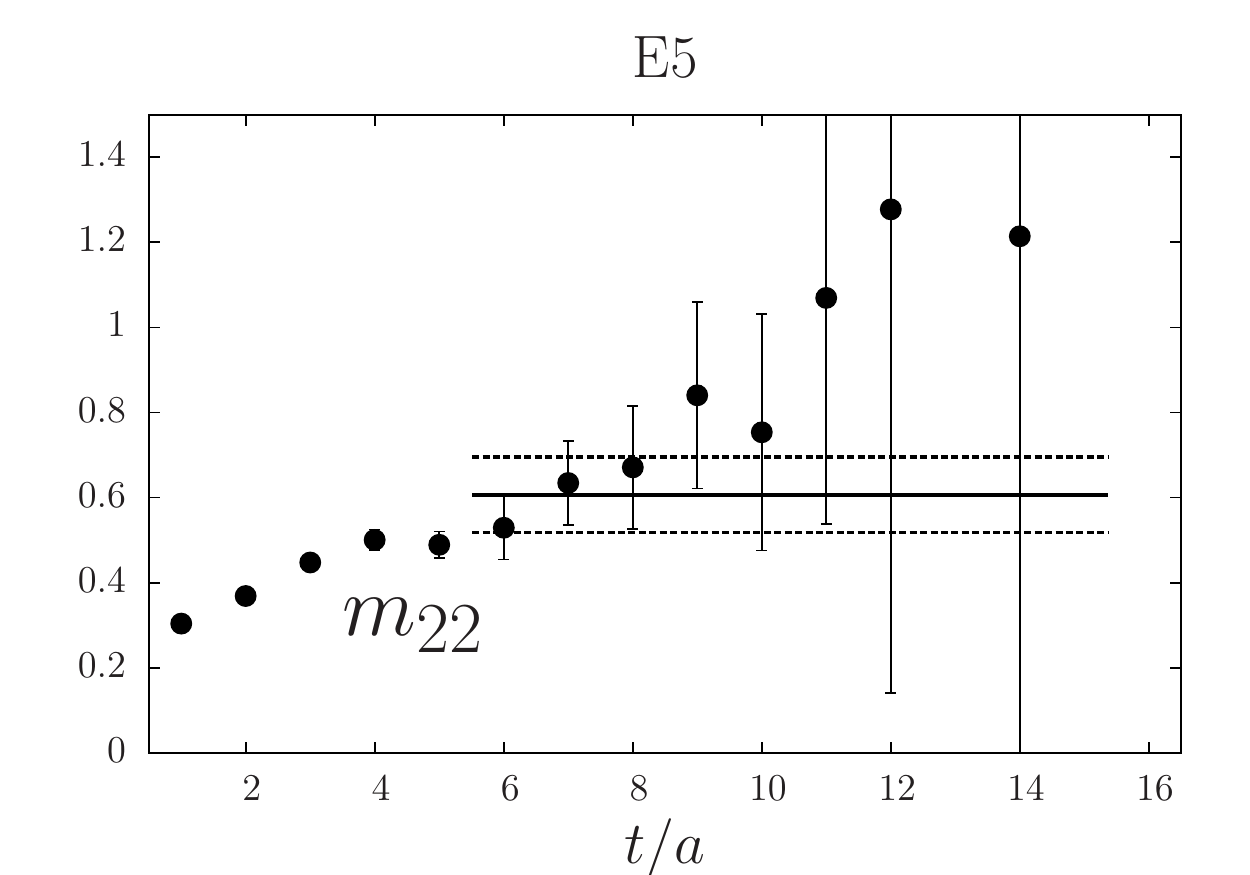}
	\end{minipage}	
	\caption{Plateaus of total vector charges $c_{11}$, $c_{12}$ and $c_{22}$ (top) and scalar charges $m_{11}$, $m_{12}$ and $m_{22}$ (bottom) for the lattice ensemble E5.}	
\label{fig:sv_sum_rules}
\end{figure}

Taking the sum over all values of $\vec{r}=(x,y,z)$ of the charge and matter radial distributions, one should obtain the (bare) couplings
\begin{equation*}
c_{mn} =  a^3 \sum_{\vec{r}} f_{\gamma_0}^{(mn)}( \vec{r} ) \quad \,, \quad m_{mn} = a^3 \sum_{\vec{r}} f_{\mathbb{1}}^{(mn)}( \vec{r} )  \,.
\end{equation*}
For charge distributions, one expects, in the continuum limit and after renormalisation, $Z_V(g_0^2)\,c_{11} = Z_V(g_0^2)\,c_{22} = 1$ for the diagonal couplings and $c_{12} = 0$ for the off-diagonal coupling. Plateaus are depicted in Figs.~\ref{fig:sv_sum_rules}  and results are collected in Table~\ref{tab:sum_rule_charge}. The total charges $\overline{c}_{11}$ and $\overline{c}_{22}$ are close to unity and deviations from unity are probably due to lattice artefacts. Moreover, $c_{12}$ is compatible with zero, which confirms that $c_{11}$ can be interpreted as a wave function of the ground state. For matter distributions, one would also expect, in the continuum limit and after renormalisation (including also the renormalisation constant of the quark mass), ${\cal Z}(g_0^2)\, m_{11} = {\cal Z}(g_0^2)\, m_{22} = 1$ for the diagonal couplings and $m_{12} = 0$ for the off-diagonnal couplings. However, our computation does not take into account disconnected contributions and the interpretation of results is not clear. In particular, $m_{12} \neq 0$ indicates that the disconnected diagram probably has a significant contribution.

\renewcommand{\arraystretch}{1.3}
\begin{table}[t]
	\begin{center}
	\begin{tabular}{c@{\quad}c@{\quad}c@{\quad}c@{\quad}c@{\quad}c@{\quad}c}
	\hline
	$ij$		&	$11$			&	$22$			&	$33$ &	$12$			&	$13$			&	$23$ \\ 
	\hline
	$c_{ij}$	&	$1.311(17)$	&	$1.212(38)$	&	$1.153(33)$ &	$0.015(32)$	&	$-0.062(49)$	&	$-0.010(35)$ \\ 
	\hline
	$\overline{c}_{ij}$	&	$0.983(13)$	&	$0.909(29)$	&	$0.865(25)$		&	$0.011(24)$	&	$-0.047(37)$	&	$-0.008(26)$ \\ 
	\hline
	$m_{ij}$  &	$1.177(55)$	&	$0.602(88)$	&	$0.249(57)$ &	$-0.833(67)$	&	$0.318(40)$	&	$-0.338(29)$ \\ 
	\hline
	\end{tabular}
	\end{center}	
	\vspace{-0.2cm}
	\caption{Bare and renormalized couplings associated to the charge and matter densities for the CLS ensemble E5. We use the nonperturbative estimate $Z_V = 0.750(5)$ extracted from~\cite{DellaMorteRD, Fritzsch:2012wq}. }
	\label{tab:sum_rule_charge}
\end{table}

\subsection{Radial distributions} 

\begin{figure}[t] 
	\begin{minipage}[c]{0.28\linewidth}
	\centering 
	\includegraphics*[width=\linewidth]{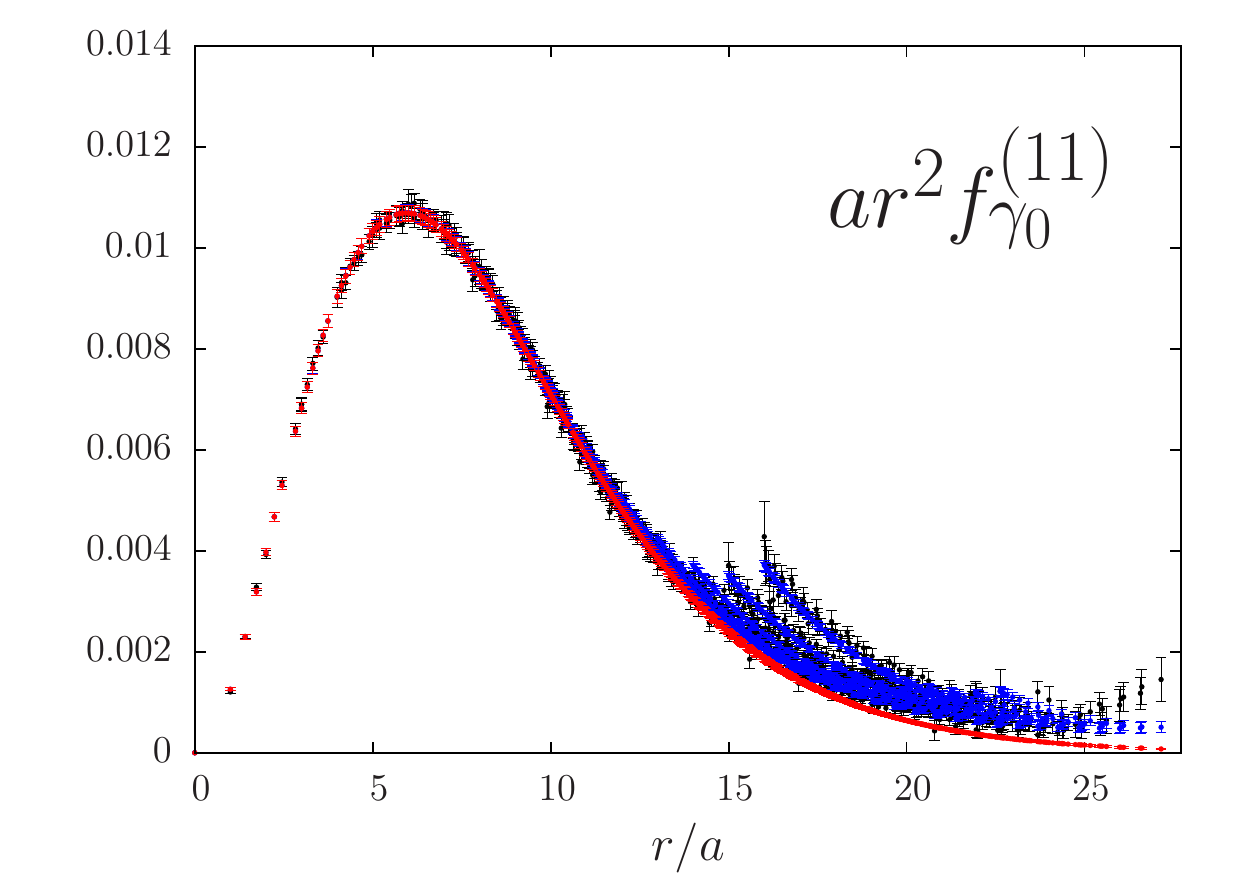}
	\end{minipage}
	\begin{minipage}[c]{0.28\linewidth}
	\centering 
	\includegraphics*[width=\linewidth]{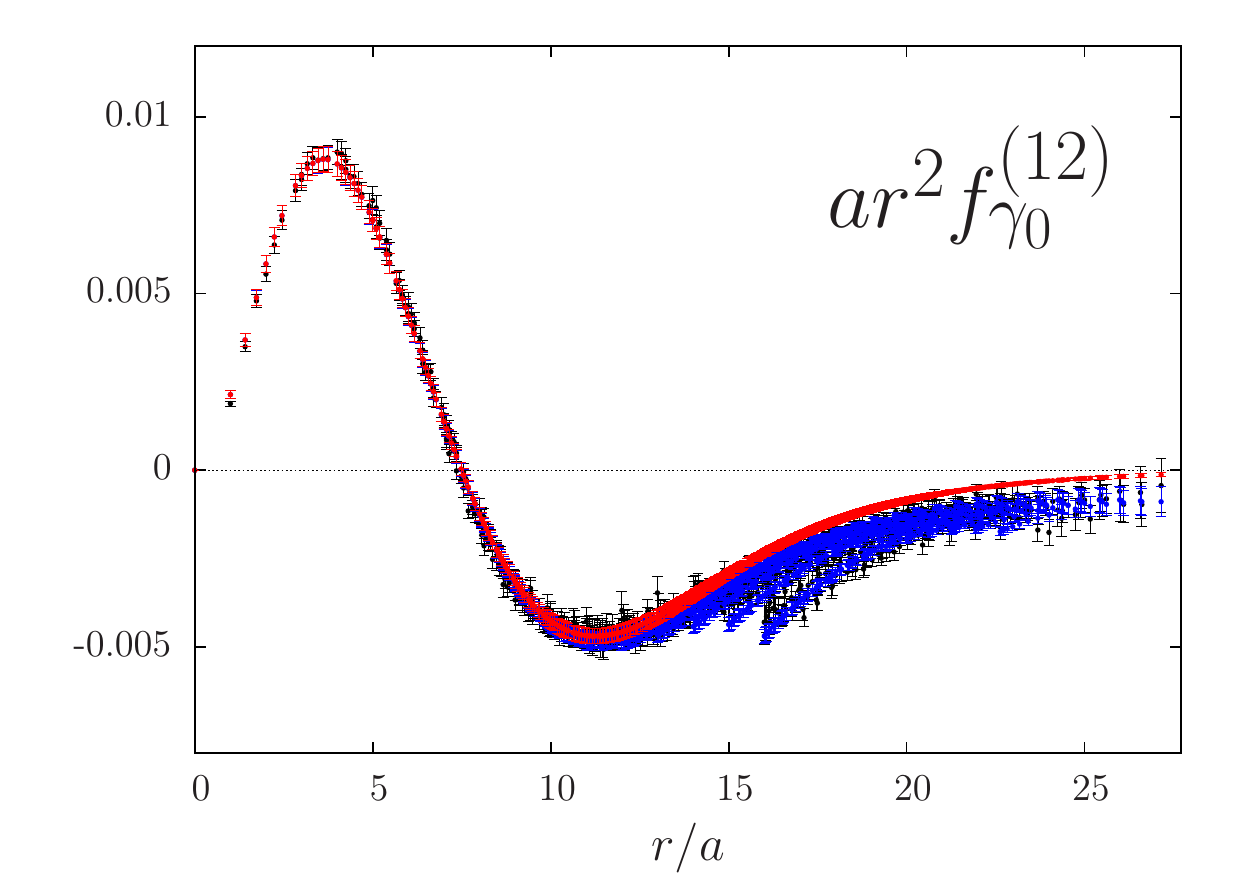}
	\end{minipage}
	\begin{minipage}[c]{0.28\linewidth}
	\centering 
	\includegraphics*[width=\linewidth]{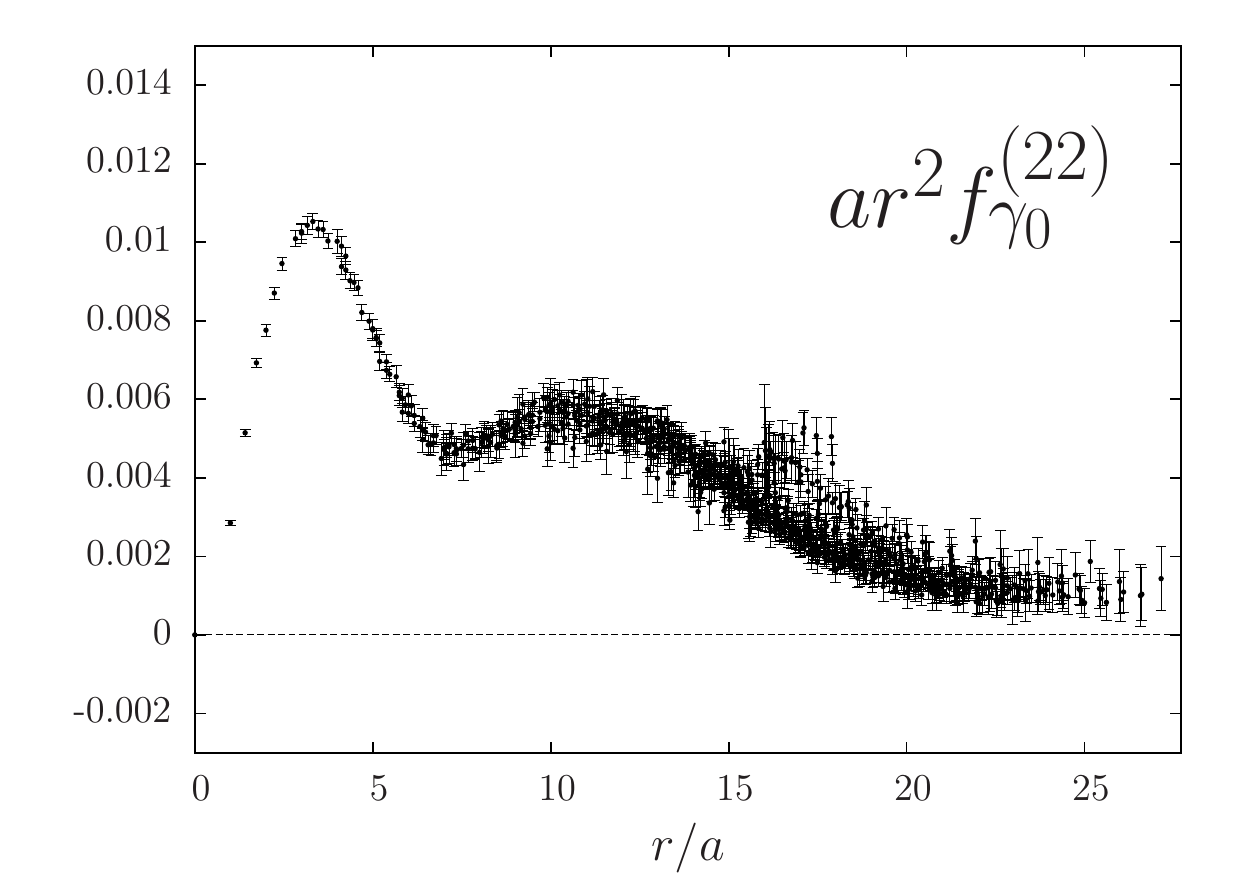}
	\end{minipage}
	\begin{minipage}[c]{0.28\linewidth}
	\centering 
	\includegraphics*[width=\linewidth]{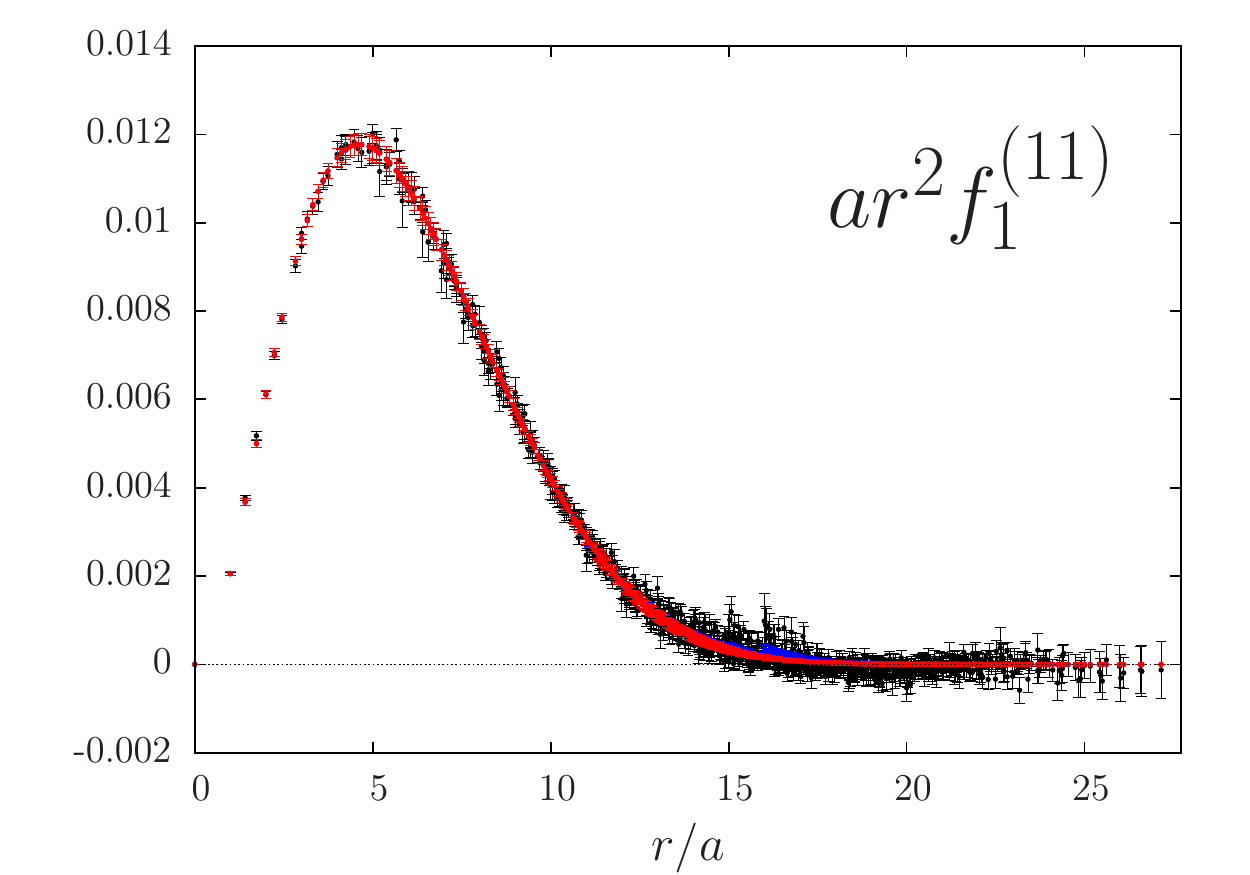}
	\end{minipage}
	\begin{minipage}[c]{0.28\linewidth}
	\centering 
	\includegraphics*[width=\linewidth]{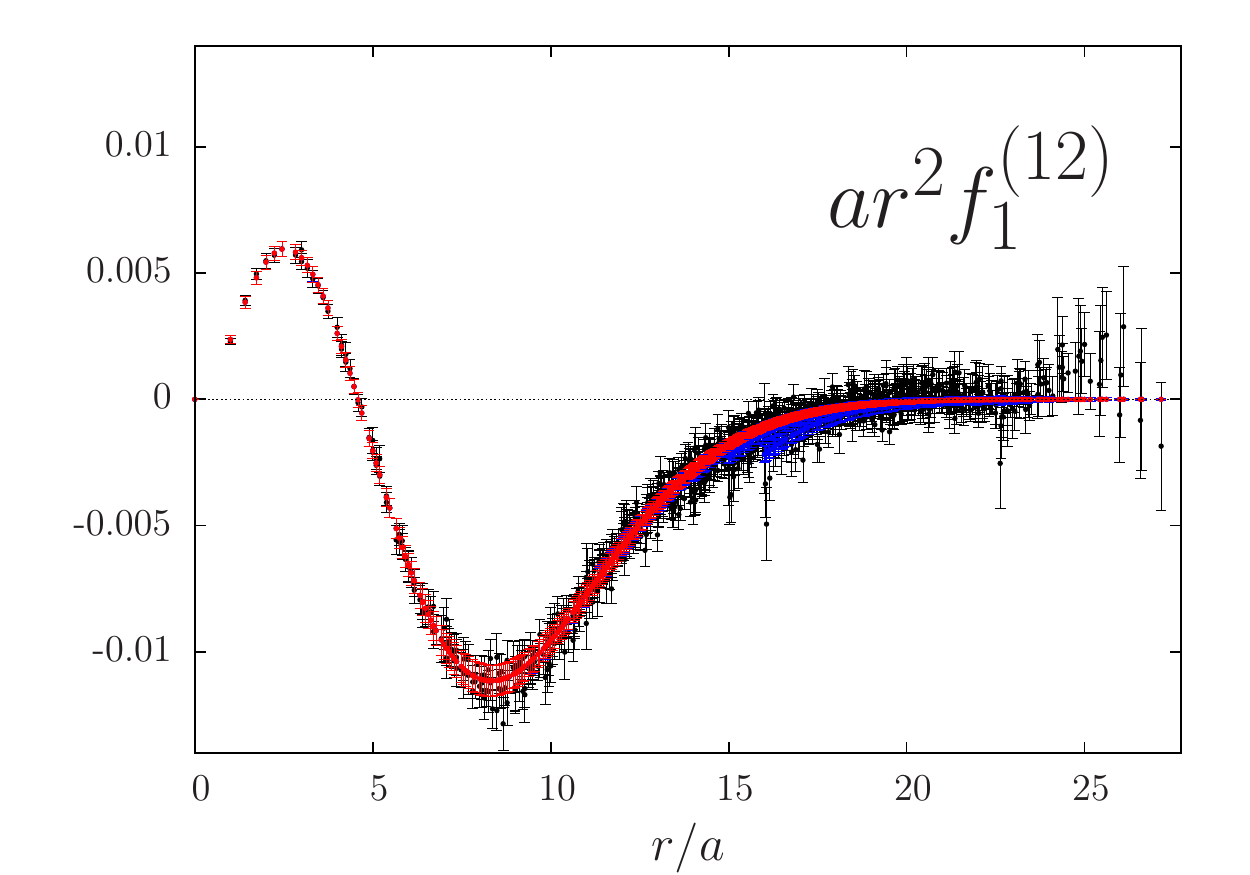}
	\end{minipage}
	\begin{minipage}[c]{0.28\linewidth}
	\centering 
	\includegraphics*[width=\linewidth]{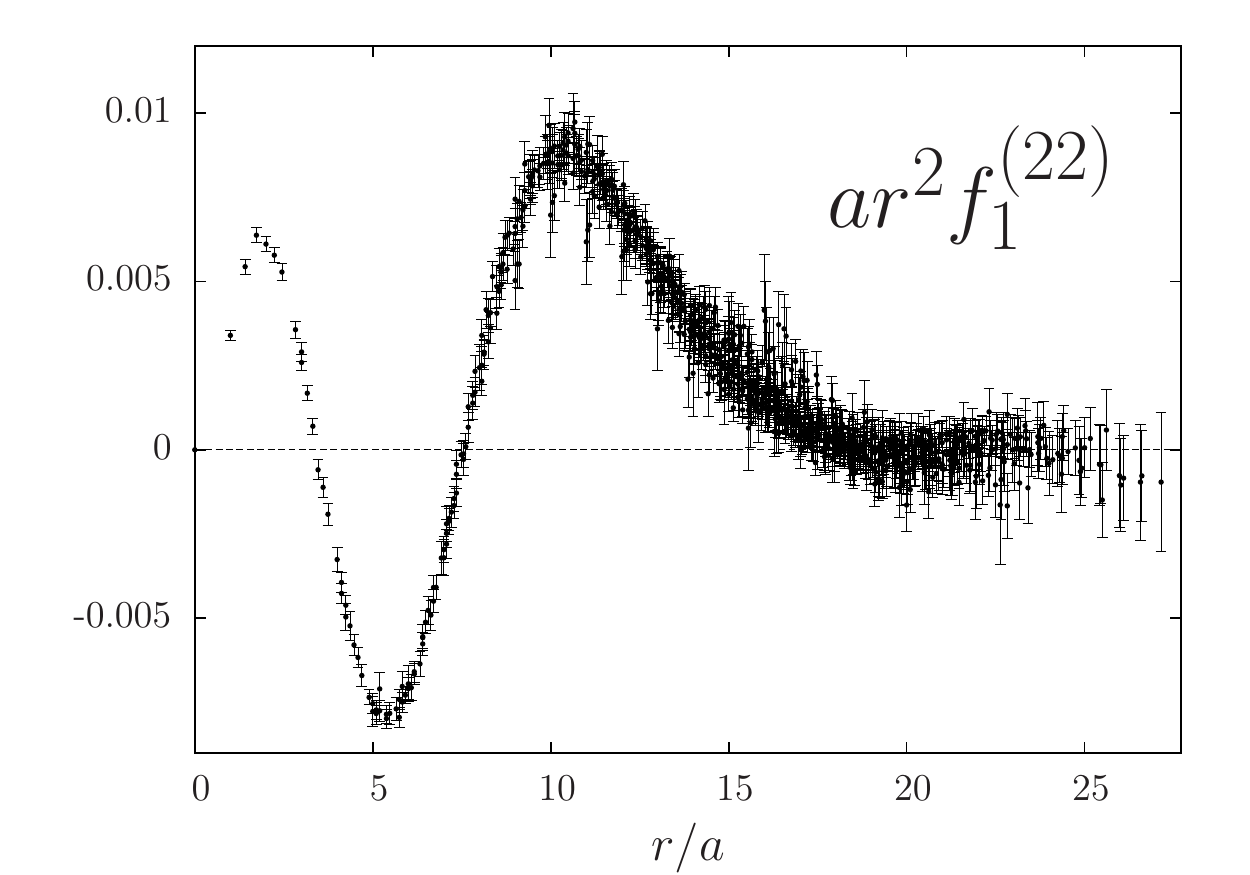}
	\end{minipage}
	
	\caption{(\textit{top}) Charge densities $ar^2 f^{(11)}_{\gamma_0}(r/a)$, $ar^2 f^{(12)}_{\gamma_0}(r/a)$ and $ar^2 f^{(22)}_{\gamma_0}(r/a)$ for the CLS ensemble E5. (\textit{bottom}) Matter densities $ar^2 f^{(11)}_{\mathbb{1}}(r/a)$, $ar^2 f^{(12)}_{\mathbb{1}}(r/a)$ and $ar^2 f^{(22)}_{\mathbb{1}}(r/a)$ for the CLS ensemble E5}	
\label{fig:densities_mc}
\end{figure}

\begin{figure}[t]
	\centering 
	\includegraphics*[width=0.5\linewidth]{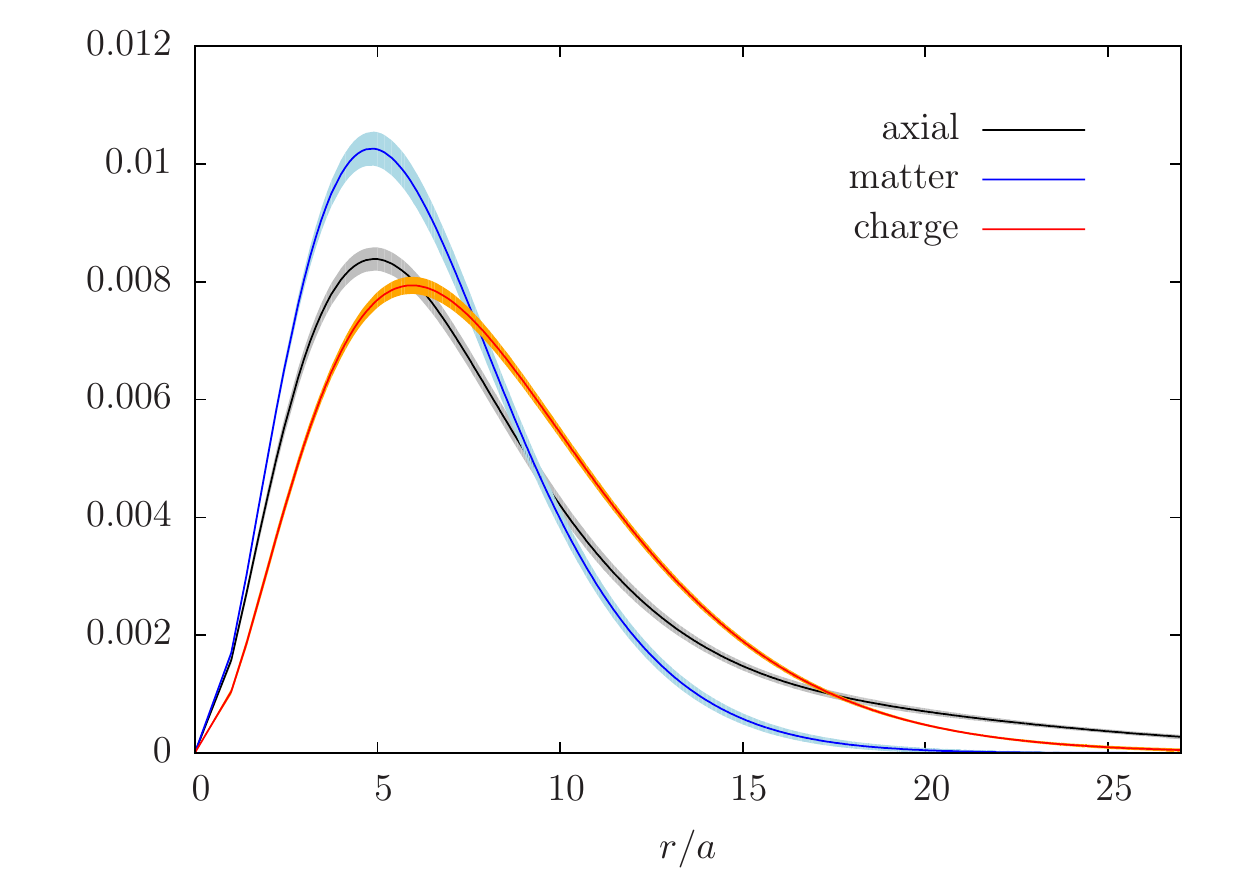}
	\caption{Axial, matter and charge densities $a r^2 f^{(11)}_{\alpha}(r/a)$ obtained after the subtraction of volume effects for the CLS ensemble E5. The normalisation is such that the area below the curve is one.}
\label{fig:cmp_ACM}
\end{figure}

The radial distributions obtained using the sGEVP method are plotted in Fig.~\ref{fig:densities_mc}. Matter distributions decrease faster than charge distributions and are compatible with zero at $r \approx L/2$. It explains the absence of any fishbone structure due to overlap of the tail with periodic images. We used the same method as for the axial density radial distribution to remove volume and cubic lattice artefacts. Using the fit formula (\ref{eq:fit_vol}), we obtain for the matter density $\alpha=1.3(1)$, $r_0=0.23(1)~\fm$ and for the charge density $\alpha=1.32(5)$, $r_0=0.34(1)~\fm$. Similarly to the axial distributions, we define the matter and charge square radii by
\begin{equation*}
\langle r^2 \rangle_M = \frac{ \displaystyle \int_0^{\infty} \, \mathrm{d}r \, r^4 \, f^{(11)}_{\mathbb{1}}(r)  }{  \displaystyle \int_0^{\infty} \, \mathrm{d}r \, r^2 \, f^{(11)}_{\mathbb{1}}(r) } \quad , \quad
\langle r^2 \rangle_C = \frac{ \displaystyle \int_0^{\infty} \, \mathrm{d}r \, r^4 \, f^{(11)}_{\gamma_0}(r)  }{ \displaystyle  \int_0^{\infty} \, \mathrm{d}r \, r^2 \, f^{(11)}_{\gamma_0}(r) }\,,
\end{equation*}
and results are given in Table~\ref{tab:square_radii_CM}, as well as the position of the node of the various distributions. We observe that $\langle r^2 \rangle_M < \langle r^2 \rangle_A < \langle r^2 \rangle_C$ but that the axial distribution decreases slower at large $r$, as can been seen in Fig.~\ref{fig:cmp_ACM}. It explains why volume artefacts are almost absent for the matter density but are large in the case of the axial density. It is interesting to note that the hierarchy $\langle r^2 \rangle_M < \langle r^2 \rangle_C$ had been observed in light bound systems like the pion and the proton \cite{AlexandrouQT}.\\

\renewcommand{\arraystretch}{1.3}
\begin{table}[t]
\begin{center}
\begin{tabular}{cc}
\begin{tabular}{c@{\quad}c@{\quad}c}
	\hline
	$\langle r^2 \rangle_M~[{\rm fm}^2]$	 &	$\langle r^2 \rangle_C~[{\rm fm}^2]$ &	$\langle r^2 \rangle_A~[{\rm fm}^2]$	  \\
	\hline
	$0.213(10)$   	&	$0.380(8)$	&	$0.358(15)$	\\  
	\hline
 \end{tabular} 
& \qquad \qquad
\begin{tabular}{c@{\quad}c@{\quad}c}
	\hline
	$(r_n)_M~[\fm]$	 &	$(r_n)_C~[\fm]$ 	&	$(r_n)_A~[\fm]$	  \\
	\hline 
	$0.311(3)$		&	$0.484(6)$		&	$0.369(4)$	\\  
	\hline
 \end{tabular} 
\end{tabular}
\end{center}
\caption{Left panel: Square radius of the ground state radial distributions $f^{(11)}_{\gamma_0}(r)$, $f^{(11)}_{\mathbb{1}}(r)$ and $f^{(11)}_{\gamma_i \gamma_5}(r)$ and for the lattice ensemble E5; right panel: Position of the node $r_n$ of the radial distributions $r^2\, f^{(12)}_{\gamma_0}(r)$, $r^2\, f^{(12)}_{\mathbb{1}}(r)$ and $r^2\, f^{(12)}_{\gamma_i \gamma_5}(r)$ for the lattice ensemble E5.}
\label{tab:square_radii_CM} 
\end{table}

\section{Cubic lattice artefacts}
\label{sec:artefacts}

On hypercubic lattices, the SO(3) rotational symmetry is explicitly broken down to the isometry group $H(3)$. Therefore, a function which depends only on $r$ in the continuum can, on the lattice, take different values if sites are related by an SO(3) symmetry (same $r$) but not invariant under H(3) (or, equivalently, if they belong to different orbits). These differences should vanish as the lattice spacing goes to zero. We follow the method presented in~\cite{deSotoHT,Boucaud:2003dx} to subtract such lattice artefacts. Of course, other lattice artefacts are still present after this procedure (in particular lattice artefacts that depend only on $r^2$) and they are removed by taking the continuum limit. The time component of the axial density distribution is odd in the direction $i$ and therefore is not a function of $r$ only. In this case, the value on the function is not expected to be the same for different radii belonging to the same orbit, even in the continuum limit. Therefore, the technique presented here can be applied to the spatial component only.\\

Starting with the dimensionless distribution $a^3 f^{(nm)}_{\gamma_{i} \gamma_5}(r)$ computed on the lattice by the sGEVP method and assuming that, at fixed $r$, the lattice artefacts vanish smoothly to zero when the continuum limit is taken, one can write the following Taylor expansion (valid near $a=0$) 
\begin{equation}
a^3 f_{\alpha}(r^2, r^{[4]}, r^{[6]}) = a^3 f_{\alpha}(r^2, 0, 0) + \sum_n \underbrace{r^{[2i]} r^{[2j]} \dots }_{n} \times \frac{\partial^{n} \left( a^3 f_{\alpha} \right) }{ \partial r^{[2i]} \partial r^{[2j]} \dots } \Big|_{(r^2,0,0)} \,,
\label{eq:dev}
\end{equation}
where $r^{[2k]} = \sum_{i=1}^3 r_i^{2k}$ index the set of orbits (it can be shown that any polynomial function invariant under the isometry group H(3) is a function of the three invariants $r^{[n]}$, $n=2,4,6$). Here, $a^3 f_{\alpha}(r^2, 0, 0)$ corresponds to the radial distribution free of cubic artefacts. Neglecting for the moment volume effects, the only dimensionless quantities are $r/a$, $r^{[2i]}/a^{2i}$ and $r^{[2i]}/r^{2i}$ and any (dimensionless) polynomial function in $r^{[2i]}$ can be written in term of the monomials
$$ \left( \frac{ r^{[2n]} a^{2k} }{ r^{2n+2k} }  \right)^m \,,$$
where $n=1,2,3$ and $k>0$ since we want lattice artefacts to vanish in the continuum limit.
Thanks to $\mathcal{O}(a)$-improvement, one expects first lattice artefacts to be proportional to $a^2$ and there are only two dimensionless terms:
\begin{equation}
\frac{ a^{2} r^{[4]} }{ r^{6}} \,,\ \frac{ a^{2} r^{[6]} }{ r^{8}}  \,.
\label{eq:dim}
\end{equation}
At order $a^4$, the new terms would be
\begin{equation}
\frac{ a^{4} r^{[4]} }{ r^{8}} \,,\ \frac{ a^{4} r^{[6]} }{ r^{10}} \,, \ \left(\frac{ a^{2} r^{[4]} }{ r^{6}}\right)^2 \,,\ \left(\frac{ a^{2} r^{[6]} }{ r^{8}}\right)^2  \,.
\label{eq:dim2}
\end{equation}
Therefore, based on Eqs.~(\ref{eq:dev}) and (\ref{eq:dim}), and considering only $a^2$ artefacts for the moment, we expect
\begin{equation*}
a^3 f_{\alpha}(r^2, r^{[4]}, r^{[6]}) = a^3 f_{\alpha}(r^2, 0, 0) + r^{[4]} \ \frac{\partial (a^3 f_{\alpha})}{ \partial r^{[4]}} \Big|_{(r^2,0,0)} + r^{[6]} \ \frac{\partial (a^3 f_{\alpha)}}{ \partial r^{[6]}} \Big|_{(r^2,0,0)} \,,
\end{equation*}
where, based on the previous dimensional analysis,
\begin{align*}
\frac{\partial (a^3 f_{\alpha})}{ \partial r^{[4]}} \Big|_{(r^2,0,0)} &\sim \frac{a^2}{r^6} \,,\\ 
\frac{\partial (a^3 f_{\alpha})}{ \partial r^{[6]}} \Big|_{(r^2,0,0)} &\sim \frac{a^2}{r^{8}} \,. 
\end{align*}
When different orbits exist with the same value of $r$, the previous derivatives can be estimated numerically by making a linear regression. Then, based on dimensional arguments, the derivative is extended to all values of $r$ by fitting the result in $b/r^\alpha$ where $b$ is a constant and $\alpha=6,8$ respectively for $r^{[4,6]}$ cubic artefacts. However, since only a small subset of our $r^2$ orbits contain more than three points, the numerical estimation of the derivatives is difficult. This difficulty is even worse since artefacts $r^{[6]}$ already appear at leading order in $a$ compared to what happens in momentum space \cite{Boucaud:2003dx} where $p^{[6]}$ and $p^{[8]}$ artefacts appear only at order $a^4$ and $a^6$ respectively. To circumvent this problem, a more powerful method was proposed in \cite{Boucaud:2003dx}: the idea is to fit the full data sample by a function having the form
\begin{equation}
a^3 f_{\alpha}(r^2, r^{[4]}, r^{[6]}) = a^3 f_{\alpha}(r^2, 0, 0) + A \times \frac{ a^2 r^{[4]} }{r^{6}} + B \times \frac{ a^2 r^{[6]} }{r^{8}} \,,
\label{eq:fit_artefacts}
\end{equation}
where $a^3 f_{\alpha}(r^2, 0, 0)$ is taken as a free parameter which is fitted with the other dimensionless coefficients $A$ and $B$. In particular, no assumption is made on the functional form of $f_{\alpha}(r^2, 0, 0)$. Of course, when only one point belongs to an orbit, it does not contribute to the fit since, in this case, $f_{\alpha}(r^2, 0, 0)$ can always be adjusted freely. In Table~\ref{tab:fit_info}, the number of fit parameters and the number of data available in the fit is given for each lattice resolution used here.
\renewcommand{\arraystretch}{1.3}
\begin{table}[t]
	\begin{center}
	\begin{tabular}{l@{\quad}c@{\quad}c@{\quad}c@{\quad}c}
	\hline	\multicolumn{1}{c}{}			&	$\#$~r	&	$\#$~orbits	&	$\#$~fitted orbits 	& \#~data	\\ 
	\hline
	$L=32$			&	969		&	464			&	284				&	789\\ 
	\hline 
	$L=48$			&	2925		&	1057			&	768				&	2636\\
	\hline
	\end{tabular}
	\end{center}	
	\caption{Number of independent radii $\#r=(N+1)(N+2)(N+3)/6$ where $N=L/2$, number of H(3) orbits, number of H(3) orbits which contain more than one point and contribute to the fit formula given by Eq.~(\ref{eq:fit_artefacts}), number of data point used in the fit.}
	\label{tab:fit_info}
\end{table}

That analysis holds as long as periodic images do not contribute: if this is not the case the assumption that the function is in $r$ only breaks down. However, it is still acceptable if the deformation of the tail due to interaction with periodic images is negligible, which is the case at small $r$ in the case of density distributions.

\end{document}